\documentclass[twoside]{article}

\usepackage{PRIMEarxiv}

\usepackage[utf8]{inputenc}
\usepackage[T1]{fontenc}
\usepackage{soul}
\usepackage{enumitem}
\usepackage{hyperref}
\usepackage{url}
\usepackage{booktabs}
\usepackage{amsmath, amssymb, amsfonts}
\usepackage{nicefrac} 
\usepackage{microtype}
\usepackage{fancyhdr} 
\usepackage{graphicx}
\graphicspath{{images/}} 

\usepackage[round, authoryear]{natbib}

\usepackage{pdflscape}
\usepackage{enumitem}
\usepackage{array,tabularx}
\usepackage[table]{xcolor}
\usepackage[most]{tcolorbox}
\usepackage[bf]{caption}
\usepackage{color, colortbl}
\usepackage{longtable}
\newcolumntype{Y}{>{\centering\arraybackslash}X}
\usepackage{cases}

\usepackage{todonotes}



\title{What's the Weight?\\ Estimating Controlled Outcome Differences in Complex Surveys for Health Disparities Research
  \thanks{\textit{\underline{Citation}}: 
  \textbf{Authors. Title. Pages.... DOI:000000/11111.}} 
}

\author{
  Stephen Salerno \\ 
    Division of Public Health Sciences, Biostatistics \\
    Fred Hutchinson Cancer Center \\ 
    Seattle, WA \\ 
    \texttt{ssalerno@fredhutch.org} \\
  \And
  Emily K.~Roberts \\
    Department of Biostatistics \\
    University of Iowa \\
    Iowa City, IA \\
    \texttt{emily-roberts-1@uiowa.edu} \\
  \And
  Belinda L.~Needham \\
    Department of Epidemiology \\
    University of Michigan \\
    Ann Arbor, MI \\
    \texttt{needhamb@umich.edu} \\
  \And
  Tyler H.~McCormick \\
    Department of Statistics \\
    Department of Sociology \\
    University of Washington \\ 
    Seattle, WA \\
    \texttt{tylermc@uw.edu} \\
  \And
  Fan Li \\
    Department of Biostatistics \\
    Department of Cardiovascular Medicine \\
    Yale University \\
    New Haven, CT \\
    \texttt{fan.f.li@yale.edu} \\    
  \And
  Bhramar Mukherjee \\
    Department of Biostatistics \\
    Department of Epidemiology \\
    Department of Statistics and Data Science \\
    Yale University \\
    New Haven, CT \\
    \texttt{bhramar.mukherjee@yale.edu} \\
  \And
  Xu Shi \\
    Department of Biostatistics \\
    University of Michigan \\
    Ann Arbor, MI \\
    \texttt{shixu@umich.edu} \\
}

\begin{document}

\maketitle

\vspace{-4ex}

\begin{abstract}
    In this work, we are motivated by the problem of estimating racial disparities in health outcomes, specifically the average controlled difference (ACD) in telomere length between Black and White individuals, using data from the National Health and Nutrition Examination Survey (NHANES). To do so, we build a {\it propensity for race} to properly adjust for other social determinants while characterizing the controlled effect of race on telomere length. Propensity score methods are broadly employed with observational data as a tool to achieve covariate balance, but how to implement them in complex surveys is less studied -- in particular, when the survey weights depend on the group variable under comparison (as the NHANES sampling scheme depends on self-reported race). We propose identification formulas to properly estimate the ACD in outcomes between Black and White individuals, with appropriate weighting for both {\it covariate imbalance} across the two racial groups and {\it generalizability}. Via extensive simulation, we show that our proposed methods outperform traditional analytic approaches in terms of bias, mean squared error, and coverage when estimating the ACD for our setting of interest. In our data, we find that evidence of racial differences in telomere length between Black and White individuals attenuates after accounting for confounding by socioeconomic factors and utilizing appropriate propensity score and survey weighting techniques. Software to implement these methods and code to reproduce our results can be found in the {\tt R} package {\tt svycdiff}, available through the Comprehensive R Archive Network (CRAN) at \url{cran.r-project.org/web/packages/svycdiff/}, or in a development version on GitHub at \url{github.com/salernos/svycdiff}.
\end{abstract}


\newpage

\section{Introduction}
\label{sec:introduction}

\vspace{-1ex}

There is a deep and broad literature in the United States (US) documenting racial disparities along multiple socioeconomic axes, such as income and education~\citep{needham2013socioeconomic, rehkopf2019impact, zota2021integrating}. These factors are interrelated, making it difficult to separate the effect of race from other sources. Further, these complex relationships between race and socioeconomic status (SES), and the porous collection of such data in the US, make it challenging to draw definitive conclusions about race differences in health outcomes~\citep{li2023using}. Observational studies can offer a wealth of information, but they are often limited by confounding, covariate imbalance, and a lack of representation~\citep{imai2013estimating, pattanayak2011propensity}. Nationally representative surveys, using techniques such as stratified and clustered sampling, allow us to study diverse populations, but generalizing results while accounting for confounding can be difficult due to their complex designs. Our motivating question focuses on the relationship between race, SES, and telomere length, a biomarker for health, using data from the National Health and Nutrition Examination Survey (NHANES): {\it if we could hypothetically balance SES between Black and White individuals in a nationally representative sample, would we still see significant Black/White differences in telomere length?}

This question is statistically complex because the population characteristic of interest (i.e., race) is correlated with SES, and both factors influence the probability of being selected into the sample. For NHANES, the sampling mechanism is partly determined by race and socioeconomic status, with an oversampling of non-Hispanic Black individuals and those at or below 130\% of the poverty level~\citep{curtin2012national}. To address this, we propose a strategy that appropriately accounts for both  confounding and  selection bias when estimating the controlled difference in outcomes between groups, where an individual's probability of selection depends on their group membership and its confounders.

\citet{rosenbaum1983central} proposed using propensity scores to balance the distributions of covariates between comparison groups in the presence of  confounding bias. In traditional causal inference, these comparison groups are analogous to randomized `treatments' or `exposures,' and using propensity scores to reduce confounding bias relies on several assumptions, namely exchangeability of the treatment/exposure, positivity of the treatment/exposure, and the stable unit treatment value assumption. However, certain observational scenarios preclude randomizing individuals due to moral or biological restrictions. These include practical examples such as the timing of transplant waitlisting or the inability to `assign' race to an individual~\citep{peng2017impact, li2023using}. Rather than interpreting differences in outcomes across these groups as causal effects, we seek a controlled descriptive comparison. Propensity scores can still be used to compute what \citet{li2013propensity} call an {\it average controlled difference} (ACD). This quantity achieves covariate balance under the weaker assumption of positivity, but it does not render a causal interpretation to the estimated group differences. We will refer to race as the `group' (rather than `treatment') of interest hereafter.

While \citet{li2023using} showed that propensity scores can be used in this way to study racial disparities, a gap remains in determining the ideal practice when the sampling design is complex. Many studies have discussed using propensity score methods to generalize inference from randomized trials~\citep{stuart2011use, lim2018minimizing, zeng2021propensity}, and complex surveys~\citep{austin2015moving, lee2011weight, dugoff2014generalizing, fuentes2022causal}, however, not when sample selection depends on the group of interest, such as race. For example, \citet{rossen2014social} used propensity scores to assess disparities in weight among US children and adolescents from 2001 to 2010 with data from NHANES. However, there was no consensus on whether to incorporate survey weights when estimating the propensity score and/or outcome model. Our paper addresses this knowledge gap by proposing a strategy for incorporating survey weights from complex survey designs with propensity score methods to achieve covariate balance and generalizability, specifically when survey selection depends on the group of interest. We offer four approaches for estimating the ACD, which rely on multi-stage modeling techniques such as inverse probability weighting (IPW) or direct standardization via `g-methods'~\citep[see][]{richardson2014causal} and only require the assumptions of positivity, selection positivity, and weak selection exchangeability. While we later show that our proposed methods also target a causal quantity such as the  population average treatment effect (PATE) under stronger assumptions, we do not require this for validity of our proposed estimation procedure, as one or more of these assumptions may be violated in observational data settings -- particularly when studying racial differences in health outcomes. 

In this work, we introduce a novel framework that enhances existing approaches for estimating racial disparities in health outcomes in complex survey settings. Our contributions are threefold. First, we derive new identification formulas that provide theoretical insight into the estimation of controlled group differences under conditions of confounding and selection bias, addressing a critical gap in the literature where selection depends on the group of interest, such as race. Second, we present a comprehensive set of simulation settings that consider different data-generating mechanisms and allow for a thorough examination of the proposed methodology's performance under various real-world scenarios. Lastly, we implement these methods within a comprehensive R package, enabling researchers to rigorously and efficiently apply these techniques to their own data. Together, these contributions offer a robust framework for analyzing health disparities using complex survey data.

The rest of this manuscript is structured as follows. Section \ref{sec:background} reviews related work in this area. Section \ref{sec:method} introduces and justifies our proposed methods for estimating the quantity of interest, which we will refer to as the ACD in the spirit of~\citet{li2023using}, and discusses differences with other common approaches. Section \ref{sec:simulations} explores how the proposed methods compare to current analytic strategies and provides simulation studies to support our recommendations. Section \ref{sec:nhanes} introduces the data motivating this work and provides results for this example. We then present an auxiliary example where we estimate the PATE for an environmental exposure. We close with a summary discussion of the concepts outlined here and some final recommendations to researchers who may be working under similar considerations.

\section{Related Work}
\label{sec:background}

To our knowledge,~\citet{zanutto2006comparison} and~\citet{dugoff2014generalizing} were among the first to discuss using propensity score methods to address confounding in observational studies with survey weights. Both authors suggest that, intuitively, since the purpose of propensity scores weighting is to achieve in-sample balance, survey weighting the propensity score is unnecessary. The authors also recommend including survey weights as covariates in the propensity model, drawing a parallel between this intuition to the recommendation by~\citet{lumley2011complex} of including these covariates to account for non-response in surveys. However, this simple inclusion is often problematic in practice, as it may lead to separation in the data or large biases~\citep{ridgeway2015propensity}. \citet{dugoff2014generalizing} also addressed limitations regarding how to handle extreme weights, model misspecification, nonlinear propensity score estimation, and additional model covariates. Such recommendations are largely based on intuition for our particular setup and require more rigorous study. Subsequently,~\citet{ridgeway2015propensity} began to formalize the discussion of propensity score analysis for complex survey designs. The authors demonstrated through proof and simulation that not appropriately accounting for the selection mechanism can potentially lead to biased estimates. A natural question is in what stages of the analysis (i.e., the propensity score and/or outcome model stage) should the survey weights be included? The authors recommended that researchers use survey weights in both stages based on their results. However, the simulation results pertaining to our scenario are inconclusive as to which approach achieves the lowest bias, and the derivations of consistency are not specific to this scenario. Further, they consider only the population average treatment effect on the treated.

Similarly, others have addressed the task of combining causal inference and survey methodology. \citet{austin2018propensity}, for example, focused primarily on propensity score matching techniques in simulation. However, these results did not provide a clear answer with respect to which method of estimating the propensity score model, survey-weighted or not, was best. \citet{dong2020using} explored the performance of different propensity and outcome weighted models and suggested the use of survey weights at both stages, though they concluded the estimates from different strategies were not statistically different in their data example. \citet{lenis2018measuring} further studied the causal inference framework for estimating population versus sample effects and explored the impact of non-response mechanisms with matching techniques. Their simulation setting followed~\citet{austin2018propensity} and concluded that the degree of misspecification impacted their analysis results and population balance. More recently,~\citet{yang2024propensity} conducted simulations, also following the design of~\citet{austin2018propensity}, to compare various propensity score weighting approaches for estimating population-level treatment effects on a binary outcome with survey weighted data. The authors found that two-stage methods, with predicted outcomes weighted by survey weights, outperformed the other methods for estimating the population average treatment effect on the treated, but for population average treatment effects, the best performing method depended on the degree of model misspecification and propensity score overlap. \citet{mccaffrey2024estimating} further extended the use of survey and attrition weights from binary treatment propensity score analyses to the continuous exposure setting. They first derive the conditions under which sampling or nonresponse weights must be incorporated into both the generalized propensity score (GPS) estimation and the subsequent outcome model to obtain unbiased estimates of a continuous dose-response curve. Through an extensive simulation study, they show that while including weights in both stages generally ensures against bias, in some settings (e.g., when the same covariates drive both treatment assignment and sample selection), unbiased estimation is possible without weighting the GPS itself. Conversely, naively omitting weights can lead to bias or inefficiency when selection and/or attrition depend on treatment or covariates. They further demonstrate that, although weighting both stages often reduces bias, it can inflate variance when weights are highly variable. In our setting, the sample selection weights are supplied by NHANES, and race is part of the selection mechanism in the NHANES design, which complicates the relationships underlying our motivating research question. 

\section{Methods}
\label{sec:method}

\subsection{Notation}
\label{sec:notation}

Let $Y$ denote our outcome, here telomere length. Our goal is to estimate the deconfounded, or  controlled difference in mean telomere length between non-Hispanic White and non-Hispanic Black-identifying individuals in the US. To that end, let $\mathcal{A}$ be the discrete set of potential comparison groups of interest, such that $a\in \mathcal{A}$ denotes a specific group. We are interested in comparing the values of $Y$ across levels of $A$. Our motivating analysis departs from the traditional causal literature in that $\mathcal{A}$ represents race, with $a = 1, 0$ denoting Black versus White race, respectively, rather than a potential `treatment,' per se. Given the nature of these variables, our contrast of interest will be viewed as a controlled comparison of the observed outcomes after balancing the two groups with respect to potential confounding variables~\citep{li2013propensity, li2023using}. As the lived experience of Black and White individuals differs significantly in terms of social determinants of health, we believe a {\it controlled descriptive comparison} can help inform the dynamics of these complex relationships~\citep{li2023using}. While race may not be interpreted as a modifiable treatment, the methods described herein are equally applicable in settings where sample selection depends on a treatment in a more traditional sense (see Section \ref{sec:identification}).

\begin{figure*}[t]
    \centering
    \centerline{\includegraphics[width=\textwidth]{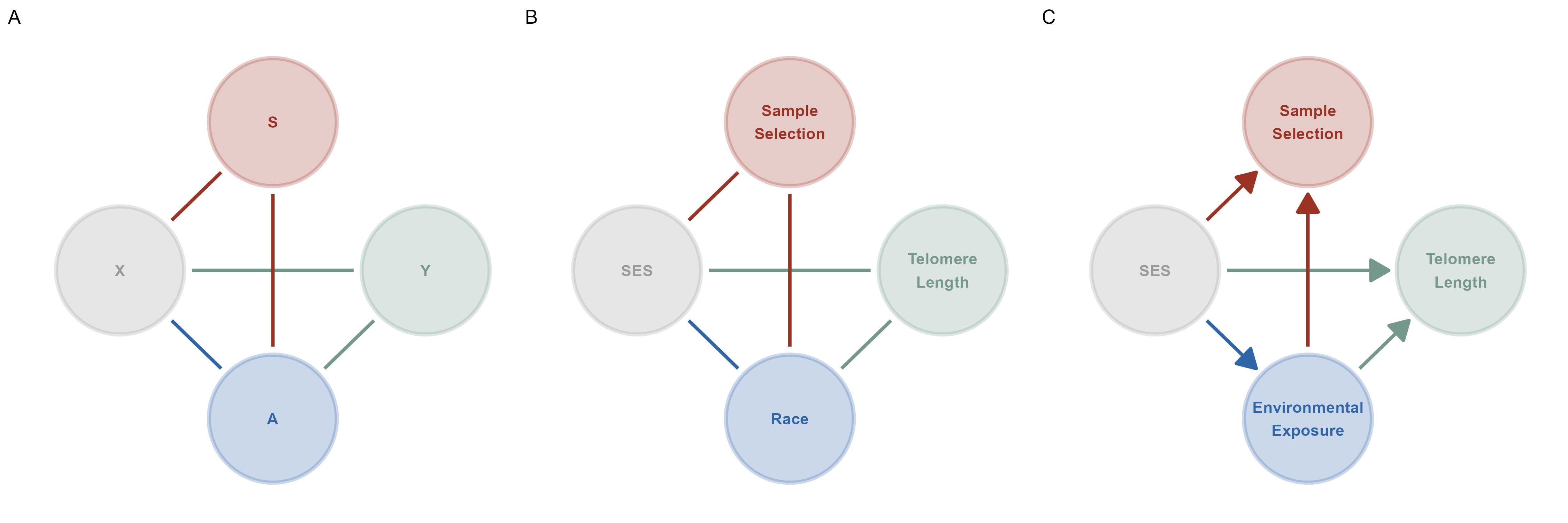}}
    \caption{Conceptual diagrams. (A) Assumed variable relationships: $S$ = sample selection indicator; $A$ = group membership variable of interest; $\boldsymbol{X}$ = covariates; $Y$ = observed outcome. (B) Conceptual diagram for our motivating data example: SES = Socioeconomic Status. (C) Directed acyclic graph for our auxiliary data example.}
    \label{fig:diagram}
\end{figure*}

Our observed data are $\mathcal{O} =\{(Y_i, A_i, \boldsymbol{X}_i); i = 1, \ldots, n\}\overset{\text{i.i.d.}}{\sim} \mathcal{Q} := \Pr(\ \cdot \mid S = 1)$, where $\mathcal{Q}$ is the sample law for units selected into the analytic sample ($S_i = 1$), $S\in \{0,1\}$ indicates selection, and $\boldsymbol{X}\in\mathbb{R}^p$ denotes our (potentially imbalanced) covariates. The underlying population law is $\mathcal{P} := \Pr(\cdot)$ on $N$ individuals. In our motivating data, survey weights, and hence the design probabilities, $\Pr(S = 1 \mid \cdot)$, are known from the sampling plan. There is also the possibility of having missing outcome values among sampled individuals, but this situation will not be explored here and is instead left to future work. Our assumed variable relationships are given in Figure \ref{fig:diagram}. Here our outcome ($Y$), the variable of interest ($A$), and the sample selection indicator ($S$) all depend on covariates ($\boldsymbol{X}$), and sample selection also depends on $A$. Unlike in a causal setting, where $\boldsymbol{X}$ consists of variables which affect both the treatment assignment and the outcome (for example, consider the DAG in the third panel of Figure \ref{fig:diagram}), in the ACD setting, $\boldsymbol{X}$ consists of covariates which are unbalanced across $A$, and will bias the ACD estimate if not properly accounted for. Further, $\boldsymbol{X}$ depends on the research question. That is, $\boldsymbol{X}$ is the set of covariates for which we wish to balance to estimate the controlled difference between outcomes. To some extent, this answers the question of, if we balance the distribution of $\boldsymbol{X}$ between groups, what is the ``residual difference'' in outcomes? Here, if we do not account for the contribution of $\boldsymbol{X}$ to the group differences, then we cannot estimate a balanced, marginal effect of the comparison group membership.

\subsection{Proposed Estimators}
\label{sec:identification}

We wish to draw inference on the population from which our analytic sample was selected. For $a \in \{0,1\}$, we define 

\begin{equation*}
    \mu(a) = \mathbb{E}_{\mathcal{P}}\left[\mathbb{E}(Y \mid A = a, \boldsymbol{X})\right] = \mathbb{E}_{\mathcal{Q}}\left[w(\boldsymbol{X}) g_a(\boldsymbol{X})\right],
\end{equation*}

\noindent where $g_a(\boldsymbol{X}) = \mathbb{E}_{\mathcal{Q}}(Y \mid A = a, \boldsymbol{X})$ and $w(\boldsymbol{X})$ is a known design selection factor. The target of inference, the ACD, is given by ${\rm ACD} = \mu(1) - \mu(0)$. We make the following assumptions:

\begin{enumerate}[label={(\roman*)}]
    \item {\it Positivity:} $e_a^{\mathcal{Q}}(\boldsymbol{X}) := \mathcal{Q}(A = a \mid \boldsymbol{X}) > 0$ $\mathcal{Q}$-a.s. $\forall a\in\mathcal A$
    \item {\it Selection Positivity:} $\pi_a(\boldsymbol X) := \Pr(S = 1 \mid A = a, \boldsymbol{X}) > 0$ on the support of $A, \boldsymbol{X}$
    \item {\it Weak Selection Exchangeability:} $\mathbb{E}_{\mathcal{P}}(Y \mid A = a, \boldsymbol{X}) = \mathbb{E}_{\mathcal{Q}}(Y \mid A = a, \boldsymbol{X}) = g_a(\boldsymbol{X})$.
\end{enumerate}

We require (i) and (ii) to ensure that our observed data functional, $\mu(a)$, is well-defined. Condition (iii) allows us to generalize the conditional mean of the outcome from the study sample to the target population. Specifically, the weak selection exchangeability assumption posits that the expected outcome, conditional on confounders, $\boldsymbol{X}$, and the exposure/group membership variable of interest, $A$, is the same between the sample and the target population. This assumption is less stringent than exchangeability in distribution (i.e, $Y \perp S \mid A, \boldsymbol{X}$), as it does not require the entire conditional distribution of the outcomes to be independent of selection, only their conditional expectations. Under survey sampling, this assumption automatically holds when the survey weights fully encode the (known) design and response adjustments. Further, when the parameter of interest is a conditional quantity, incorporation of the survey weights will not affect the bias or standard error if the model is correctly specified. However, we are interested in ACD, which is a marginal quantity, thus the weighting is needed to standardize to the target population. In other words, even though the survey weights adjust for the sampling design, incorporating them into the ACD is necessary to correctly weight the final estimator to generalize and deconfound estimation. How the survey weights are incorporated depends on the factorization of the joint probability of exposure/group membership and selection.

Depending on how we factorize the joint probability of being sampled $(S = 1)$ and belonging to a particular group $(A = a)$, $\mu(a)$ admits the following equivalent identification formulas under the sample law, $\mathcal{Q}$, given by

\begin{align}
    \mu(a) &= \mathbb{E}_{\mathcal{Q}}\left[\mathbb{E}_{\mathcal{Q}}(Y \mid A = a, \boldsymbol{X}) \cdot \frac{\Pr(S = 1)}{\Pr(S = 1 \mid \boldsymbol{X})}\right] = \mathbb{E}_{\mathcal{Q}}\left[w(\boldsymbol{X}) g_a(\boldsymbol{X})\right],\quad \text{or}
    \label{eq:id1}  \\
    \mu(a) &= \mathbb{E}_{\mathcal{Q}}\left[\frac{\mathbb{I}(A = a) Y \Pr(S = 1)}{\Pr(S = 1, A = a \mid \boldsymbol{X})} \right],      
    \label{eq:id2}
\end{align}

\noindent where $w(\boldsymbol{X}) = \tfrac{\Pr(S = 1)}{\Pr(S = 1 \mid \boldsymbol{X})} = \tfrac{\Pr(S = 1)}{\pi(\boldsymbol{X})}$, $\mathbb{I}(\cdot)$ is the indicator function, and Equation \eqref{eq:id2} is equivalent to the following two factored forms:

\begin{align}
    \mu(a) &= \mathbb{E}_{\mathcal{Q}}\left[\frac{\mathbb{I}(A = a) Y}{\mathcal{P}(A = a \mid \boldsymbol X)} \cdot \frac{\Pr(S = 1)}{\Pr(S = 1 \mid A = a, \boldsymbol{X})}\right] = \mathbb{E}_{\mathcal{Q}}\left[w_a(\boldsymbol{X})\frac{\mathbb{I}(A = a) Y}{e_a^{\mathcal{P}}(\boldsymbol{X})}\right],\quad \text{or}      
    \label{eq:id2a} \\
    \mu(a) &= \mathbb{E}_{\mathcal{Q}}\left[\frac{\mathbb{I}(A = a) Y}{\mathcal{Q}(A = a \mid \boldsymbol{X})}\cdot \frac{\Pr(S = 1)}{\Pr(S = 1 \mid \boldsymbol{X})}\right] = \mathbb{E}_{\mathcal{Q}}\left[w(\boldsymbol{X})\frac{\mathbb{I}(A = a) Y}{e_a^{\mathcal{Q}}(\boldsymbol{X})}\right],   
    \label{eq:id2b}
\end{align}

\noindent where $e_a^{\mathcal{Q}}(\boldsymbol{X}) := \mathcal{Q}(A = a \mid \boldsymbol{X})$ is the within-sample propensity, $e_a^{\mathcal{P}}(\boldsymbol{X}) := \mathcal{P}(A = a \mid \boldsymbol{X})$ is the population propensity, and $w_a(\boldsymbol{X}) = \tfrac{\Pr(S = 1)}{\Pr(S = 1 \mid A = a, \boldsymbol{X})}$. We refer to \eqref{eq:id1}, \eqref{eq:id2a}, and \eqref{eq:id2b} as Outcome Modeling (OM), Inverse Probability Weighting 1 (IPW1), and IPW2, respectively. For the full derivation of \eqref{eq:id1} - \eqref{eq:id2b}, see Appendix A.1.

The assumptions required to estimate the ACD are relatively weaker than those required to identify a causal quantity. Figure \ref{fig:diagram}, Panel (C) gives an example of a directed acyclic graph (DAG) where a causal target may be warranted, such as the case where $A$ is an environmental exposure, like lead, which can also alter telomere length. We provide a supplementary analysis in which we estimate the causal effect of lead exposure on telomere length for illustration, as a practitioner may also be interested in such causally interpretable estimated differences. If the group variable under study is a treatment or exposure, we show under stronger assumptions that Formulas \eqref{eq:id1} - \eqref{eq:id2b} can equivalently target the population potential outcome means, denoted by $\mathbb{E}_\mathcal{P}[Y^{a}]\ \forall a\in\mathcal{A}$. Along with the previous assumptions (i) and (ii), the following conditions,
\begin{enumerate}[label={(\roman*)}]
    \setcounter{enumi}{3}
    \item[(iii')] {\it Weak Selection Exchangeability:} $\mathbb{E}_\mathcal{P}[Y^a \mid A = a, \boldsymbol{X}] = \mathbb{E}_\mathcal{Q}[Y^a \mid A = a, \boldsymbol{X}]$
    \item {\it Stable Unit Treatment Value Assumption (SUTVA):} $Y_i = Y_i^{a}\ \forall i,\ A_i = a \in \mathcal{A}$ 
    \item {\it Weak Exchangeability:} $\mathbb{E}_\mathcal{P}[Y^a \mid \boldsymbol{X}] = \mathbb{E}_\mathcal{P}[Y^a \mid A = a, \boldsymbol{X}]$,
\end{enumerate}
are sufficient to identify $\mathbb{E}_\mathcal{P}[Y^{a}]$. SUTVA underpins the existence of the counterfactual outcomes, $Y^a$. The compact representation in (iv) captures two aspects of SUTVA -- {\it no interference}, which states that the $i$th observation's outcome is unaffected by the treatment of other observations and {\it consistency}, which states that the potential outcome is well-defined for each treatment level, $a$~\citep{rubin1980randomization}. The weak exchangeability assumption allows us to represent the conditional distribution of the unobserved potential outcome using that of the observed potential outcome. As before, we only require a weaker version of this assumption, though {\it strong exchangeability}, i.e., $Y^{a} \perp A \mid \boldsymbol{X}$, is often assumed. For additional details and the full identification proof, see Appendix A.2. This is thematically similar to~\citet{dahabreh2020extending}, who study the generalization of causal inferences from individuals in randomized trials. The authors also differentiate between mean exchangeability in $A$ for the randomized population and mean exchangeability in $S$ for the target population in the context of randomized trials.

\subsection{Comparisons to Existing Approaches}

Formulas \eqref{eq:id1} - \eqref{eq:id2b} differ from standard approaches one might take in practice, which warrants a brief discussion here. Table \ref{tab:methods} compares our proposed approaches to such existing methods. If our assumed conceptual diagram holds, there are two sources of bias: bias from covariate imbalance between comparison groups (i.e., {\it confounding bias}) and unequal selection induced by the design of the complex survey (i.e., {\it selection bias}), which impacts the generalizability of our within-sample results to a larger, target population. Existing approaches may use balancing and/or generalizability weights to address these biases, but omitting either can result in biased estimates of our target quantity. For instance, the effect of race can be either exaggerated or attenuated depending on the direction of the true association between confounders, race, and the outcome in simple regression~\citep[Approach 1 in Table \ref{tab:methods}; ][]{williams2019assumptions}. Multiple regression (Approach 2) adjusts for covariates but will suffer from bias due to model misspecification unless the outcome model includes, for example, appropriate interaction terms~\citep{rao1971some}. With inverse probability weighting (Approach 3), the race-covariate relationship is modeled, rather than the covariate-outcome relationship, foregoing the need to specify any race $\times$ covariate interactions~\citep{mansournia2016inverse}. However, with all three of these approaches, the target of inference is the within-sample difference in outcomes, which cannot be generalized to a larger population if a probability sample was taken~\citep{shi2023data,degtiar2023review}. Survey-weighted regression (Approach 4) accounts for selection bias by accounting for how many individuals in the population that one individual in the sample represents~\citep{lumley2017fitting}, allowing for generalizability, but will be biased if there is covariate imbalance~\citep{dugoff2014generalizing}.

\begin{table*}[!ht]%
    \tiny
    \caption{Comparison of existing approaches for analysis to our proposed approach. Dashes (--) denote that the model is not used.}
    \label{tab:methods}
    \begin{tabular*}{\textwidth}{@{\extracolsep\fill}lccccc@{}}
        \toprule
        & \multicolumn{3}{@{}c}{\textbf{Model}} & \multicolumn{2}{@{}c}{\textbf{Bias Being Addressed}} \\
        \cmidrule{2-4} \cmidrule{5-6}
        \textbf{Approach} & \textbf{Outcome}  & \textbf{Propensity}  & {\textbf{Selection}}  & \textbf{Confounding} & \textbf{Selection} \\
        \midrule
        \multicolumn{6}{@{}l}{\textbf{Existing Approaches}} \\
        \midrule
        \hphantom{0}1.~Simple Regression                                   & $\mathbb{E}[Y \mid A, S = 1]$    & --                         & --                                  & No  & No  \\
        \hphantom{0}2.~Multiple Regression                                 & $\mathbb{E}[Y \mid A, \boldsymbol{X}, S = 1]$ & --                         & --                                  & No  & No  \\
        \hphantom{0}3.~IPTW Estimator$^{{\bf a}}$                    & --                               & $\Pr(A = 1 \mid \boldsymbol{X}, S = 1)$ & --                                  & Yes & No  \\
        \hphantom{0}4.~Survey-Weighted Multiple Regression                 & $\mathbb{E}[Y \mid A, \boldsymbol{X}]$        & --                         & $\Pr(S = 1 \mid A, \boldsymbol{X})$              & No  & Yes \\
        \hphantom{0}5.~IPTW Multiple Regression                            & $\mathbb{E}[Y \mid A, \boldsymbol{X}, S = 1]$ & $\Pr(A = 1 \mid \boldsymbol{X}, S = 1)$ & --                                  & Yes & No  \\
        \hphantom{0}6.~IPTW + Survey-Weighted Multiple Regression          & $\mathbb{E}[Y \mid A, \boldsymbol{X}]$        & $\Pr(A = 1 \mid \boldsymbol{X}, S = 1)$ & $\Pr(S = 1 \mid A, \boldsymbol{X})$              & Yes & Yes \\
        \hphantom{0}7.~Weighted IPTW + Survey-Weighted Multiple Regression & $\mathbb{E}[Y \mid A, \boldsymbol{X}]$        & $\Pr(A = 1 \mid \boldsymbol{X})$        & $\Pr(S = 1 \mid A, \boldsymbol{X})$              & Yes & Yes \\
        \midrule
        \multicolumn{6}{@{}l}{\textbf{Proposed Approaches}} \\
        \midrule
        \hphantom{0}8.~Outcome Modeling and Direct Standardization         & $\mathbb{E}[Y \mid A, \boldsymbol{X}, S = 1]$ & --                         & $\Pr(S = 1 \mid \boldsymbol{X})^{{\bf b}}$ & Yes & Yes \\
        \hphantom{0}9.~Inverse Probability Weighting 1                     & --                               & $\Pr(A = 1 \mid \boldsymbol{X})$        & $\Pr(S = 1 \mid A, \boldsymbol{X})$              & Yes & Yes \\
        10.~Inverse Probability Weighting 2                                & --                               & $\Pr(A = 1 \mid \boldsymbol{X}, S = 1)$ & $\Pr(S = 1 \mid \boldsymbol{X})$                 & Yes & Yes \\
        11.~Augmented Inverse Probability Weighting                        & $\mathbb{E}[Y \mid A, \boldsymbol{X}, S = 1]$ & $\Pr(A = 1 \mid \boldsymbol{X}, S = 1)$ & $\Pr(S = 1 \mid \boldsymbol{X})$                 & Yes & Yes \\
        \bottomrule
    \end{tabular*}
    $^{\rm a}$ IPTW: Inverse Probability of Treatment Weighting. \\
    $^{\rm b}$ $\Pr(S = 1 \mid \boldsymbol{X}) = \Pr(S = 1 \mid A = 1, \boldsymbol{X})\Pr(A = 1 \mid \boldsymbol{X}) + \Pr(S = 1 \mid A = 0, \boldsymbol{X})\Pr(A = 0 \mid \boldsymbol{X})$.
\end{table*}

In our research question, we have an intersecting consideration of both selection and confounding bias; thus, neither source of bias should be disregarded. A more principled approach would be one that accounts for both confounding and selection bias, as given by Approaches 6 and 7, which utilize a combination of propensity score and survey weighting. These approaches, advocated by \citet{zanutto2006comparison}, \citet{dugoff2014generalizing}, and \citet{ridgeway2015propensity} are considered the current analytic standard. They suggest survey-weighted generalizations of IPTW, either at the propensity model stage and/or the outcome model stage. Although it may seem that either extension addresses both sources of bias, there is little theoretical justification for how weighting one or multiple models within IPTW achieves both tasks in our setting. Additionally, propensity scores are often seen as in-sample balancing scores~\citep{zanutto2006comparison, dugoff2014generalizing}. From this perspective, their generalizability (and corresponding selection bias correction based on survey weights) is often ignored. Survey-weighting both the propensity score and outcome model is reasonable~\citep{ridgeway2015propensity}, however directly combining survey-weighted IPWs and survey weights in a second stage outcome regression may also lead to biased results. 

From a practical standpoint, one question is: `should we implement survey-weighted propensity score estimation in IPTW or not?' Our analytic results show that for estimation of the ACD, the propensity score does not have to be survey-weighted. We see this in Equations \eqref{eq:id2a} and \eqref{eq:id2b}, which differ with respect to the factorization of the joint distribution into propensity and survey weight models. If the propensity score is not survey-weighted, based on Equation \eqref{eq:id2b}, we need $\Pr(S = 1 \mid \boldsymbol{X})$, i.e., one may need to marginalize $\Pr(S=1\mid A,\boldsymbol{X})$ over the group membership variable, $A$. If survey weights are included in the propensity score estimation, then we must know $\Pr(S = 1 \mid A, \boldsymbol{X})$, i.e., the probability of selection conditional on each level of the group membership variable and the covariates. 

Moreover, our proposed approaches differ from each other and prove useful in different settings. Formula \eqref{eq:id1}, outcome modeling and direct standardization (OM), requires knowledge of $f_X(x)$, as an outcome regression, $g_a(\boldsymbol{X})$, is fit in practice. Formulas \eqref{eq:id2a} and \eqref{eq:id2b}, IPW1 and IPW2, do not require estimation of an outcome model and may be the best options depending on the level of information available in practice. We view IPW1 and IPW2 as simple corrections to what one might do in practice, namely the traditional inverse probability of treatment weighted (IPTW) estimator, and thus may be a better starting point for an analysis of this type as they additionally account for selection bias. While IPW1 does align better with the NHANES survey weighting structure, IPW1 versus IPW2 show that the incorporation of the survey weights into each stage of modeling depends on two alternate factorizations for the joint distribution of being sampled and belonging to the exposure/group of interest. The choice of IPW1 versus IPW2 is also driven by design, as IPW1 is better suited for retrospective designs, where IPW2 is better suited for prospective designs. Further, IPW1 and IPW2 may be more robust if the outcome model for \eqref{eq:id1} is likely misspecified, however, the additional modeling of the two propensity score models for IPW2 may result in efficiency loss. While OM, IPW1, and IPW2 are all consistent, OM is most efficient if correctly specified. Thus, depending on the availability of the data or the ability to estimate certain key quantities, a researcher may prefer one approach over the others. A doubly robust estimator, by combining both an outcome model and an inverse‐probability weighting component, requires only one of these two models to be correctly specified in order to yield a consistent ACD estimate (see Section \ref{sec:dr}). In practice, this flexibility eliminates the need for a choice strictly between IPW1 and IPW2. If either the propensity model or the outcome model are well-specified, the doubly robust approach remains consistent and often achieves smaller variance than either IPTW estimator alone. For these reasons, we generally recommend the doubly robust estimator as a practical default.

Regarding the incorporation of propensity and sample selection weights, our weights differ from those in traditional g-computation or IPTW frameworks. We derived a new estimator with weights for both within-sample inference and generalizability to our target population. Specifically, note the term $\Pr(S = 1 \mid \boldsymbol{X})$ in the denominators of Formulas \eqref{eq:id1} and \eqref{eq:id2b}. If selection does not depend on $A$, identification simplifies to the standard IPTW identification formula. However, in our setting, selection depends on the participant's self-reported race, requiring the probability of selection to be marginalized over $A$, where 
\[
  \Pr(S = 1 \mid \boldsymbol{X}) = \Pr(S = 1 \mid A = 1, \boldsymbol{X})\Pr(A = 1 \mid \boldsymbol{X}) + \Pr(S = 1 \mid A = 0, \boldsymbol{X})\Pr(A = 0 \mid \boldsymbol{X}).
\]
As with OM, IPW2 uses the marginalized $\Pr(S = 1 \mid \boldsymbol{X})$ above, rather than directly modeling $\Pr(S = 1 \mid \boldsymbol{X})$ by omitting $A$~\citep{wilms2021omitted}. Thus, IPW2 requires that both the propensity model be fit without incorporating selection weights, i.e., $e_a^\mathcal{Q}(\boldsymbol{X})$, for inverse probability weighting, and the survey-weighted propensity model be fit to calculate $\Pr(S = 1 \mid \boldsymbol{X})$. In contrast, IPW1 requires the selection mechanism be known (e.g., in the case of survey weights), conditional on $A$, i.e., $\Pr(S = 1 \mid A = a, \boldsymbol{X})$, and that a survey-weighted propensity model be fit, i.e., $e_a^\mathcal{P}(\boldsymbol{X})$. The key difference here comes with respect to how the joint propensity and selection mechanisms are factorized.

\subsection{Estimation}
\label{sec:estimation}

To estimate the ACD based on the three formulas presented above (OM, IPW1, and IPW2), we propose the following. Let $\hat{g}_a(\boldsymbol{X})$ be an estimator of $g_a(\boldsymbol{X}) = \mathbb{E}_\mathcal{Q}(Y \mid A = a, \boldsymbol{X})$. In practice, $\hat{g}_a(\boldsymbol{X})$ can be estimated, for example, using a parametric outcome model with a finite-dimensional parameter vector. Our outcome model-based estimator for $\mu(a)$ represented in Equation \eqref{eq:id1} is 
\begin{equation}
\label{eq:id1_est}
    \hat{\mu}_{\rm OM}(a) = \frac{1}{n}\sum_{i = 1}^n w(\boldsymbol{X}_i) \hat{g}_a(\boldsymbol{X}_i).
\end{equation}
When the outcome model is correctly specified, $\hat{g}_a(\boldsymbol{X}) \overset{p}{\rightarrow} g_a(\boldsymbol{X})$, leading to the result that $\hat{\mu}_{\rm OM}(a) \overset{p}{\rightarrow} \mu(a)$. Alternatively, IPW1 and IPW2 rely on inverse probability (IP) weighting to control for confounding~\citep{horvitz1952generalization}. Let $\hat{e}^\mathcal{Q}_a(\boldsymbol{X})$ denote the within-sample propensity model, which provides an estimator for $e_a^\mathcal{Q}(\boldsymbol{X})$. Similarly, let $\hat{e}^\mathcal{P}_a(\boldsymbol{X})$ denote the survey-weighted estimator of $e_a^\mathcal{P}(\boldsymbol{X})$. These are both estimable, as we assume a probability sample with known survey weights. A common approach may be to model the propensity via logistic regression, though many non-parametric approaches have been proposed~\citep{lee2010improving}. Depending on how we factorize the joint probability of sample selection and group membership, we can write the IPW estimators as
\begin{align}
    \hat{\mu}_{\rm IPW1}(a) &= \frac{1}{n}\sum_{i = 1}^n w_a(\boldsymbol{X}_i) \frac{\mathbb{I}(A_i = a) Y_i}{\hat{e}^\mathcal{P}_a(\boldsymbol{X}_i)}, \quad {\rm or} \\[2ex]
    \hat{\mu}_{\rm IPW2}(a) &= \frac{1}{n}\sum_{i = 1}^n w(\boldsymbol{X}_i)\frac{\mathbb{I}(A_i = a) Y_i}{\hat{e}^\mathcal{Q}_a(\boldsymbol{X}_i)}.
\end{align}
If the propensity model is correct, that is $\hat{e}^\mathcal{P}_a(\boldsymbol{X}) \pi_a(\boldsymbol{X}) \overset{p}{\rightarrow} \Pr(S = 1, A = a \mid \boldsymbol{X})$ and $\hat{e}^\mathcal{Q}_a(\boldsymbol{X}) \pi(\boldsymbol{X}) \overset{p}{\rightarrow} \Pr(S = 1, A = a \mid \boldsymbol{X})$, then $\hat{\mu}_{\rm IPW1}(a)$ and $\hat{\mu}_{\rm IPW2}(a)$ are consistent.

\subsubsection{Augmented Inverse Probability Weighted Estimator}
\label{sec:dr}

Following \citet{dahabreh2020extending}, we can also derive the efficient influence function (EIF) for our setting. Specifically, the EIF is the canonical gradient for the parameter $\mu(a)$. Here, the EIF for $\mu(a)$ is given by
\[
    \phi_a(Y, A, \boldsymbol{X}) = \frac{\Pr(S = 1)}{\Pr(S = 1 \mid \boldsymbol{X})}\left\{\frac{\mathbb{I}(A=a)}{e^\mathcal{Q}_a(\boldsymbol{X})}\left[Y - g_a(\boldsymbol{X})\right] + g_a(\boldsymbol{X})\right\} - \mu(a) = w(\boldsymbol{X})\left\{\frac{\mathbb{I}(A=a)}{e^\mathcal{Q}_a(\boldsymbol{X})}\left[Y - g_a(\boldsymbol{X})\right] + g_a(\boldsymbol{X})\right\} - \mu(a).
\]
This also provides us the form of the doubly robust estimator, given by
\begin{equation}
\label{eq:dr_est}
    \hat{\mu}_{DR}(a) = \frac{1}{n} \sum_{i=1}^{n} w(\boldsymbol{X}_i)\left\{\hat{g}_a(\boldsymbol{X}_i) + \frac{\mathbb{I}(A_i = a)}{\hat{e}^\mathcal{Q}_a(\boldsymbol{X}_i)}\left[Y_i - \hat{g}_a(\boldsymbol{X}_i)\right]\right\},
\end{equation}
where $\hat{g}_a(\boldsymbol{X}_i)$ is an estimator for $g_a(\boldsymbol{X})$ and $\hat{e}^\mathcal{Q}_a(\boldsymbol{X}_i)$ is an estimator for the within-sample propensity score. For the ACD comparing groups 1 and 0, the final estimator is given by
\[
    \widehat{ACD}_{DR}=\hat{\mu}_{DR}(1)-\hat{\mu}_{DR}(0).
\]
We provide the full derivation and proof in Appendix A.3.

\subsubsection{Estimation of Uncertain Sample Selection or Non-Probability Samples}
\label{sec:nonprob}

Throughout this work, we are motivated by settings where the sample arises from a known probability survey design. That is, we know $\pi_a(\boldsymbol{X}) = \Pr(S = 1 \mid A = a, \boldsymbol{X})$. In practice, however, the sample selection mechanism may be unknown. For example, in non-probability samples, such as those arising from the study of electronic health records, we must estimate $\pi_a(\boldsymbol{X})$ with an assumed model \citep[(e.g., through beta regression, as recommended by ][]{elliot2009combining}. In these settings, we replace $\pi_a(\boldsymbol{X})$ by a consistent estimator, $\hat{\pi}_a(\boldsymbol{X})$. Further, we can replace $\pi(\boldsymbol{X}) = \Pr(S = 1 \mid \boldsymbol{X})$, the sample selection probability marginalized over $A$, by $\hat{\pi}(\boldsymbol{X})$, and $\Pr(S = 1)$ by its estimate, $\hat{\pi} = n/N$. Here, our asymptotic results hold if the propensity and selection models are {\it both} correctly specified, then we have that $\hat{e}^\mathcal{P}_a(\boldsymbol{X}) \hat{\pi}_a(\boldsymbol{X}) \overset{p}{\rightarrow} \Pr(S = 1, A = a \mid \boldsymbol{X})$ and $\hat{e}^\mathcal{Q}_a(\boldsymbol{X}) \hat{\pi}(\boldsymbol{X}) \overset{p}{\rightarrow} \Pr(S = 1, A = a \mid \boldsymbol{X})$, leading to $\hat{\mu}(a)$ being consistent estimators of $\mu(a)$. As consistency depends on correctly specifying the selection model, alternative approaches, such as simplex regression or nonparametric methods, may provide more robust estimates~\citep{kundu2023framework, salvatore2024weight}. If selection depends on additional, unmeasured covariates, $U$, that do not affect the group membership or the outcome, then our procedure remains unbiased, but we may underestimate the analytic standard errors. For inference, we must account for this additional step to obtain correct standard errors (see below).

\subsection{Inference}
\label{sec:inference}

Inference can be carried out analytically via Taylor linearization or numerically by recognizing our estimators as M-estimators and using the empirical sandwich estimator~\citep{stefanski2002calculus, boos2013essential}. As before, let $\mathcal{Q}$ denote the sample law for $\mathcal{O} = (Y, A, \boldsymbol{X})$ with $O_i \overset{\rm i.i.d.}{\sim} \mathcal{Q}(S_i = 1)$, and let $\hat{\mathcal{Q}_n}$ be its empirical counterpart based on a sample of $n$ observations. Each estimator $\hat{\boldsymbol{\theta}} = \boldsymbol{\theta}(\hat{\mathcal{Q}_n})$ solves a system of unbiased estimating equations
\[
    \sum_{i = 1}^n \psi(O_i; \boldsymbol{\theta}) = 0,
\]
where $\psi(O; \boldsymbol{\theta})$ stacks the estimating functions for the nuisance models (e.g., outcome and/or propensity, and selection where applicable; see Section \ref{sec:nonprob}) and the target components, and $\boldsymbol{\theta} = \boldsymbol{\theta}(\mathcal{Q})$ denotes the probability limit under $\mathcal{Q}$. Under standard regularity conditions, $\hat{\boldsymbol{\theta}}$ is asymptotically linear,
\[
    \hat{\boldsymbol{\theta}} - \boldsymbol{\theta}(\mathcal{Q}) = \frac{1}{n}\sum_{i = 1}^n \phi(O_i) + o_p(n^{-1/2}),
\]
with influence function
\[
  \phi^{\mathcal{Q}}(O) = -M^{-1}\psi[O; \boldsymbol{\theta}(\mathcal{Q})]; \quad M = -\mathbb{E}_\mathcal{Q}[\nabla_\theta \psi(O; \boldsymbol{\theta})]_{\boldsymbol{\theta} = \boldsymbol{\theta}(\mathcal{Q})}.
\]
Here, $M$ is the Jacobian matrix of the estimating equations~\citep{huber1996robust}. The population asymptotic covariance matrix is then given by
\[
    V(\theta) = \frac{1}{n}\Sigma(\theta); \quad \Sigma(\theta) = \mathbb{E}_{\mathcal{Q}}\left[\left(\phi^{\mathcal{Q}}\right)\left(\phi^{\mathcal{Q}}\right)^\top\right] = \left[M(\theta)^{-1}\right]B(\theta)\left[M(\theta)^{-1}\right]^\top; \quad B(\theta) = \mathbb{E}_{\mathcal{Q}}\left[\psi\psi^\top\right].
\]
To estimate this quantity, we define $\hat{\psi}_i = \psi(O_i; \hat{\theta})$, $\hat{M} = \frac{1}{n} \sum_{i=1}^n \nabla_{\theta} \psi(O_i; \hat{\theta})$, and $\hat{\phi}_i = -\hat{M}^{-1} \hat{\psi}_i$. A consistent estimator of influence function covariance is then given by
\[
    \hat{\Sigma}_n = \frac{1}{n}\sum_{i=1}^n \hat{\phi}_i \hat{\phi}_i^\top = \left[\hat{M}^{-1}\right]\left\{\frac{1}{n}\sum_{i=1}^n \hat{\psi}_i\hat{\psi}_i^\top\right\}\left[\hat{M}^{-1}\right]^\top.
\]
Consequently, the estimated covariance matrix of $\hat{\theta}$ is
\[
    \hat{V}(\hat{\theta}) = \frac{1}{n}\hat{\Sigma}_n = \frac{1}{n^2} \sum_{i=1}^n \hat{\phi}_i \hat{\phi}_i^\top = \frac{1}{n} \left[\hat{M}^{-1}\right]\left\{\frac{1}{n} \sum_{i=1}^n \hat{\psi}_i \hat{\psi}_i^\top\right\}\left[\hat{M}^{-1}\right]^\top.
\]
When an explicit $\mathcal{Q}$-law EIF is available (e.g., in our DR construction), we may compute $\hat{\phi}_i = \phi^{\mathcal{Q}}(O_i; \hat{\theta})$ directly and use the outer-product form $\hat{V}(\hat{\theta}) = \tfrac{1}{n^2}\sum_i \hat{\phi}_i \hat{\phi}_i^\top$.

\subsubsection{Inference in Stratified, Clustered Multi-Stage Survey Samples}
\label{ref:nhanesinference}

NHANES employs a complex stratified, clustered sampling design. While bias incurred from the differential sample selection probabilities (encoded in the survey weights) is accounted for in the estimation procedure, inference is affected by the fact that individuals within strata and primary sampling units (PSUs) are correlated. To account for  stratified, clustered multistage designs, we employ the usual cluster-robust version by replacing individual sums with primary sampling unit (PSU) sums within strata and applying a finite PSU correction. That is, we modify the sandwich estimator by summing the contributions of the $i$th individual $(i = 1, \ldots, n_{jk})$, nested in the $j$th primary sampling unit (PSU; $j = 1, \ldots, J_k$), nested in the $k$th sampling stratum $(k = 1, \ldots, K)$, with finite PSU correction, $\tfrac{J_k}{J_k - 1}$. We write $O_{ijk} = (Y_{ijk}, A_{ijk}, \boldsymbol{X}_{ijk})$ for units in PSU $j$, stratum $k$, with $n_{jk}$ observations and total analytic sample size $n = \sum_{k} \sum_{j} n_{jk}$. We further define the PSU total of influence function contributions as $\hat{U}_{jk} = \sum_{i=1}^{n_{jk}} \hat{\phi}^{\mathcal{Q}}(O_{ijk}; \hat{\theta})$ and its stratum mean as $\bar{U}_k = \tfrac{1}{J_k} \sum_{j=1}^{J_k} \hat{U}_{jk}$. Then, the estimated covariance matrix of $\hat{\theta}$ is given by
\begin{equation}
\label{eq:vcov_strat}
    \hat{V}(\hat{\theta}) = \frac{1}{n^2} \sum_{k=1}^{K} \frac{J_k}{J_k - 1} \sum_{j=1}^{J_k} \left(\hat{U}_{jk} - \bar{U}_k\right)\left(\hat{U}_{jk} - \bar{U}_k\right)^\top.
\end{equation}

\subsection{Software}

Software for these methods can be found in the {\tt R} package {\tt svycdiff}, available through the Comprehensive R Archive Network (CRAN) at \url{cran.r-project.org/web/packages/svycdiff}, or in a development version on GitHub at \url{github.com/salernos/svycdiff}. As our proposed approach uses an estimating equations framework, the {\tt svycdiff} accommodates various outcome types, including continuous, binary, and count outcomes. This includes the main functions for estimation and inference, as well as a function for simulating population-level data under a wide range of scenarios. Moreover, these accommodate the setting described in Section \ref{sec:nonprob}. Lastly, the codes necessary to reproduce our later simulations and data analysis are also available at the GitHub link.

\section{Simulation Studies}
\label{sec:simulations}

We conduct a series of simulations based on the setup of \citet{austin2018propensity}, and subsequently \citet{zeng2025moving}, to understand the finite sample performance of our proposed estimators compared to those typically used in practice. We systematically vary the associations between $A$, $\boldsymbol{X}$, and $S$ in our data-generating mechanism to understand how these relationships impact estimation and inference. We quantify the bias and variance of our proposed approach compared to existing approaches.

\subsection{Data Generation}
\label{sec:simdat}

We first generate population-level data for $N = 1,000,000$ individuals. We organize the individuals into 10 strata $(j = 1, \ldots, 10)$ that are further subdivided into 20 clusters, so that there are 200 unique clusters $(k = 1, \ldots, 200)$ and 5,000 individuals per cluster $(i = 1, \ldots, 5,000)$. For each individual, we generate two baseline covariates $(l =1,2)$ from normal distributions, allowing the means of these covariates to vary at both the stratum and cluster levels. That is, we have
\[
    X_{l,ijk} \sim \mathcal{N}\left(\nu_{l,j}^s + \nu_{l,j}^c, 1\right); \quad \nu_{l,j}^s \sim \mathcal{N}\left(0, \sigma_l^s\right), \quad \nu_{l,j}^c \sim \mathcal{N}\left(0, \sigma_l^c\right),
\]
where $\nu_{l,j}^s$ and $\nu_{l,j}^c$ are stratum- and cluster-specific random effects for the $l$th covariate. We set $\sigma_l^s = 0.35$ and $\sigma_l^c = 0.15$ across all simulation settings. Note that $\boldsymbol{X}_i = \{X_{1,i}, X_{2,i}\}$ relate to our outcome, $Y_i$, group variable of interest, $A_i$, and sample selection indicator, $S_i$. We then generate $A_i \in \{0,1\}$ from a logistic model, so the propensity model for $A_i = 1$ is given by
\[
    \Pr(A_i = 1 \mid \boldsymbol{X}_i) = {\rm logit}^{-1}\left[\tau_0 + \tau_A \left(\tau_{X_1} X_{1,i} + \tau_{X_2} X_{2,i} + \tau_{X_1X_2} X_{1,i}X_{2,i}\right)\right],
\]
where ${\rm logit}^{-1}(u) = [1 + \exp\{-u\}]^{-1}$ is the inverse logit function. Here, $\tau_0$ controls the overall group prevalence, $\tau_A$ reflects the strength of confounding and, correspondingly, the degree of overlap in the propensity score, $\tau_{X_1}$ and $\tau_{X_2}$ are coefficients for $X_{1}$ and $X_{2}$, respectively, and $\tau_{X_1X_2}$ an interaction effect between $X_{1}$ and $X_{2}$. For this main simulation, we fix $\tau_0 = 0$, $\tau_A = 0.5$, and $\tau_{X_1X_2} = 0$, and we vary $\tau_{X_1}$ and $\tau_{X_2} \in \{0,1\}$. We then simulate the $A_i$ from a Bernoulli distribution with $A_i \sim {\rm Bernoulli}[\Pr(A_i = 1 \mid \boldsymbol{X}_i)]$. We next generate a survey weight for each individual, $\omega_i$, by specifying a logistic selection model, with
\[
    \Pr(S_i = 1 \mid A_i, \boldsymbol{X}_i) = {\rm logit}^{-1}(\beta_{0} + \beta_{A} A_i + \beta_{X_1} X_{1,i} + \beta_{X_2} X_{2,i} + \varepsilon_{S});\quad \varepsilon_{S} \sim \mathcal{N}(0, 0.1),
\]
where $\beta_{0}$ controls the marginal sampling fraction, $\beta_{A}$, $\beta_{X_1}$, and $\beta_{X_2}$ are coefficients for $A$, $X_{1}$ and $X_{2}$, respectively, and $\varepsilon_{S}$ is small amount of random noise added to the generation of the survey weights, to support later sensitivity analyses. We fix $\beta_{0} = -8$ and vary $\beta_{A}$, $\beta_{X_1}$, and $\beta_{X_2} \in \{0,1\}$, and we set $\omega_i = \Pr(S_i = 1 \mid A_i, \boldsymbol{X}_i)^{-1}$. Lastly, we simulate continuous outcomes, $Y_i$, for each individual in the population as a linear combination of $A_i$, $\boldsymbol{X}_i$, and interaction terms. Specifically, we let
\[
  Y_i = \alpha_{0} + \alpha_{X_1} X_{1,i} + \alpha_{X_2} X_{2,i} + A_i \left[\alpha_{A} + \alpha_{AX}\left(\alpha_{X_1} X_{1,i} + \alpha_{X_2} X_{2,i}\right)\right] + \varepsilon_O;\quad \varepsilon_O \sim \mathcal{N}(0, 1),
\]
where we set $\alpha_{0} = 1$ and $\alpha_{X_1} = \alpha_{X_2} = 1$ for the covariate coefficients. We further set $\alpha_{A} = 1$ to reflect the true effect of $A$ on $Y$ and $\alpha_{AX} = 0.1$ to induce a heterogeneous group/exposure effect which varies with $\boldsymbol{X}$.

For this main simulation study, we consider a fully factorial design where we vary $\tau_{X_1} = \tau_{X_2} \in \{0,1\}$, $\beta_A \in \{0,1\}$, and $\beta_{X_1} = \beta_{X_2} \in \{0,1\}$, yielding eight parameter settings, where each setting corresponds to a progressively more complex relationship among the variables: two with no confounding or selection bias, two with confounding bias but no selection bias, two with selection bias but no confounding bias, and two with both selection and confounding bias (see Table \ref{tab:simulation_results}). 

Within each simulation setting, we draw 500 random samples from our generated population by repeatedly simulating sample selection indicators, $S_i$. As the focus of the current work is in setting of known survey weights, which can be stratified or multistage, we extract the probability of selection for the $i$th individual and generate $S_i$ from a Bernoulli distribution with $S_i \sim {\rm Bernoulli}[\Pr(S_i = 1 \mid  A_i, \boldsymbol{X}_i)]$, retaining those observations with $S_i = 1$ in each replicate. Note that this deviates slightly from a multistage stratified, clustered design. However, we opted for this simplification to focus on the more important aspect of the problem, namely that the final weights need to appropriately factorize the joint distribution of being sampled and belonging to a particular group. This holds over a range of sampling designs (see our sensitivity analyses below).

We compare our proposed \eqref{eq:id1_est} -- \eqref{eq:dr_est} to the standard analytic approaches in Table \ref{tab:methods}. Across all methods where survey weights are considered, we apply the corresponding appropriate standard error calculations. For example, for our proposed approaches, we use the variance formula we have derived for stratified, clustered designs \eqref{eq:vcov_strat}. For comparison methods such as survey-weighted regression, we use design-based standard errors~\citep{lumley2011complex}. The oracle estimator, which knows the full data generating process, serves as our benchmark throughout. Our target quantity of interest is the ACD. We compare the relative bias, average of the estimated standard error (ASE), Monte Carlo standard error (MCSE), mean squared error (MSE), and coverage probability at a significance level of 0.05 for each method.

\subsection{Results}
\label{sec:simresult}

Results are given in Table \ref{tab:simulation_results}. Here, we discuss these results in the context of each simulation setting, where the data are generated to vary what sources of bias we expect to encounter when estimating the ACD.

\vspace{2ex}

\noindent {\bf Settings 1 and 3: No Bias Expected.} When there is neither confounding nor selection bias in the data generating mechanism, most estimators recover the true ACD with negligible bias and nominal 95\% coverage. Specifically, the relative bias among our four proposed estimators ranges from -0.002 to -0.001, as compared to -0.003 to 0.005\% among the existing methods, as expected. The proposed approaches also have the lowest MSEs (0.012), as compared to MSEs ranging from 0.012-0.069 for the existing approaches, as well as standard errors that are comparable to the more efficient approaches (ASE: 0.109-0.113). Simple linear regression (SLR) has larger standard errors, though the ASE is within 5\% of the MCSE, whereas the ASE of the traditional IPTW estimator is less than 50\% that of its MCSE, suggesting the variance is underestimated. Lastly, the proposed methods have coverage rates of 95.6-96.2\%, as compared to 95.6-96.4\% for the existing methods except the IPTW estimator (58.6\%). In contrast, the traditional IPTW estimator, though, unbiased on average, is particularly susceptible to underestimating the true variance when ignoring the stratified, clustered sampling design. in Setting 3, where selection depends on $A$, all methods except IPTW perform well. The relative biases for the proposed methods are all -0.005\%, equal to the other methods, except IPTW (1.228). Mean squared errors are low (0.008) for the proposed methods, similar to MSEs of 0.008 to 0.024 among the existing methods, except IPTW (1.659). Further, all the methods had nominal coverage (95.2-96.4\%), except IPTW (0\%).

\vspace{2ex}

\noindent {\bf Settings 2 and 4: Confounding Bias Expected.} When we generate $A$ conditional on $X_1$ and $X_2$, with selection also depending on $A$ in Setting 4, we now expect there to be confounding bias. We see that methods which account for $A$ and $\boldsymbol{X}$ perform comparatively well, whereas SLR performs poorly, as $\Pr(A \mid X_1, X_2)$ is no longer uniform over the range of $\boldsymbol{X}$ values. Specifically, the relative bias in the SLR estimator increases from 0.002-0.005 in Settings 1 and 3 to 2.021-2.123 in Settings 2 and 4, whereas the other estimators (including the proposed methods) have biases ranging in magnitude from -0.004 to 0 in Setting 2 and 0.001 to 0.027 in Setting 4. We further see a higher in MSE for the SLR (1.089-1.357) and IPTW (0.078-1.421) estimators, while the MSE for the other existing methods (0.010-0.015) and the proposed estimators (0.010-0.022) remain low. Coverage for the proposed (93.4-94.4\%) and other existing (91.4-93.6\%) are also at the nominal level, whereas SLR (0\%) and IPTW (0-63.4\%) do not cover the true ACD. Among our proposed estimators, outcome modeling/direct standardization and AIPW deliver unbiased estimates with slightly lower ASE/MCSE than the purely IPW‑based estimators, indicating better efficiency in this confounding scenario.

\vspace{2ex}

\noindent {\bf Settings 5 and 7: Selection Bias Expected.} When the $\boldsymbol{X}$ variables influence the probability of inclusion but do not confound the relationship between $Y$ and $A$, we expect there to be selection bias. In Setting 5, the relative bias among the existing approaches ranges from 0.001 to 0.231, with unweighted methods having higher bias (0.217-0.231) than the survey-weighted approaches (0.001-0.008). The bias among the proposed methods ranges from 0.000 to 0.027, demonstrating empirically that accounting for the survey weights corrects this selection bias when estimating the target ACD. These results persist in Setting 7, where selection also depends on $A$, with the relative bias among the existing unweighted approaches (0.173-3.560) being higher than existing survey-weighted approaches (0.001-0.005) and the proposed approaches (0.003-0.029). Similar patterns are seen in the MSEs of the existing approaches (0.019-0.162 in Setting 5 and 0.008-0.111 in Setting 7) versus the proposed methods (0.043-0.128 in Setting 5 and 0.018-12.637 in Setting 7). However, only the proposed methods have nominal coverage (94.2-97.4\% in Setting 5 and 93.4-97.8\% in Setting 7), while all other methods have coverage ranging from 19.4-85.8\% and 0.0-90.2\% in Settings 5 and 7, respectively. Among our proposed estimators, Outcome modeling again shows highest efficiency (ASE = 0.154 vs.~MCSE = 0.137), as it is correctly specified, but AIPW approach strikes a balance between bias control and efficiency.

\vspace{2ex}

\noindent {\bf Settings 6 and 8: Both Confounding and Selection Bias Expected.} Lastly, Settings 6 and 8 are the most complex, with Setting 8 mirroring the conceptual diagram for our motivating data. In these settings, our proposed estimators outperform all current approaches with respect to the relative bias, MSE, and coverage for estimating the ACD. In particular, the relative biases for existing approaches that not use survey weights range from 0.203-1.198 in Setting 6 and 0.230-2.872 in Setting 8, and their coverages range from 0-44.6\% and 0-14.4\%, respectively. Further, while survey-weighting reduces bias (-0.004 to 0.021), the existing methods still under-cover (87.4-92.0\%), indicating that design adjustments alone are insufficient when confounding and selection bias co-occur. In contrast, the bias among the proposed estimators ranges from -0.005 to 0.025, with coverages ranging from 94.6-98.2\%. This is expected, as the proposed methods have the correct balancing and generalizability weights under these settings, whereas the existing approaches fail to incorporate one or both weights, or fail to incorporate them correctly into the form of the estimator. Among our proposed approaches outcome modeling/direct standardization is most efficient, while AIPW again achieves low bias, correct SEs, and nominal coverage, while offering better efficiency than the IPW-based approaches.

\newpage

{\small
\begin{longtable}{lrrrrrrr}
\caption{Simulation results comparing different analytic methods for calculating the population average controlled difference (ACD) in $Y$ across $A$ for different data generating mechanism settings. Results presented are the true ACD (ACD) estimated ACD (Est.~ACD), relative bias (Rel.~Bias), average of the estimated standard error (ASE), Monte Carlo standard error (MCSE), mean squared error (MSE) and coverage probability (Cov.) averaged over 500 samples from the population generated under each setting.} \\
\label{tab:simulation_results} \\
    \toprule
    Method & ACD & Est.~ACD & Rel.~Bias & ASE & MCSE & MSE & Cov. \\ 
    \midrule
    \multicolumn{8}{l}{\bf Setting 1: No Bias Expected ($\tau_X = 0$, $\beta_A = 0$, $\beta_X = 0$)} \\
    \midrule
    Oracle Estimator & 1.003 & 1.000 & -0.003 & 0.109 & 0.107 & 0.012 & 0.964 \\ 
    Simple Reg. & 1.003 & 1.004 & 0.002 & 0.204 & 0.206 & 0.042 & 0.948 \\ 
    Multiple Reg. & 1.003 & 1.001 & -0.002 & 0.109 & 0.107 & 0.012 & 0.958 \\ 
    IPTW Estimator & 1.003 & 1.008 & 0.005 & 0.110 & 0.263 & 0.069 & 0.586 \\ 
    Survey-Weighted Multiple Reg. & 1.003 & 1.001 & -0.002 & 0.109 & 0.108 & 0.012 & 0.956 \\ 
    IPTW Multiple Reg. & 1.003 & 1.001 & -0.002 & 0.109 & 0.108 & 0.012 & 0.958 \\ 
    IPTW + Survey-Weighted Multiple Reg. & 1.003 & 1.001 & -0.002 & 0.109 & 0.108 & 0.012 & 0.956 \\ 
    Weighted IPTW + Survey-Weighted Multiple Reg. & 1.003 & 1.001 & -0.002 & 0.109 & 0.108 & 0.012 & 0.956 \\ 
    Outcome Modeling and Direct Standardization & 1.003 & 1.001 & -0.002 & 0.109 & 0.107 & 0.012 & 0.962 \\ 
    Inverse Probability Weighting 1 & 1.003 & 1.001 & -0.002 & 0.110 & 0.108 & 0.012 & 0.956 \\ 
    Inverse Probability Weighting 2 & 1.003 & 1.002 & -0.001 & 0.113 & 0.110 & 0.012 & 0.960 \\ 
    Augmented Inverse Probability Weighting & 1.003 & 1.001 & -0.002 & 0.110 & 0.108 & 0.012 & 0.958 \\ 
    \midrule
    \multicolumn{8}{l}{\bf Setting 2: Confounding Bias Expected ($\tau_X = 1$, $\beta_A = 0$, $\beta_X = 0$)} \\ 
    \midrule
    Oracle Estimator & 0.996 & 0.995 & 0.000 & 0.115 & 0.120 & 0.014 & 0.930 \\ 
    Simple Reg. & 0.996 & 2.021 & 1.030 & 0.195 & 0.194 & 1.089 & 0.000 \\ 
    Multiple Reg. & 0.996 & 0.995 & -0.001 & 0.115 & 0.121 & 0.015 & 0.932 \\ 
    IPTW Estimator & 0.996 & 0.982 & -0.014 & 0.135 & 0.280 & 0.078 & 0.634 \\ 
    Survey-Weighted Multiple Reg. & 0.996 & 0.995 & -0.001 & 0.116 & 0.122 & 0.015 & 0.932 \\ 
    IPTW Multiple Reg. & 0.996 & 0.996 & 0.000 & 0.116 & 0.123 & 0.015 & 0.936 \\ 
    IPTW + Survey-Weighted Multiple Reg. & 0.996 & 0.995 & -0.001 & 0.117 & 0.124 & 0.015 & 0.928 \\ 
    Weighted IPTW + Survey-Weighted Multiple Reg. & 0.996 & 0.995 & 0.000 & 0.117 & 0.124 & 0.015 & 0.928 \\ 
    Outcome Modeling and Direct Standardization & 0.996 & 0.995 & -0.001 & 0.116 & 0.121 & 0.015 & 0.934 \\ 
    Inverse Probability Weighting 1 & 0.996 & 0.993 & -0.003 & 0.136 & 0.146 & 0.021 & 0.942 \\ 
    Inverse Probability Weighting 2 & 0.996 & 0.992 & -0.004 & 0.138 & 0.147 & 0.022 & 0.940 \\ 
    Augmented Inverse Probability Weighting & 0.996 & 0.996 & 0.000 & 0.118 & 0.124 & 0.015 & 0.934 \\ 
    \midrule
    \multicolumn{8}{l}{\bf Setting 3: No Bias Expected ($\tau_X = 0$, $\beta_A = 1$, $\beta_X = 0$)} \\
    \midrule
    Oracle Estimator & 0.991 & 0.986 & -0.005 & 0.090 & 0.090 & 0.008 & 0.962 \\ 
    Simple Reg. & 0.991 & 0.996 & 0.005 & 0.169 & 0.167 & 0.028 & 0.962 \\ 
    Multiple Reg. & 0.991 & 0.986 & -0.005 & 0.090 & 0.089 & 0.008 & 0.964 \\ 
    IPTW Estimator & 0.991 & 2.267 & 1.288 & 0.105 & 0.177 & 1.659 & 0.000 \\ 
    Survey-Weighted Multiple Reg. & 0.991 & 0.986 & -0.005 & 0.090 & 0.089 & 0.008 & 0.954 \\ 
    IPTW Multiple Reg. & 0.991 & 0.986 & -0.005 & 0.090 & 0.089 & 0.008 & 0.966 \\ 
    IPTW + Survey-Weighted Multiple Reg. & 0.991 & 0.986 & -0.005 & 0.090 & 0.089 & 0.008 & 0.952 \\ 
    Weighted IPTW + Survey-Weighted Multiple Reg. & 0.991 & 0.986 & -0.005 & 0.090 & 0.089 & 0.008 & 0.952 \\ 
    Outcome Modeling and Direct Standardization & 0.991 & 0.986 & -0.005 & 0.093 & 0.089 & 0.008 & 0.962 \\ 
    Inverse Probability Weighting 1 & 0.991 & 0.986 & -0.005 & 0.093 & 0.089 & 0.008 & 0.962 \\ 
    Inverse Probability Weighting 2 & 0.991 & 0.986 & -0.005 & 0.095 & 0.090 & 0.008 & 0.954 \\ 
    Augmented Inverse Probability Weighting & 0.991 & 0.986 & -0.005 & 0.093 & 0.089 & 0.008 & 0.960 \\ 
    \midrule
    \multicolumn{8}{l}{\bf Setting 4: Confounding Bias Expected ($\tau_X = 1$, $\beta_A = 0$, $\beta_X = 1$)} \\
    \midrule
    Oracle Estimator & 0.969 & 0.969 & 0.001 & 0.094 & 0.100 & 0.010 & 0.938 \\ 
    Simple Reg. & 0.969 & 2.123 & 1.191 & 0.167 & 0.161 & 1.357 & 0.000 \\ 
    Multiple Reg. & 0.969 & 0.995 & 0.027 & 0.097 & 0.103 & 0.011 & 0.930 \\ 
    IPTW Estimator & 0.969 & 2.146 & 1.215 & 0.148 & 0.188 & 1.421 & 0.000 \\ 
    Survey-Weighted Multiple Reg. & 0.969 & 0.970 & 0.001 & 0.094 & 0.100 & 0.010 & 0.940 \\ 
    IPTW Multiple Reg. & 0.969 & 0.995 & 0.027 & 0.098 & 0.106 & 0.012 & 0.914 \\ 
    IPTW + Survey-Weighted Multiple Reg. & 0.969 & 0.995 & 0.027 & 0.098 & 0.107 & 0.012 & 0.920 \\ 
    Weighted IPTW + Survey-Weighted Multiple Reg. & 0.969 & 0.969 & 0.001 & 0.095 & 0.102 & 0.010 & 0.934 \\ 
    Outcome Modeling and Direct Standardization & 0.969 & 0.970 & 0.001 & 0.096 & 0.100 & 0.010 & 0.942 \\ 
    Inverse Probability Weighting 1 & 0.969 & 0.980 & 0.011 & 0.111 & 0.112 & 0.013 & 0.944 \\ 
    Inverse Probability Weighting 2 & 0.969 & 0.976 & 0.008 & 0.118 & 0.123 & 0.015 & 0.934 \\ 
    Augmented Inverse Probability Weighting & 0.969 & 0.970 & 0.001 & 0.098 & 0.102 & 0.010 & 0.942 \\ 
    \midrule
    \multicolumn{8}{l}{\bf Setting 5: Selection Bias Expected ($\tau_X = 0$, $\beta_A = 0$, $\beta_X = 1$)} \\
    \midrule
    Oracle Estimator & 0.971 & 0.976 & 0.006 & 0.133 & 0.137 & 0.019 & 0.946 \\ 
    Simple Reg. & 0.971 & 1.187 & 0.222 & 0.136 & 0.127 & 0.063 & 0.650 \\ 
    Multiple Reg. & 0.971 & 1.182 & 0.217 & 0.074 & 0.074 & 0.050 & 0.194 \\ 
    IPTW Estimator & 0.971 & 1.195 & 0.231 & 0.075 & 0.279 & 0.128 & 0.280 \\ 
    Survey-Weighted Multiple Reg. & 0.971 & 0.973 & 0.002 & 0.163 & 0.214 & 0.046 & 0.858 \\ 
    IPTW Multiple Reg. & 0.971 & 1.182 & 0.218 & 0.074 & 0.074 & 0.050 & 0.194 \\ 
    IPTW + Survey-Weighted Multiple Reg. & 0.971 & 0.972 & 0.001 & 0.163 & 0.216 & 0.046 & 0.866 \\ 
    Weighted IPTW + Survey-Weighted Multiple Reg. & 0.971 & 0.978 & 0.008 & 0.159 & 0.207 & 0.043 & 0.858 \\ 
    Outcome Modeling and Direct Standardization & 0.971 & 0.979 & 0.008 & 0.154 & 0.137 & 0.019 & 0.974 \\ 
    Inverse Probability Weighting 1 & 0.971 & 0.971 & 0.000 & 0.197 & 0.211 & 0.044 & 0.942 \\ 
    Inverse Probability Weighting 2 & 0.971 & 0.944 & -0.027 & 0.327 & 0.402 & 0.162 & 0.954 \\ 
    Augmented Inverse Probability Weighting & 0.971 & 0.971 & 0.000 & 0.209 & 0.227 & 0.051 & 0.946 \\ 
    \midrule
    \multicolumn{8}{l}{\bf Setting 6: Both Confounding and Selection Bias Expected ($\tau_X = 1$, $\beta_A = 0$, $\beta_X = 1$)} \\ 
    \midrule
    Oracle Estimator & 1.002 & 1.020 & 0.018 & 0.116 & 0.118 & 0.014 & 0.946 \\ 
    Simple Reg. & 1.002 & 2.201 & 1.198 & 0.128 & 0.121 & 1.454 & 0.000 \\ 
    Multiple Reg. & 1.002 & 1.225 & 0.223 & 0.079 & 0.078 & 0.056 & 0.192 \\ 
    IPTW Estimator & 1.002 & 1.205 & 0.203 & 0.133 & 0.336 & 0.154 & 0.446 \\ 
    Survey-Weighted Multiple Reg. & 1.002 & 1.015 & 0.013 & 0.144 & 0.168 & 0.028 & 0.898 \\ 
    IPTW Multiple Reg. & 1.002 & 1.230 & 0.228 & 0.080 & 0.081 & 0.059 & 0.202 \\ 
    IPTW + Survey-Weighted Multiple Reg. & 1.002 & 1.018 & 0.016 & 0.147 & 0.187 & 0.035 & 0.874 \\ 
    Weighted IPTW + Survey-Weighted Multiple Reg. & 1.002 & 1.023 & 0.021 & 0.143 & 0.176 & 0.031 & 0.868 \\ 
    Outcome Modeling and Direct Standardization & 1.002 & 1.021 & 0.019 & 0.136 & 0.119 & 0.015 & 0.978 \\ 
    Inverse Probability Weighting 1 & 1.002 & 1.003 & 0.002 & 0.187 & 0.183 & 0.033 & 0.956 \\ 
    Inverse Probability Weighting 2 & 1.002 & 1.026 & 0.025 & 0.261 & 0.307 & 0.095 & 0.982 \\ 
    Augmented Inverse Probability Weighting & 1.002 & 1.022 & 0.020 & 0.199 & 0.244 & 0.060 & 0.960 \\ 
    \midrule
    \multicolumn{8}{l}{\bf Setting 7: Selection Bias Expected ($\tau_X = 0$, $\beta_A = 1$, $\beta_X = 1$)} \\
    \midrule
    Oracle Estimator & 0.997 & 0.991 & -0.007 & 0.094 & 0.091 & 0.008 & 0.956 \\ 
    Simple Reg. & 0.997 & 1.170 & 0.173 & 0.097 & 0.093 & 0.038 & 0.562 \\ 
    Multiple Reg. & 0.997 & 1.205 & 0.208 & 0.053 & 0.051 & 0.046 & 0.030 \\ 
    IPTW Estimator & 0.997 & 4.548 & 3.560 & 0.095 & 0.171 & 12.637 & 0.000 \\ 
    Survey-Weighted Multiple Reg. & 0.997 & 1.000 & 0.002 & 0.122 & 0.137 & 0.019 & 0.902 \\ 
    IPTW Multiple Reg. & 0.997 & 1.205 & 0.208 & 0.053 & 0.051 & 0.046 & 0.030 \\ 
    IPTW + Survey-Weighted Multiple Reg. & 0.997 & 0.998 & 0.001 & 0.122 & 0.140 & 0.020 & 0.898 \\ 
    Weighted IPTW + Survey-Weighted Multiple Reg. & 0.997 & 1.003 & 0.005 & 0.120 & 0.134 & 0.018 & 0.902 \\ 
    Outcome Modeling and Direct Standardization & 0.997 & 0.992 & -0.005 & 0.110 & 0.092 & 0.008 & 0.978 \\ 
    Inverse Probability Weighting 1 & 0.997 & 1.002 & 0.005 & 0.148 & 0.140 & 0.020 & 0.954 \\ 
    Inverse Probability Weighting 2 & 0.997 & 1.026 & 0.029 & 0.262 & 0.332 & 0.111 & 0.934 \\ 
    Augmented Inverse Probability Weighting & 0.997 & 1.001 & 0.003 & 0.158 & 0.153 & 0.023 & 0.968 \\ 
    \midrule
    \multicolumn{8}{l}{\bf Setting 8: Both Confounding and Selection Bias Expected ($\tau_X = 1$, $\beta_A = 1$, $\beta_X = 1$)} \\ 
    \midrule
    Oracle Estimator & 1.001 & 0.991 & -0.010 & 0.097 & 0.093 & 0.009 & 0.956 \\ 
    Simple Reg. & 1.001 & 2.210 & 1.207 & 0.117 & 0.110 & 1.474 & 0.000 \\ 
    Multiple Reg. & 1.001 & 1.232 & 0.230 & 0.073 & 0.079 & 0.059 & 0.126 \\ 
    IPTW Estimator & 1.001 & 3.877 & 2.872 & 0.173 & 0.172 & 8.301 & 0.000 \\ 
    Survey-Weighted Multiple Reg. & 1.001 & 0.994 & -0.007 & 0.118 & 0.134 & 0.018 & 0.920 \\ 
    IPTW Multiple Reg. & 1.001 & 1.233 & 0.231 & 0.074 & 0.081 & 0.060 & 0.144 \\ 
    IPTW + Survey-Weighted Multiple Reg. & 1.001 & 1.016 & 0.015 & 0.109 & 0.123 & 0.015 & 0.912 \\ 
    Weighted IPTW + Survey-Weighted Multiple Reg. & 1.001 & 0.998 & -0.004 & 0.122 & 0.145 & 0.021 & 0.912 \\ 
    Outcome Modeling and Direct Standardization & 1.001 & 0.993 & -0.008 & 0.114 & 0.095 & 0.009 & 0.982 \\ 
    Inverse Probability Weighting 1 & 1.001 & 0.996 & -0.005 & 0.157 & 0.155 & 0.024 & 0.954 \\ 
    Inverse Probability Weighting 2 & 1.001 & 1.017 & 0.016 & 0.226 & 0.264 & 0.070 & 0.968 \\ 
    Augmented Inverse Probability Weighting & 1.001 & 0.992 & -0.010 & 0.154 & 0.158 & 0.025 & 0.946 \\ 
    \bottomrule
\end{longtable}
}

\newpage

\subsection{Sensitivity Analyses}
\label{sec:sensitivity}

We carry out five sensitivity analyses to better understand the robustness of our approach to the following various real-world situations. In the following, we briefly describe the setup of each sensitivity analysis and our key takeaways. Full sensitivity analysis results can be found in Appendix B.

\subsubsection{Single-Stage Unequal Probability Sampling}

We compare our estimators to the current standard methods, but with the data drawn from a single-stage unequal probability sample, to examine how the performances of the  methods change under different sampling designs. In this sensitivity analysis, we set the number of strata and clusters to be $n_{strata} = n_{cluster} = 1$ and retain all other parameter settings from our main simulation.

When the data are generated under a single-stage unequal probability sample, all estimators show reduced Monte Carlo variability compared to our main, stratified clustered complex design setting, but their relative performances remain the same. The unweighted and survey-weighted regression methods are unbiased across all eight scenarios, with analytic SEs matching MCSEs and coverage near 95\% in the earlier settings, with performance decreasing under settings with confounding and/or selection bias. In contrast, the four proposed methods have low bias, within $\pm$ 1\% of the truth, ASEs closely aligned with the MCSEs, and coverages of 94-98\% across all settings. Outcome Modeling often achieves the lowest ASE and MCSE, while IPW1 and IPW2 remain robust with minimal bias and nominal coverage, and AIPW combines these results to offer reliable inference even when either the propensity or outcome model alone is correctly specified. Full results are given in Supplemental Table B.1.

\subsubsection{Propensity/Outcome Model Misspecification}

In a second sensitivity analysis, we study our proposed approaches under increasing misspecification of either/both the propensity and outcome models. We consider nine scenarios by crossing three levels of propensity score model misspecification (via the interaction effect, $\tau_{X12} \in \{0, 0.1, 0.5\}$) with three levels of outcome model misspecification (via the group heterogeneity effect, $\alpha_{AX} \in \{0, 0.1, 0.5\}$). In each case, we estimate only main effects models, thereby omitting the true interaction terms when they are non-zero. The results of this sensitivity analysis show that outcome modeling alone is vulnerable to both mild and moderate outcome model misspecification, as expected. Further, IPW1 is fairly robust to moderate propensity score misspecification but loses efficiency, while IPW2 is unstable under propensity score misspecification. The AIPW estimator remains the only estimator with low bias in most settings, reasonable variance, and coverage closest to 95\% across the fully factorial design.

Specifically, when neither model is misspecified, all four proposed methods match the oracle's performance (bias $\approx 0$, ASE $\approx$ MCSE, and coverage at the nominal level), with outcome modeling/direct standardization being most efficient, and AIPW and IPW1 being slightly more efficient than IPW2. Under mild outcome model misspecification, outcome modeling incurs 17\% bias and has only 68\% coverage, while IPW1, IPW2, and AIPW are unbiased and having nominal coverage. These results are magnified in the moderate outcome misspecification setting. When only the propensity model is mildly or moderately misspecified, outcome modeling and AIPW remain unbiased, and IPW1 is fairly robust, but IPW2 shows large bias under moderate misspecification (-20\%) despite nominal coverage (94\%). In scenarios where both models are mildly or mixed misspecified, AIPW consistently has bias near 2-3\% and coverage $\geq$ 88\%, IPW1 maintains < 3\% bias and $\geq$ 91\% coverage, whereas outcome modeling and IPW2 break down under moderate misspecification. Finally, when both models are moderately misspecified, AIPW's bias is 7.9\% (coverage 83\%) and IPW1's bias is 6.1\% (coverage 85\%), while outcome modeling and IPW2 have bias > 20\%. Overall, the doubly robust approach offers the best balance of bias control and inference validity across all misspecification patterns, as expected. Full results are given in Supplemental Table B.2.

\subsubsection{Limited Overlap}

\begin{figure}[!ht]
    \centering
    \includegraphics[width=\linewidth]{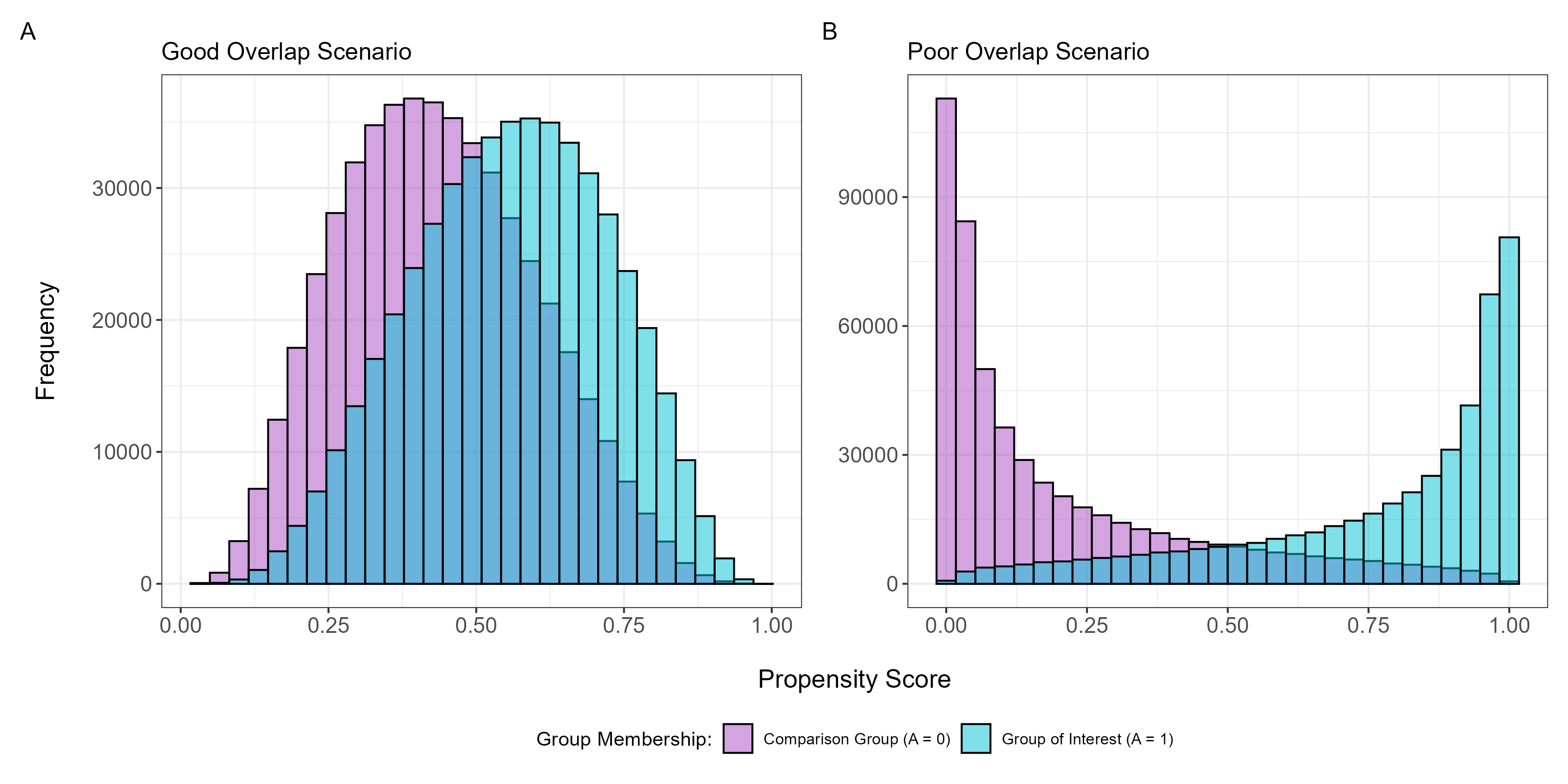}
    \caption{Distribution of estimated propensity scores by group membership under two overlap scenarios. A. Good Overlap $(\tau_A = 0.5)$. Propensity scores for the comparison group (purple, $A = 0$) and group of interest (teal, $A = 1$) show good overlap around 0.5, supporting stable inverse‐probability weighting. B. Poor Overlap $(\tau_A = 2.0)$. Propensity scores concentrate near 0 for the comparison group and near 1 for the group of interest, indicating limited common support.}
    \label{fig:overlap}
\end{figure}

In practice, the positivity assumption may be violated due to limited overlap in propensity scores between groups. To study the impact of this violation, we conduct a sensitivity analysis exploring scenarios with different degrees of overlap. Following \citet{zeng2025moving}, we tune the parameters in the propensity model generating mechanism to simulate propensity scores with good versus poor overlap. Specifically, we vary $\tau_A \in \{0.5, 2.0\}$, in the true propensity model to emulate settings with good versus poor overlap in the propensity scores, respectively. Figure \ref{fig:overlap} illustrates the distributions of the propensity scores under these scenarios.

Under good overlap ($\tau_A = 0.5$), all four proposed methods yield unbiased ACD estimates and nominal coverage. Outcome modeling has the smallest bias (-0.2\%) and ASE (0.113), while IPW1 and IPW2 have modestly higher bias (0.7\% and 2.8\%) and (0.153, 0.213). AIPW balances these outcomes, with 0.8\% bias, ASE = 0.150, and 96.8\% coverage. When overlap is poor ($\tau_A = 2.0$), outcome modeling remains unbiased (1.4\%), has low ASE (0.129) that aligns with its MCSE (0.109), and maintains nominal coverage (97.2\%). Similarly, AIPW maintains low bias (-1.0\%) and nominal coverage (96.2\%). In contrast, IPW1 and IPW2 have higher bias (-9.5\% and 2.5\%), underestimated standard errors (ASE of 0.542 and 0.352 versus and MCSE of 0.966 and 0.714, respectively), reducing their coverage to 91.8\% and 90.6\%. These results suggest that while the IPW-based estimators are affected by poorer overlap, outcome modeling and AIPW are robust and provide more stable performance in terms of bias, variability, and coverage. For the full simulation results, see Supplemental Table B.3.

\subsubsection{Estimated Selection Mechanism}

As discussed in Section \ref{sec:nonprob}, if sample selection mechanism is not known, we need to model $\Pr(S \mid A, \boldsymbol{X})$. In this supplementary analysis, we treat  the probability of selection given our variable of interest and our confounders as unknown and estimate $\Pr(S \mid A, \boldsymbol{X})$ via a beta regression, as recommended by \citet{elliot2009combining}. Note that this deviates from our motivating example, but is relevant for applications with non-probability samples. Here, we replicate all the settings of our main simulation, but for our proposed methods we treat the selection probabilities as unknown and estimate them.

These results show that our proposed approaches still have low bias and again are the only methods to achieve nominal coverage in our setting of interest. Specifically, outcome modeling remains nearly unbiased (relative bias within $\pm$ 1\%) with ASEs that align with the Monte Carlo SEs and 94-96\% coverage. Both IPW1 and IPW2 also maintain minimal bias ($\leq$ 1\%) and nominal coverage (92-96\%), although their Monte Carlo variability is occasionally slightly larger than their ASEs in more complex settings. AIPW delivers the most consistent performance, combining low bias ($\pm$ 1\%) with ASEs within 10\% of the MCSEs and coverage around 95\% in every scenario. Moreover, while the standard survey-weighted regressions also achieve low bias, they all have coverage $\leq$ 90.4\% in our setting of interest. Full results are given in Supplemental Table B.4.

\subsubsection{Sampling Covariate Omission}

Lastly, we extend our main simulation by introducing two additional, unobserved covariates, $U_1$ and $U_2$, that vary by stratum and cluster and drive the selection mechanism, but do not enter the true propensity or outcome models. In generating each sample, we first simulate $U_1$ and $U_2$ with stratum‐ and cluster‐specific means and variances, then let the probability of inclusion depend on $A$, $X_1$, $X_2$, $U_1$, and $U_2$, with coefficients 
$\beta_{U_1} = \beta_{U_2} = 1$ across all settings. All other parameter settings followed from the main simulation.  In our analysis, however, we treat the selection mechanism as unknown and estimate $\Pr(S = 1 \mid A, \boldsymbol{X})$, omitting $U_1$ and $U_2$ to understand if this induces bias. We then apply our four proposed estimators using the estimated selection model, and compare their performance to the oracle estimator and the existing methods.

Despite this misspecification, all four proposed methods retain low bias and nominal coverage in most settings. Outcome modeling incurs a modest negative bias (ranging from –3.3\% to –5.7\% across all scenarios), but it remains efficient (MCSE from 0.06 to 0.12) and maintains coverage at the nominal level. IPW1, which uses a single estimated selection model, shows slightly larger bias (-1.7\% to -10.6\%) with coverage generally between 86\% and 96\%. IPW2, which estimates more models, exhibits bias within $\pm$ 3.6\% in most scenarios and preserves coverage around 95-97\%, though its variance inflation is substantial in some settings. Lastly, AIPW strikes the best balance, with bias remaining under 8.1\% (-8.1\% to 6.7\%), ASEs aligning closely with the MCSEs, and coverage between 90.8\% and 95.4\%. Overall, even when key design covariates are omitted from the estimated selection model, the proposed approaches have reasonable performance, with the AIPW estimator being particularly robust to this form of misspecification. Full results are given in Supplemental Table B.5.

\subsection{Summary Takeaways}
\label{sec:simtakeaways}

In summary, the simulation and sensitivity analyses described above vary the associations between $Y$, $A$, $\boldsymbol{X}$, $\boldsymbol{U}$, and $S$ to introduce varying degrees of confounding and/or selection bias into the data generating mechanism. In our main simulation, across all eight simulation scenarios, the drivers of performance are (1) whether the estimator properly adjusts for the source(s) of bias (confounding, selection, or both), and (2) whether its variance calculation accounts for the stratified, clustered survey design and any weight variability. In settings without confounding or selection bias, most existing and proposed methods perform similarly well, providing estimates of the ACD with low bias, low MSE, and nominal coverage. In settings with confounding bias (but no selection bias), methods that account for both $A$ and $\boldsymbol{X}$, such as multiple regression and IPTW-based estimators, tend to perform well. Methods that only model $A$ often show high bias and negligible coverage. When $\boldsymbol{X}$ is introduced into the generation of either/both $S$ and $A$, that is, when selection or selection and confounding bias exists, we see that standard analytic approaches lead to biased estimates and do not attain nominal coverage, whereas the proposed methods perform empirically well, supporting theoretical results derived in the previous section.

Our sensitivity analyses reveal several key points that have implications in practice. First, in settings of good propensity score overlap, all proposed methods are unbiased with nominal coverage, outcome modeling is most efficient, IPW2 most variable, and AIPW intermediate. In poor overlap scenarios, outcome modeling and AIPW remain relatively stable, while IPW1 and IPW2 have larger bias or variance. Researchers should examine the degree of overlap and possible violations of the positivity assumption when conducting analyses. Second, when either the outcome or propensity model alone is mildly or moderately misspecified, AIPW and IPW1/IPW2 maintain low bias and nominal coverage. In contrast, outcome modeling incurs bias and low coverage under outcome model misspecification, and IPW2 can be unstable under propensity model misspecification. When both models are misspecified, AIPW is the most robust. Third, when the selection mechanism is unknown, estimating $\Pr(S=1\mid A,\boldsymbol{X})$ may lead to slight increases in bias and under coverage for IPW1/IPW2, but AIPW remains robust. Lastly, even when key sampling covariates are omitted from the estimated selection model, IPW1, IPW2, and AIPW maintains lower bias and good coverage, while outcome modeling shows modest negative bias but retains high coverage due to conservative SEs.

Thus, in real-world survey analyses, where both confounding and selection mechanisms may be complex or partially unknown, overlap can be limited, and model misspecification is a concern, the doubly robust augmented IPW estimator provides the best balance of bias protection, valid variance estimation under complex sampling, and efficiency. We therefore recommend our proposed AIPW estimator as the default analytic approach for estimating controlled differences or causal contrasts in complex survey settings such as the one described in our motivating example.

\section{NHANES: Race, SES, and Telomere Length}
\label{sec:nhanes}

We now present our motivating example, which comes from NHANES. NHANES is a nationally representative survey sample collected by the Centers for Disease Control, which contains a wealth of demographic, dietary, physical examination, laboratory, and questionnaire data on individuals. Our group variable of interest is the survey participant's self-reported race. Racial groups are categories of people treated as a social entity by virtue of physical characteristics. While an individual’s race is often self-reported, racial groups may represent commonalities such as geographic origin, genetic ancestry, cultural norms and traditions, social history, or current social circumstances~\citep{williams2010race}. From a statistical perspective, race is often a proxy for other sources of shared variation which must be accounted for when making inferential statements. We focus on non-Hispanic participants who self-identify as Black or White as, particularly in the US, it is known that Black individuals tend to have poorer health outcomes than White individuals due to structural inequalities~\citep{arias2022united}.

\subsection{Motivating Question}
\label{sec:motivation}

Telomeres are regions of DNA at the ends of chromosomes that protect against cellular aging and senescence~\citep{lin2016systematic}. Telomere shortening is associated with disease incidence and many health outcomes~\citep{shammas2011telomeres}. Variation in telomere lengths is attributable to age and shortening during mitosis, in addition to genetics, social history, exposure to environmental contaminants, and psychosocial stressors~\citep{jiang2007telomere, blackburn2015human}. Paradoxically, Black individuals typically have longer telomeres than White individuals~\citep{needham2013socioeconomic, venkateswaran2024robustly}, with plausible explanations citing leukocyte cell composition, population stratification, or genetic differences~\citep{freedman1997black, lin2016systematic, hansen2016shorter}. Others posit that environmental exposures associated with longer telomere length underlie this so-called `race effect,' as these exposures differ significantly between Blacks and Whites~\citep{needham2020black, roberts2022persistent}. Further, studies have found telomere length to be comparable between Black and White individuals in socioeconomically homogeneous populations~\citep{geronimus2015race}, calling into question purely genetic explanations. Our motivating question is this: if we could hypothetically balance SES ($\boldsymbol{X}$) between Black and White individuals ($A$) in a nationally representative sample ($S$), would we still see significant Black/White differences in telomere length ($Y$)?

\subsection{NHANES Sampling Design}
\label{sec:nhanesdesign}

The stratified, clustered sampling design of the NHANES survey is designed to over-represent Black individuals and underrepresented groups in particular, to achieve a more nationally representative sample. Namely, NHANES samples in four stages. In the first stage, primary sampling units (PSUs; i.e., counties) are selected. Counties are stratified and selected with probability proportional to their weighted average population counts, where the weights give a higher probability of selection to counties with a higher proportion of underrepresented subgroups (e.g., certain age groups, marginalized racial groups, lower-income households). In the second stage, the PSUs are further divided (e.g., into city blocks) and subsampled, and in the third stage, households are sampled from these sample segments. Again, households within segments that have a greater proportion of underrepresented age, ethnic, or income groups are given additional weight for being selected. Lastly, individuals within households are sampled according to designated age-sex-race/ethnicity screening subdomains. This complex sampling scheme motivates our proposed methods, as sample selection depends on self-reported race and potential confounders (i.e., SES).

\subsection{Study Sample}
\label{sec:sample}

The 1999-2002 NHANES waves consisted of 7,839 participants with recorded laboratory measures, representing our survey sample $(S = 1)$. Our primary endpoint, $Y$, was the log-transformed mean ratio of an individual's telomere length to a standard reference DNA sample across all leukocyte cell types (mean T/S). Telomere length was assayed via quantitative polymerase chain reaction~\citep[PCR; ][]{cawthon2002telomere, needham2013socioeconomic}. We focus on the 1999-2002 NHANES waves, as they featured 4-year adjusted survey weights designed for aggregating data across cohorts. Among the initial 7,839 participants, 5,308 (67.7\%) self-identified as either non-Hispanic White or non-Hispanic Black, the levels of our group variable $A$. Excluding those participants without our outcome of interest, our final analytic sample contained 5,298 Non-Hispanic White or Non-Hispanic Black identifying participants with measured telomere length. As discussed previously, race was our variable of interest. We further adjusted our models for study participant age, sex, and blood cell composition, including complete blood count and five part differential consisting of the proportion of neutrophils, eosinophils, basophils, lymphocytes, and monocytes, to account for known differences in these factors~\citep{freedman1997black}, as well as twelve indicators of SES, collectively representing our confounding variables $\boldsymbol{X}$. Ten of these, namely marital status, education level, household income, insurance status, Special Supplemental Nutrition Program for Women, Infants, and Children (WIC) usage, household size, home ownership, home type, food security status, and an individual’s poverty income ratio (PIR), were extracted directly from the NHANES demographic and occupation questionnaires. Occupation category was constructed by mapping occupation group codes in the NHANES occupation questionnaire to the national statistics socioeconomic job classifications, and employment status was derived from three occupational measures: type of work done last week, hours worked last week at all jobs, and main reason for not working last week~\citep{rehkopf2008non, rose2005national}. In later modeling, we perform complete case analysis on $n = 5,270$ participants after removing 28 individuals without measured blood composition.

Full descriptive statistics for our analytic sample are presented in Supplemental Table C1, both overall and stratified by self-reported race. Non-Hispanic Black participants had a median telomere length (mean T/S ratio) of 1.05 (i.e., 0.05 times longer than the standard reference DNA), as compared to non-Hispanic White participants with a median telomere length of 0.98. With respect to the identified socioeconomic characteristics, non-Hispanic Black versus White participants differed significantly across all 12 SES indicators, with Black participants having lower educational attainment, lower household income, and lower home ownership rates, as well as higher food insecurity and WIC utilization rates, larger household sizes, and a higher proportion uninsured, unemployed, and below the federal poverty line. Further, a higher proportion of Black participants reported not working or working in low blue collar occupations, a higher proportion reported being widowed, divorced, or separated, and a higher proportion reported residing in apartments, all of which suggest lower socioeconomic status~\citep{rose2005national}. Based on these factors, we then calculated propensity scores with race as the outcome using unweighted and survey-weighted logistic regressions. The results from these models fit are given in Supplemental Table C2, and plots of the resulting covariate balance statistics are given in Supplemental Figures C1 and C2. As shown, the SES covariate balance was improved across all variables after adjustment by both the unweighted and survey-weighted propensity scores, bringing all below a threshold of 0.1 for both the absolute standardized mean differences and Kolmogorov-Smirnov statistics. The propensity scores from these models were then carried forward to our later outcome models in order to estimate the ACD of telomere length by race.

\subsection{Results}

We compared our proposed approaches to six methods that one might consider in practice, each of which successively increase in their respective adjustments for potential confounding and selection bias. Namely, we fit a multiple linear regression (Approach 2 in Table \ref{tab:methods}) of log-transformed telomere length on race, age, sex, the blood composition variables, and the socioeconomic indicators. We then fit a survey-weighted multiple regression (Approach 4), and inverse probability of treatment weighted (IPTW) multiple regression (Approach 5), a regression weighted by the product of the within sample propensity score weights and the survey weights (Approach 6), and a regression weighted by the product of the survey-weighted propensity score weights and the survey weights (Approach 7), not unlike our simulation study. We omitted the oracle regression and simple linear regression, as they were not applicable (as with the oracle regression) or inadvisable (as with the simple linear regression). We calculated
standard errors for the ACD with the proposed methods via Equation \eqref{eq:vcov_strat}, whereas for the existing methods, we calculate model- versus design-based standard errors where appropriate.

ACD estimates and corresponding 95\% confidence intervals across the various analytic approaches are presented in Table \ref{tab:nhanes_results} below. In general, the ACD estimates across different analytic strategies tend to attenuate in magnitude as we made more appropriate adjustments for confounding and selection bias. For example, the multiple linear regression had an ACD estimate of 0.0265, corresponding to approximately a 2.7\% longer telomere length, on average, among non-Hispanic Black individuals as compared to non-Hispanic White individuals, adjusting for all other factors (95\% CI: 0.0106-0.0423). In contrast, the ACD point estimates from our proposed approaches varied from 0.0122 (AIPW) to 0.0176 (OM), though these ACD estimates were not statistically different from zero. These results suggest that there is a strong confounding relationship between SES and race. Moreover, the methods which incorporated the NHANES stratified, clustered design tended to have more conservative standard error estimates, as evidenced by the wider confidence intervals (Table \ref{tab:nhanes_results}).

\begin{table*}[!ht]
    \centering
    \caption{Average controlled difference (ACD) estimates and corresponding 95\% confidence intervals across various analytic approaches for the comparison of effect of race (non-Hispanic Black versus non-Hispanic White participants) on log-transformed telomere length among $n = 5,270$ participants in the National Health and Nutrition Examination Survey (NHANES). IPTW: Inverse Probability of Treatment Weighting.}
    \vspace{2ex}
    \label{tab:nhanes_results}
    \begin{tabular*}{\textwidth}{@{\extracolsep\fill}lrr@{}}
    \toprule
    {\bf Method} & {\bf ACD Estimate} & {\bf 95\% Confidence Interval} \\
    \midrule
    Multiple Regression                                  & 0.0265 &  0.0106, 0.0424 \\ 
    IPTW Estimator                                       & 0.0262 &  0.0079, 0.0445 \\ 
    Survey-Weighted Multiple Regression                  & 0.0298 & -0.0010, 0.0606 \\ 
    IPTW Multiple Regression                             & 0.0183 &  0.0053, 0.0312 \\ 
    IPTW + Survey-Weighted Multiple Regression           & 0.0220 & -0.0088, 0.0527 \\ 
    Weighted IPTW + Survey-Weighted Multiple Regression  & 0.0186 & -0.0126, 0.0498 \\ 
    Proposed Outcome Modeling and Direct Standardization & 0.0176 & -0.0029, 0.0381 \\ 
    Proposed Inverse Probability Weighting 1             & 0.0141 & -0.0134, 0.0416 \\ 
    Proposed Inverse Probability Weighting 2             & 0.0131 & -0.0081, 0.0342 \\ 
    Proposed Augmented Inverse Probability Weighting     & 0.0122 & -0.0077, 0.0321 \\
    \bottomrule
    \end{tabular*}
\end{table*}

These results align with our analytic and simulation results in two ways. Firstly, we show analytically that if our conceptual diagram holds, the proposed approaches will lead to consistent estimates of the ACD, whereas methods which do not account for confounding and/or the sampling design appropriately will incur bias in estimating the ACD. This is further evidenced in our simulations, which show that the ACD is overestimated, on average, in the settings which closely reflect our assumed conceptual diagram for these data. The ACD estimates for the proposed approaches are roughly 50\% those of the existing approaches. Secondly, the methods which respect the complex survey design tend to have wider confidence intervals, appropriately, as individuals within sampling units are correlated. Lastly, in a supplementary analysis, we present a different example from NHANES which may be amenable to a causal interpretation (in the sense that counterfactual/potential outcomes and plausibly changing exposures are more meaningful). To illustrate how the proposed methods can, under stronger assumptions, identify the average difference in potential outcome means ({\it population average treatment effect}; PATE), we study the effect of exposure to a particular environmental contaminant, lead, on telomere length~\citep{zota2015associations}. Please see Supplement C for full details.

\section{Discussion}
\label{sec:discussion}

There is a growing literature on the use of propensity score weighting techniques in complex survey design settings~\citep{zanutto2006comparison, dugoff2014generalizing, ridgeway2015propensity, austin2018propensity, dong2020using, lenis2018measuring, yang2024propensity, mccaffrey2024estimating}. However, a consensus has not been reached for the setting we consider in this manuscript. Notably, propensity score methods have been developed and used primarily for simple random samples. When analyzing results from more complicated sampling schemes, such as the NHANES clustered, stratified multi-stage sample, one must make the distinction between the population and the sample as the target of inference~\citep{heckman2001instrumental}. This distinction between the target populations and target estimands is crucial, as researchers must be clear in their inferential statements at both the propensity score model and outcome model stages in their analysis. This article examines a particular consideration for drawing inference in which sample selection depends on the group variable of interest, motivated from a social epidemiology question regarding racial disparity and socioeconomic confounding in health outcomes. This work advances current literature by deriving new identification formulas and conducting new simulations for analyzing the structure of our motivating data. The previous works of \citet{ridgeway2015propensity} and \citet{yang2024propensity} studied the question of when to incorporate survey weights in multi-stage, propensity score-based modeling for continuous and binary outcomes, respectively. Empirically, the authors show that a two-stage modeling approaches where the survey weights are included in the calculation of the PATE, but not in the outcome model, yielded the best performance compared to other potential analysis strategies. However, among these two-stage approaches, including the survey weights in the propensity model versus not yielded comparable empirical performance. We note that these conclusions are consistent with other recent papers that study propensity score methods in complex survey designs, and this ambiguity in the context of our specific case is clarified by the factorization of the joint probability of being selected and belonging to the treatment/exposure group. Further, our results dovetail with \citet{mccaffrey2024estimating} on how and when to incorporate sampling weights into propensity score-based estimators. Like them, we show that the decision to weight the propensity score model hinges on the underlying ignorability assumptions, specifically, which covariates drive both treatment (or group membership) and selection. In our setting, with discrete comparison groups that influence sample selection, we further demonstrate two valid factorizations of the joint probability of group membership and selection. In the first, one estimates a survey weighted propensity score and directly uses the selection weight conditional on $A$ and $\boldsymbol{X}$. In the second, one uses an unweighted propensity model and must then marginalize the sample selection probability over $A$. This explicit marginalization is critical for identification of the marginal ACD. Further, our analytic and numeric results suggest that, in settings where our assumed structure among the variables holds, our proposed methods minimize the bias in estimating the average controlled difference and achieve nominal coverage rates, as compared to standard analytic methods. This represents a significant step forward in the capacity to infer disparities, as it explicitly incorporates the complex relationships between race, SES, and sample selection bias, providing a rigorous approach for studying health outcomes in diverse populations.

Further, we note that targeting the ACD as our estimand is particularly useful in contexts where the `exposure' group membership (in our context, race), is not manipulable. Unlike causal targets such as the ATE or ATT, which are rooted in the assumption that an intervention could, even if only hypothetically, be implemented~\citep[i.e., `no causation without manipulation,' see][]{holland1986statistics}, the ACD is meant to provide a descriptive comparison in observational studies, where the propensity score is used as a covariate balancing tool to facilitate a controlled comparison between exposure groups. Mechanically, we note that our estimators take the same form whether the target is the ACD or ATE, with the difference lying in the stronger assumptions needed to claim a causal target. These include the stable unit treatment value assumption, which underpins the existence of counterfactual potential outcomes, the addition of a (weak) treatment exchangeability assumption and a modification of the weak selection exchangeability assumption in the context of the potential outcomes. The ACD requires only the positivity assumptions and weak selection exchangeability. This distinction has been previously articulated in \citet{li2023using}, which emphasizes that ACD estimation requires fewer assumptions, even though its formula matches that of the ATE estimator. \citet{li2023using} discussed the ACD in the context of studying racial differences in healthcare expenditure. We took inspiration from this, and subsequent works, which have developed consensus around this distinction.

We believe that the context of studying racial disparities presents these particularities in such a way that should be rigorously studied for best practice recommendations, as racial disparities are of increasing interest and relevance in scientific research. Moreover, the assumed notion of `race' itself warrants discussion. If the scientific question under study seeks to understand the biological underpinnings of a so-called `race effect,' say due to differences in genetic ancestry, then we cannot posit a causal counterfactual framework, as `race' cannot be manipulated or thought of as a `cause.' However, if race is measured as a social construct, we can then discuss altering racialization as a lived experience~\citep{marcellesi2013race}. In our motivating example, we are interested in studying differences in telomere length between non-Hispanic Black and non-Hispanic White individuals. We find that evidence of racial differences in telomere length between Black and White individuals attenuates after accounting for available socioeconomic status variables in NHANES and after utilizing appropriate propensity score and survey weighting techniques. Among our proposed methods, we estimated the ACD in telomere length to be between 1.22\% and 1.76\%, though we failed to detect a statistically significant difference. This contrasts with the results from standard analytic approaches, which varied from 1.81\% to 3.0\% in this study. \citet{needham2013socioeconomic} estimated 4.8\% longer telomeres in Black individuals, on average, adjusting for education, income, health behavior, and body mass index. \citet{needham2020black} and \citet{roberts2022persistent} also suggested evidence of social and environmental factors playing a role in racial disparities in telomere length. Taken together, our proposed method contributes to the collection of strategies that can investigate the role of social determinants of health on disparities, and in particular through the use of nationally representative survey data.

Though we focus on a particular research question within a particular sampling design, these concepts readily extend to other settings where observational data are collected via an underlying sample selection mechanism. In our experience, the design does not fundamentally change the results patterns or our conclusions, however, a more systematic/comprehensive exploration of how our results may vary under a wide range of multi-stage sampling designs is important future work. More broadly, studying this type of problem is crucial to prevent research from worsening health inequities by inadequately considering confounding factors and biases arising from non-representative selection in databases. In particular, other large-scale databases, such as electronic medical records, provide valuable insights into real-world data but may be subject to their own selection biases. Areas of interest for future work include electronic health records data with unknown sample selection probabilities and inferential targets with a time-to-event outcome. Several conceptual questions about the in-sample or population generalizable propensity score have not yet been reconciled in the literature. For example, while this does not appear to be an issue in our data, another consideration may be practical violations of the positivity assumption. This may be more likely to happen under the motivating data structure, as we are estimating the joint distribution of `treatment' and selection. As we develop this method further in different contexts, we can consider calibration weighting methods for better balance in situations of non-positivity, which should be checked in practice. Further, causal principles related to this work may be extended to more complex relational diagrams or sampling designs. For instance, as certain biomedical data become inexpensive and easy to collect, but other endpoints are not, a natural question is whether this framework can be extended to two-stage sampling or sequential designs. Lastly, we can expand on the utility of the proposed method by studying deviations from the intended sampling design, such as non-response and dropout.

In conclusion, this work outlines approaches and novel estimands for using propensity score and survey weights in tandem to study controlled outcome differences when the groups under comparison influence sample selection. Though the conceptual diagram for this study is rather specific, this setup is common to many settings of health disparities research. That is, our proposed methods and investigation to estimate the ACD follow a motivating health disparities research question in which the sampling scheme and corresponding survey weights depend on the target of the propensity score model. We fill this gap in the literature by providing recommendation on the form of the appropriate estimators in this setting and provide analytic and numeric evidence of their properties. We have shown that when sample selection depends on the comparison groups of interest and/or potential confounders, we must use the direct standardization, propensity weighting, or a doubly robust combination of these strategies, as outlined here, rather than standard analytic approaches. It is our hope that this work will not only further our understanding of appropriate survey design considerations in this framework, but also make the task of analyzing such data in similar settings more accessible to researchers.

\newpage

\section*{Acknowledgments}

We thank Dr. Yajuan Si for helpful discussions on this topic. This work was supported in part by NIH/NIGMS R35 GM144128 and the Fred Hutchinson Cancer Center J.~Orin Edson Foundation Endowed Chair (SS), NIH/NIMH DP2MH122405, R01HD107015, and P2C HD042828 (THM), NSF DMS 1712933 and NIH/NCI UG3-CA267907 (BM), and NIH/NIGMS R01GM139926 (XS).


\bibliographystyle{abbrvnat}  

\bibliography{ref}  

\begin{thebibliography}{71}
\providecommand{\natexlab}[1]{#1}
\providecommand{\url}[1]{\texttt{#1}}
\expandafter\ifx\csname urlstyle\endcsname\relax
  \providecommand{\doi}[1]{doi: #1}\else
  \providecommand{\doi}{doi: \begingroup \urlstyle{rm}\Url}\fi

\bibitem[Arias and Xu(2022)]{arias2022united}
E.~Arias and J.~Xu.
\newblock United states life tables, 2019.
\newblock \emph{National Vital Statistics Reports: From the Centers for Disease
  Control and Prevention, National Center for Health Statistics, National Vital
  Statistics System}, 70\penalty0 (19):\penalty0 1--59, 2022.

\bibitem[Austin and Stuart(2015)]{austin2015moving}
P.~C. Austin and E.~A. Stuart.
\newblock Moving towards best practice when using inverse probability of
  treatment weighting (iptw) using the propensity score to estimate causal
  treatment effects in observational studies.
\newblock \emph{Statistics in medicine}, 34\penalty0 (28):\penalty0 3661--3679,
  2015.

\bibitem[Austin et~al.(2018)Austin, Jembere, and Chiu]{austin2018propensity}
P.~C. Austin, N.~Jembere, and M.~Chiu.
\newblock Propensity score matching and complex surveys.
\newblock \emph{Statistical methods in medical research}, 27\penalty0
  (4):\penalty0 1240--1257, 2018.

\bibitem[Blackburn et~al.(2015)Blackburn, Epel, and Lin]{blackburn2015human}
E.~H. Blackburn, E.~S. Epel, and J.~Lin.
\newblock Human telomere biology: a contributory and interactive factor in
  aging, disease risks, and protection.
\newblock \emph{Science}, 350\penalty0 (6265):\penalty0 1193--1198, 2015.

\bibitem[Boos et~al.(2013)Boos, Stefanski, et~al.]{boos2013essential}
D.~D. Boos, L.~A. Stefanski, et~al.
\newblock \emph{Essential statistical inference}.
\newblock Springer, 2013.

\bibitem[Cawthon(2002)]{cawthon2002telomere}
R.~M. Cawthon.
\newblock Telomere measurement by quantitative pcr.
\newblock \emph{Nucleic acids research}, 30\penalty0 (10):\penalty0 e47--e47,
  2002.

\bibitem[{Centers for Disease Control and
  Prevention}(2001{\natexlab{a}})]{cdc1999lab}
{Centers for Disease Control and Prevention}.
\newblock \emph{Laboratory Procedure Manual: Blood Cadmium and Lead, NHANES
  1999–2000}.
\newblock Atlanta, GA, 2001{\natexlab{a}}.
\newblock Method No. 1090A/02-OD.

\bibitem[{Centers for Disease Control and
  Prevention}(2001{\natexlab{b}})]{cdc2001lab}
{Centers for Disease Control and Prevention}.
\newblock \emph{Laboratory Procedure Manual: Blood Cadmium and Lead, NHANES
  2001–2002}.
\newblock Atlanta, GA, 2001{\natexlab{b}}.
\newblock Method No. 1090A/02-OD.

\bibitem[Curtin et~al.(2012)Curtin, Mohadjer, Dohrmann, Montaquila,
  Kruszan-Moran, Mirel, Carroll, Hirsch, Schober, and
  Johnson]{curtin2012national}
L.~R. Curtin, L.~K. Mohadjer, S.~M. Dohrmann, J.~M. Montaquila,
  D.~Kruszan-Moran, L.~B. Mirel, M.~D. Carroll, R.~Hirsch, S.~Schober, and
  C.~L. Johnson.
\newblock The national health and nutrition examination survey: Sample design,
  1999-2006.
\newblock \emph{Vital and health statistics}, 2\penalty0 (155):\penalty0 1--39,
  2012.

\bibitem[Dahabreh et~al.(2020)Dahabreh, Robertson, Steingrimsson, Stuart, and
  Hernan]{dahabreh2020extending}
I.~J. Dahabreh, S.~E. Robertson, J.~A. Steingrimsson, E.~A. Stuart, and M.~A.
  Hernan.
\newblock Extending inferences from a randomized trial to a new target
  population.
\newblock \emph{Statistics in medicine}, 39\penalty0 (14):\penalty0 1999--2014,
  2020.

\bibitem[Degtiar and Rose(2023)]{degtiar2023review}
I.~Degtiar and S.~Rose.
\newblock A review of generalizability and transportability.
\newblock \emph{Annual Review of Statistics and Its Application}, 10:\penalty0
  501--524, 2023.

\bibitem[Dong et~al.(2020)Dong, Stuart, Lenis, and Quynh~Nguyen]{dong2020using}
N.~Dong, E.~A. Stuart, D.~Lenis, and T.~Quynh~Nguyen.
\newblock Using propensity score analysis of survey data to estimate population
  average treatment effects: A case study comparing different methods.
\newblock \emph{Evaluation review}, 44\penalty0 (1):\penalty0 84--108, 2020.

\bibitem[DuGoff et~al.(2014)DuGoff, Schuler, and
  Stuart]{dugoff2014generalizing}
E.~H. DuGoff, M.~Schuler, and E.~A. Stuart.
\newblock Generalizing observational study results: applying propensity score
  methods to complex surveys.
\newblock \emph{Health services research}, 49\penalty0 (1):\penalty0 284--303,
  2014.

\bibitem[Elliot(2009)]{elliot2009combining}
M.~R. Elliot.
\newblock Combining data from probability and non-probability samples using
  pseudo-weights.
\newblock \emph{Survey Practice}, 2\penalty0 (6), 2009.

\bibitem[Freedman et~al.(1997)Freedman, Gates, Flanders, Van~Assendelft,
  Barboriak, Joesoef, and Byers]{freedman1997black}
D.~S. Freedman, L.~Gates, W.~D. Flanders, O.~W. Van~Assendelft, J.~J.
  Barboriak, M.~R. Joesoef, and T.~Byers.
\newblock Black/white differences in leukocyte subpopulations in men.
\newblock \emph{International journal of epidemiology}, 26\penalty0
  (4):\penalty0 757--764, 1997.

\bibitem[Fuentes et~al.(2022)Fuentes, L{\"u}dtke, and
  Robitzsch]{fuentes2022causal}
A.~Fuentes, O.~L{\"u}dtke, and A.~Robitzsch.
\newblock Causal inference with multilevel data: A comparison of different
  propensity score weighting approaches.
\newblock \emph{Multivariate Behavioral Research}, 57\penalty0 (6):\penalty0
  916--939, 2022.

\bibitem[Geronimus et~al.(2015)Geronimus, Pearson, Linnenbringer, Schulz,
  Reyes, Epel, Lin, and Blackburn]{geronimus2015race}
A.~T. Geronimus, J.~A. Pearson, E.~Linnenbringer, A.~J. Schulz, A.~G. Reyes,
  E.~S. Epel, J.~Lin, and E.~H. Blackburn.
\newblock Race-ethnicity, poverty, urban stressors, and telomere length in a
  detroit community-based sample.
\newblock \emph{Journal of health and social behavior}, 56\penalty0
  (2):\penalty0 199--224, 2015.

\bibitem[Hansen et~al.(2016)Hansen, Hunt, Stone, Horvath, Herbig, Ranciaro,
  Hirbo, Beggs, Reiner, Wilson, et~al.]{hansen2016shorter}
M.~E. Hansen, S.~C. Hunt, R.~C. Stone, K.~Horvath, U.~Herbig, A.~Ranciaro,
  J.~Hirbo, W.~Beggs, A.~P. Reiner, J.~G. Wilson, et~al.
\newblock Shorter telomere length in europeans than in africans due to
  polygenetic adaptation.
\newblock \emph{Human molecular genetics}, 25\penalty0 (11):\penalty0
  2324--2330, 2016.

\bibitem[Heckman and Vytlacil(2001)]{heckman2001instrumental}
J.~J. Heckman and E.~J. Vytlacil.
\newblock Instrumental variables, selection models, and tight bounds on the
  average treatment effect.
\newblock In \emph{Econometric Evaluation of Labour Market Policies}, pages
  1--15. Springer, 2001.

\bibitem[Holland(1986)]{holland1986statistics}
P.~W. Holland.
\newblock Statistics and causal inference.
\newblock \emph{Journal of the American statistical Association}, 81\penalty0
  (396):\penalty0 945--960, 1986.

\bibitem[Hornung and Reed(1990)]{hornung1990estimation}
R.~W. Hornung and L.~D. Reed.
\newblock Estimation of average concentration in the presence of nondetectable
  values.
\newblock \emph{Applied occupational and environmental hygiene}, 5\penalty0
  (1):\penalty0 46--51, 1990.

\bibitem[Horvitz and Thompson(1952)]{horvitz1952generalization}
D.~G. Horvitz and D.~J. Thompson.
\newblock A generalization of sampling without replacement from a finite
  universe.
\newblock \emph{Journal of the American statistical Association}, 47\penalty0
  (260):\penalty0 663--685, 1952.

\bibitem[Huber(1996)]{huber1996robust}
P.~J. Huber.
\newblock \emph{Robust statistical procedures}.
\newblock SIAM, 1996.

\bibitem[Imai and Ratkovic(2013)]{imai2013estimating}
K.~Imai and M.~Ratkovic.
\newblock {Estimating treatment effect heterogeneity in randomized program
  evaluation}.
\newblock \emph{The Annals of Applied Statistics}, 7\penalty0 (1):\penalty0 443
  -- 470, 2013.
\newblock \doi{10.1214/12-AOAS593}.
\newblock URL \url{https://doi.org/10.1214/12-AOAS593}.

\bibitem[Jiang et~al.(2007)Jiang, Ju, and Rudolph]{jiang2007telomere}
H.~Jiang, Z.~Ju, and K.~Rudolph.
\newblock Telomere shortening and ageing.
\newblock \emph{Zeitschrift f{\"u}r Gerontologie und Geriatrie}, 40\penalty0
  (5):\penalty0 314--324, 2007.

\bibitem[Kundu et~al.(2023)Kundu, Shi, Morrison, Barrett, and
  Mukherjee]{kundu2023framework}
R.~Kundu, X.~Shi, J.~Morrison, J.~Barrett, and B.~Mukherjee.
\newblock A framework for understanding selection bias in real-world healthcare
  data.
\newblock \emph{arXiv preprint arXiv:2304.04652}, 2023.

\bibitem[Lee et~al.(2010)Lee, Lessler, and Stuart]{lee2010improving}
B.~K. Lee, J.~Lessler, and E.~A. Stuart.
\newblock Improving propensity score weighting using machine learning.
\newblock \emph{Statistics in medicine}, 29\penalty0 (3):\penalty0 337--346,
  2010.

\bibitem[Lee et~al.(2011)Lee, Lessler, and Stuart]{lee2011weight}
B.~K. Lee, J.~Lessler, and E.~A. Stuart.
\newblock Weight trimming and propensity score weighting.
\newblock \emph{PloS one}, 6\penalty0 (3):\penalty0 e18174, 2011.

\bibitem[Lenis et~al.(2018)Lenis, Ackerman, and Stuart]{lenis2018measuring}
D.~Lenis, B.~Ackerman, and E.~A. Stuart.
\newblock Measuring model misspecification: application to propensity score
  methods with complex survey data.
\newblock \emph{Computational statistics \& data analysis}, 128:\penalty0
  48--57, 2018.

\bibitem[Li and Li(2023)]{li2023using}
F.~Li and F.~Li.
\newblock Using propensity scores for racial disparities analysis.
\newblock \emph{Observational Studies}, 9\penalty0 (1):\penalty0 59--68, 2023.

\bibitem[Li et~al.(2013)Li, Zaslavsky, and Landrum]{li2013propensity}
F.~Li, A.~M. Zaslavsky, and M.~B. Landrum.
\newblock Propensity score weighting with multilevel data.
\newblock \emph{Statistics in medicine}, 32\penalty0 (19):\penalty0 3373--3387,
  2013.

\bibitem[Lim et~al.(2018)Lim, Walley, Yuan, Liu, Dabral, Best, Grieve, Hampson,
  Wolfram, Woodward, et~al.]{lim2018minimizing}
J.~Lim, R.~Walley, J.~Yuan, J.~Liu, A.~Dabral, N.~Best, A.~Grieve, L.~Hampson,
  J.~Wolfram, P.~Woodward, et~al.
\newblock Minimizing patient burden through the use of historical subject-level
  data in innovative confirmatory clinical trials: review of methods and
  opportunities.
\newblock \emph{Therapeutic innovation \& regulatory science}, 52:\penalty0
  546--559, 2018.

\bibitem[Lin et~al.(2016)Lin, Cheon, Brown, Coccia, Puterman, Aschbacher,
  Sinclair, Epel, and Blackburn]{lin2016systematic}
J.~Lin, J.~Cheon, R.~Brown, M.~Coccia, E.~Puterman, K.~Aschbacher, E.~Sinclair,
  E.~Epel, and E.~H. Blackburn.
\newblock Systematic and cell type-specific telomere length changes in subsets
  of lymphocytes.
\newblock \emph{Journal of immunology research}, 2016, 2016.

\bibitem[Lumley(2011)]{lumley2011complex}
T.~Lumley.
\newblock \emph{Complex surveys: a guide to analysis using R}.
\newblock John Wiley \& Sons, 2011.

\bibitem[Lumley and Scott(2017)]{lumley2017fitting}
T.~Lumley and A.~Scott.
\newblock Fitting regression models to survey data.
\newblock \emph{Statistical Science}, pages 265--278, 2017.

\bibitem[Mansournia and Altman(2016)]{mansournia2016inverse}
M.~A. Mansournia and D.~G. Altman.
\newblock Inverse probability weighting.
\newblock \emph{Bmj}, 352, 2016.

\bibitem[Marcellesi(2013)]{marcellesi2013race}
A.~Marcellesi.
\newblock Is race a cause?
\newblock \emph{Philosophy of Science}, 80\penalty0 (5):\penalty0 650--659,
  2013.

\bibitem[McCaffrey et~al.(2024)McCaffrey, Griffin, Robbins, Chakraborti,
  Coffman, and Vegetabile]{mccaffrey2024estimating}
D.~F. McCaffrey, B.~A. Griffin, M.~Robbins, Y.~Chakraborti, D.~L. Coffman, and
  B.~Vegetabile.
\newblock Estimating generalized propensity scores with survey and attrition
  weighted data.
\newblock \emph{Statistics in Medicine}, 43\penalty0 (11):\penalty0 2183--2202,
  2024.

\bibitem[Navas-Acien et~al.(2007)Navas-Acien, Guallar, Silbergeld, and
  Rothenberg]{navas2007lead}
A.~Navas-Acien, E.~Guallar, E.~K. Silbergeld, and S.~J. Rothenberg.
\newblock Lead exposure and cardiovascular disease—a systematic review.
\newblock \emph{Environmental health perspectives}, 115\penalty0 (3):\penalty0
  472--482, 2007.

\bibitem[Navas-Acien et~al.(2009)Navas-Acien, Tellez-Plaza, Guallar, Muntner,
  Silbergeld, Jaar, and Weaver]{navas2009blood}
A.~Navas-Acien, M.~Tellez-Plaza, E.~Guallar, P.~Muntner, E.~Silbergeld,
  B.~Jaar, and V.~Weaver.
\newblock Blood cadmium and lead and chronic kidney disease in us adults: a
  joint analysis.
\newblock \emph{American journal of epidemiology}, 170\penalty0 (9):\penalty0
  1156--1164, 2009.

\bibitem[Needham et~al.(2013)Needham, Adler, Gregorich, Rehkopf, Lin,
  Blackburn, and Epel]{needham2013socioeconomic}
B.~L. Needham, N.~Adler, S.~Gregorich, D.~Rehkopf, J.~Lin, E.~H. Blackburn, and
  E.~S. Epel.
\newblock Socioeconomic status, health behavior, and leukocyte telomere length
  in the national health and nutrition examination survey, 1999--2002.
\newblock \emph{Social science \& medicine}, 85:\penalty0 1--8, 2013.

\bibitem[Needham et~al.(2020)Needham, Salerno, Roberts, Boss, Allgood, and
  Mukherjee]{needham2020black}
B.~L. Needham, S.~Salerno, E.~Roberts, J.~Boss, K.~L. Allgood, and
  B.~Mukherjee.
\newblock Do black/white differences in telomere length depend on socioeconomic
  status?
\newblock \emph{Biodemography and social biology}, 65\penalty0 (4):\penalty0
  287--312, 2020.

\bibitem[Pattanayak et~al.(2011)Pattanayak, Rubin, and
  Zell]{pattanayak2011propensity}
C.~W. Pattanayak, D.~B. Rubin, and E.~R. Zell.
\newblock Propensity score methods for creating covariate balance in
  observational studies.
\newblock \emph{Revista Espa{\~n}ola de Cardiolog{\'\i}a (English Edition)},
  64\penalty0 (10):\penalty0 897--903, 2011.

\bibitem[Peng et~al.(2017)Peng, Qu, McDonald, Hollander, Bernstein, Maeda,
  Kaufman, Rosenthal, McElhinney, and Almond]{peng2017impact}
D.~M. Peng, Q.~Qu, N.~McDonald, S.~A. Hollander, D.~Bernstein, K.~Maeda, B.~D.
  Kaufman, D.~N. Rosenthal, D.~B. McElhinney, and C.~S. Almond.
\newblock Impact of the 18th birthday on waitlist outcomes among young adults
  listed for heart transplant: A regression discontinuity analysis.
\newblock \emph{The Journal of Heart and Lung Transplantation}, 36\penalty0
  (11):\penalty0 1185--1191, 2017.

\bibitem[Rao(1971)]{rao1971some}
P.~Rao.
\newblock Some notes on misspecification in multiple regressions.
\newblock \emph{The American Statistician}, 25\penalty0 (5):\penalty0 37--39,
  1971.

\bibitem[Rehkopf and Needham(2019)]{rehkopf2019impact}
D.~H. Rehkopf and B.~L. Needham.
\newblock The impact of race and ethnicity in the social epigenomic regulation
  of disease.
\newblock In \emph{Nutritional Epigenomics}, pages 51--65. Elsevier, 2019.

\bibitem[Rehkopf et~al.(2008)Rehkopf, Berkman, Coull, and
  Krieger]{rehkopf2008non}
D.~H. Rehkopf, L.~F. Berkman, B.~Coull, and N.~Krieger.
\newblock The non-linear risk of mortality by income level in a healthy
  population: Us national health and nutrition examination survey mortality
  follow-up cohort, 1988--2001.
\newblock \emph{BMC Public Health}, 8\penalty0 (1):\penalty0 1--11, 2008.

\bibitem[Richardson and Rotnitzky(2014)]{richardson2014causal}
T.~Richardson and A.~Rotnitzky.
\newblock Causal etiology of the research of james m. robins.
\newblock \emph{Statistical Science}, 29\penalty0 (4):\penalty0 459--484, 2014.

\bibitem[Ridgeway et~al.(2015)Ridgeway, Kovalchik, Griffin, and
  Kabeto]{ridgeway2015propensity}
G.~Ridgeway, S.~A. Kovalchik, B.~A. Griffin, and M.~U. Kabeto.
\newblock Propensity score analysis with survey weighted data.
\newblock \emph{Journal of causal inference}, 3\penalty0 (2):\penalty0
  237--249, 2015.

\bibitem[Roberts et~al.(2022)Roberts, Boss, Mukherjee, Salerno, Zota, and
  Needham]{roberts2022persistent}
E.~K. Roberts, J.~Boss, B.~Mukherjee, S.~Salerno, A.~Zota, and B.~L. Needham.
\newblock Persistent organic pollutant exposure contributes to black/white
  differences in leukocyte telomere length in the national health and nutrition
  examination survey.
\newblock \emph{Scientific Reports}, 12\penalty0 (1):\penalty0 19960, 2022.

\bibitem[Rose et~al.(2005)Rose, Pevalin, and O'Reilly]{rose2005national}
D.~Rose, D.~J. Pevalin, and K.~O'Reilly.
\newblock \emph{The National Statistics Socio-economic Classification: origins,
  development and use}.
\newblock Palgrave Macmillan, 2005.

\bibitem[Rosenbaum and Rubin(1983)]{rosenbaum1983central}
P.~R. Rosenbaum and D.~B. Rubin.
\newblock The central role of the propensity score in observational studies for
  causal effects.
\newblock \emph{Biometrika}, 70\penalty0 (1):\penalty0 41--55, 1983.

\bibitem[Rossen and Talih(2014)]{rossen2014social}
L.~M. Rossen and M.~Talih.
\newblock Social determinants of disparities in weight among us children and
  adolescents.
\newblock \emph{Annals of epidemiology}, 24\penalty0 (10):\penalty0 705--713,
  2014.

\bibitem[Rubin(1980)]{rubin1980randomization}
D.~B. Rubin.
\newblock Randomization analysis of experimental data: The fisher randomization
  test comment.
\newblock \emph{Journal of the American statistical association}, 75\penalty0
  (371):\penalty0 591--593, 1980.

\bibitem[Rubin(2005)]{rubin2005causal}
D.~B. Rubin.
\newblock Causal inference using potential outcomes: Design, modeling,
  decisions.
\newblock \emph{Journal of the American Statistical Association}, 100\penalty0
  (469):\penalty0 322--331, 2005.

\bibitem[Salvatore et~al.(2024)Salvatore, Kundu, Shi, Friese, Lee, Fritsche,
  Mondul, Hanauer, Pearce, and Mukherjee]{salvatore2024weight}
M.~Salvatore, R.~Kundu, X.~Shi, C.~R. Friese, S.~Lee, L.~G. Fritsche, A.~M.
  Mondul, D.~A. Hanauer, C.~L. Pearce, and B.~Mukherjee.
\newblock To weight or not to weight? studying the effect of selection bias in
  three large ehr-linked biobanks.
\newblock \emph{medRxiv}, pages 2024--02, 2024.

\bibitem[Shammas(2011)]{shammas2011telomeres}
M.~A. Shammas.
\newblock Telomeres, lifestyle, cancer, and aging.
\newblock \emph{Current opinion in clinical nutrition and metabolic care},
  14\penalty0 (1):\penalty0 28, 2011.

\bibitem[Shi et~al.(2023)Shi, Pan, and Miao]{shi2023data}
X.~Shi, Z.~Pan, and W.~Miao.
\newblock Data integration in causal inference.
\newblock \emph{Wiley Interdisciplinary Reviews: Computational Statistics},
  15\penalty0 (1):\penalty0 e1581, 2023.

\bibitem[Stefanski and Boos(2002)]{stefanski2002calculus}
L.~A. Stefanski and D.~D. Boos.
\newblock The calculus of m-estimation.
\newblock \emph{The American Statistician}, 56\penalty0 (1):\penalty0 29--38,
  2002.

\bibitem[Stuart et~al.(2011)Stuart, Cole, Bradshaw, and Leaf]{stuart2011use}
E.~A. Stuart, S.~R. Cole, C.~P. Bradshaw, and P.~J. Leaf.
\newblock The use of propensity scores to assess the generalizability of
  results from randomized trials.
\newblock \emph{Journal of the Royal Statistical Society Series A: Statistics
  in Society}, 174\penalty0 (2):\penalty0 369--386, 2011.

\bibitem[Venkateswaran et~al.(2024)Venkateswaran, Sankar, Chandrasekhar, and
  McCormick]{venkateswaran2024robustly}
A.~Venkateswaran, A.~Sankar, A.~G. Chandrasekhar, and T.~H. McCormick.
\newblock Robustly estimating heterogeneity in factorial data using rashomon
  partitions.
\newblock \emph{arXiv preprint arXiv:2404.02141}, 2024.

\bibitem[Williams et~al.(2010)Williams, Mohammed, Leavell, and
  Collins]{williams2010race}
D.~R. Williams, S.~A. Mohammed, J.~Leavell, and C.~Collins.
\newblock Race, socioeconomic status, and health: complexities, ongoing
  challenges, and research opportunities.
\newblock \emph{Annals of the new York Academy of Sciences}, 1186\penalty0
  (1):\penalty0 69--101, 2010.

\bibitem[Williams et~al.(2019)Williams, Grajales, and
  Kurkiewicz]{williams2019assumptions}
M.~N. Williams, C.~A.~G. Grajales, and D.~Kurkiewicz.
\newblock Assumptions of multiple regression: Correcting two misconceptions.
\newblock \emph{Practical Assessment, Research, and Evaluation}, 18\penalty0
  (1):\penalty0 11, 2019.

\bibitem[Wilms et~al.(2021)Wilms, M{\"a}thner, Winnen, and
  Lanwehr]{wilms2021omitted}
R.~Wilms, E.~M{\"a}thner, L.~Winnen, and R.~Lanwehr.
\newblock Omitted variable bias: A threat to estimating causal relationships.
\newblock \emph{Methods in Psychology}, 5:\penalty0 100075, 2021.

\bibitem[Wu et~al.(2012)Wu, Liu, Ni, Bao, Zhang, and Lu]{wu2012high}
Y.~Wu, Y.~Liu, N.~Ni, B.~Bao, C.~Zhang, and L.~Lu.
\newblock High lead exposure is associated with telomere length shortening in
  chinese battery manufacturing plant workers.
\newblock \emph{Occupational and environmental medicine}, 69\penalty0
  (8):\penalty0 557--563, 2012.

\bibitem[Yang et~al.(2024)Yang, Cuerden, Zhang, Aldridge, and
  Li]{yang2024propensity}
C.~Yang, M.~S. Cuerden, W.~Zhang, M.~Aldridge, and L.~Li.
\newblock Propensity score weighting with survey weighted data when outcomes
  are binary: a simulation study.
\newblock \emph{Health Services and Outcomes Research Methodology}, 24\penalty0
  (3):\penalty0 327--347, 2024.

\bibitem[Zanutto(2006)]{zanutto2006comparison}
E.~L. Zanutto.
\newblock A comparison of propensity score and linear regression analysis of
  complex survey data.
\newblock \emph{Journal of data Science}, 4\penalty0 (1):\penalty0 67--91,
  2006.

\bibitem[Zeng et~al.(2021)Zeng, Li, Wang, and Li]{zeng2021propensity}
S.~Zeng, F.~Li, R.~Wang, and F.~Li.
\newblock Propensity score weighting for covariate adjustment in randomized
  clinical trials.
\newblock \emph{Statistics in medicine}, 40\penalty0 (4):\penalty0 842--858,
  2021.

\bibitem[Zeng et~al.(2025)Zeng, Li, and Tong]{zeng2025moving}
Y.~Zeng, F.~Li, and G.~Tong.
\newblock Moving toward best practice when using propensity score weighting in
  survey observational studies.
\newblock \emph{arXiv preprint arXiv:2501.16156}, 2025.

\bibitem[Zota and VanNoy(2021)]{zota2021integrating}
A.~R. Zota and B.~N. VanNoy.
\newblock Integrating intersectionality into the exposome paradigm: a novel
  approach to racial inequities in uterine fibroids.
\newblock \emph{American Journal of Public Health}, 111\penalty0 (1):\penalty0
  104--109, 2021.

\bibitem[Zota et~al.(2015)Zota, Needham, Blackburn, Lin, Park, Rehkopf, and
  Epel]{zota2015associations}
A.~R. Zota, B.~L. Needham, E.~H. Blackburn, J.~Lin, S.~K. Park, D.~H. Rehkopf,
  and E.~S. Epel.
\newblock Associations of cadmium and lead exposure with leukocyte telomere
  length: findings from national health and nutrition examination survey,
  1999--2002.
\newblock \emph{American journal of epidemiology}, 181\penalty0 (2):\penalty0
  127--136, 2015.

\end{thebibliography}

\newpage


\appendix

\setcounter{table}{0}
\renewcommand{\thetable}{A\arabic{table}}


\section{Additional Analytic Results}
\label{sec:a}

\setcounter{table}{0}
\renewcommand{\thetable}{A\arabic{table}}

\setcounter{figure}{0}
\renewcommand{\thefigure}{A\arabic{figure}}

\vspace{2ex}

\subsection{Identification of the Average Controlled Difference}
\label{sec:a:acd}

\subsubsection{Setup and Assumptions}

We work under the sample law $\mathcal{Q} = \Pr(\ \cdot \mid S = 1)$, rather than the population law, $\mathcal{P}$, as $(Y, A, \boldsymbol{X})$ are observed only when $S = 1$. We write $e^{\mathcal{Q}}_a(\boldsymbol{X}) = \mathcal{Q}(A = a \mid \boldsymbol{X})$, $g_a(\boldsymbol{X}) = \mathbb{E}_{\mathcal{Q}}[Y \mid A = a, \boldsymbol{X}]$, and let $\Pr(S = 1 \mid \boldsymbol{X})$ denote the known selection probability marginalized over $A$. For $a \in \{0,1\}$, the averaged controlled difference (ACD) is ${\rm ACD} = \mu(1) - \mu(0)$ with
\[
    \mu(a) = \mathbb{E}_{\mathcal{P}}\left[\mathbb{E}(Y \mid A = a, \boldsymbol{X})\right],
\]
which we will show is equal to $\mathbb{E}_{\mathcal{Q}}\left[w(\boldsymbol{X}) g_a(\boldsymbol{X})\right]$, with known $w(\boldsymbol{X}) = \frac{\Pr(S = 1)}{\Pr(S = 1\mid \boldsymbol{X})}$. The assumptions required to estimate $\mu(a)$, and consequently the ACD, are relatively weaker than those required to identify the causal potential outcome means (see below). In particular, the following assumptions are sufficient for showing that our functionals, Equations (1), (3), and (4) in the main text, estimate the $\mu(a)$:
\begin{enumerate}[label={(\roman*)}]
    \item {\it Positivity:} $e^{\mathcal{Q}}_a(\boldsymbol{X}) := \mathcal{Q}(A = a \mid \boldsymbol{X}) > 0$ $\mathcal{Q}$-a.s. $\forall a\in\mathcal A$
    \item {\it Selection Positivity:} $\pi_a(\boldsymbol{X}) := \Pr(S = 1 \mid A = a, \boldsymbol{X}) > 0$ on the support of $A, \boldsymbol{X}$
    \item {\it Weak Selection Exchangeability:} $\mathbb{E}_{\mathcal{P}}(Y \mid A = a, \boldsymbol{X}) = \mathbb{E}_{\mathcal{Q}}(Y \mid A = a, \boldsymbol{X}) = g_a(\boldsymbol{X})$.
\end{enumerate}

\subsubsection{Change of Measure Under the Sample Law}

For any integrable $h(\boldsymbol{X})$,
\[
\begin{aligned}
    \mathbb{E}_{\mathcal{Q}}\left[w(\boldsymbol{X}) h(\boldsymbol{X})\right] &= \int w(\boldsymbol{x}) h(\boldsymbol{x}) f_{\mathcal{Q}}(\boldsymbol{x}) d\boldsymbol{x}; \quad \text{(definition of expectation)} \\
    &= \int \frac{\Pr(S = 1)}{\Pr(S = 1 \mid \boldsymbol{x})} h(\boldsymbol{x}) \frac{\Pr(S = 1 \mid \boldsymbol{x}) f_{\mathcal{P}}(\boldsymbol{x})}{\Pr(S = 1)} d\boldsymbol{x}; \quad \text{(Bayes: } f_{\mathcal{Q}}(\boldsymbol{x}) = f_{\mathcal{P}}(\boldsymbol{x} \mid S = 1)\text{)} \\
    &= \int h(\boldsymbol{x}) f_{\mathcal{P}}(\boldsymbol{X}) d\boldsymbol{x} \\
    &= \mathbb{E}_{\mathcal{P}}\left[h(\boldsymbol{X})\right]. 
\end{aligned}
\]
Hence $\mathbb{E}_{\mathcal{P}}\left[h(\boldsymbol{X})\right] = \mathbb{E}_{\mathcal{Q}}\left[w(\boldsymbol{X}) h(\boldsymbol{X})\right]$.

\subsubsection{Outcome Modeling and Direct Standardization} 

We first show that our Outcome Modeling and Direct Standardization estimator (OM) identifies $\mu(a)$. Assumptions (i) and (ii) ensure that $\mu(a)$ is well-defined. Focusing on $\mu(a) = \mathbb{E}_{\mathcal{P}}\left[\mathbb{E}(Y \mid A = a, \boldsymbol{X})\right]$, we have that
\[
\begin{aligned}
    \mu(a) &= \mathbb{E}_{\mathcal{P}}\left[\mathbb{E}(Y \mid A = a, \boldsymbol{X})\right]; \quad & \text{(definition of $\mu(a)$)} \\
    &= \mathbb{E}_{\mathcal{P}}\left[g_a(\boldsymbol{X})\right]; \quad & \text{(assumption (iii))} \\
    &= \mathbb{E}_\mathcal{Q}\left[w(\boldsymbol{X}) g_a(\boldsymbol{X})\right]; \quad & \text{(change of measure under $\mathcal{Q}$).}
\end{aligned}
\]
This shows the equivalence between Main Eq.~(1) and $\mathbb{E}_{\mathcal{P}}\left[\mathbb{E}(Y \mid A = a, \boldsymbol{X})\right]$.

\subsubsection{Inverse Probability Weighting 2}

We next show that our Inverse Probability Weighting 2 (IPW2) estimator identifies $\mu(a)$, as this follows from the above. Again, assumptions (i) and (ii) ensure that $\mu(a)$ is well-defined. Consider $\mathbb{E}_{\mathcal{Q}}\left[ w(\boldsymbol{X}) \frac{\mathbb{I}(A = a)}{e^{\mathcal{Q}}_a(\boldsymbol{X})} Y\right]$. By iterated expectation,
\[
\begin{aligned}
    \mathbb{E}_{\mathcal{Q}}\left[w(\boldsymbol{X}) \frac{\mathbb{I}(A = a)}{e^{\mathcal{Q}}_a(\boldsymbol{X})} Y \right] &= \mathbb{E}_{\mathcal{Q}}\left[w(\boldsymbol{X}) \mathbb{E}_{\mathcal{Q}}\left\{\left.\frac{\mathbb{I}(A = a)}{e^{\mathcal{Q}}_a(\boldsymbol{X})} Y \right| \boldsymbol{X}\right\}\right]; \quad & \text{(law of iterated expectations)} \\
    &= \mathbb{E}_{\mathcal{Q}}\left[w(\boldsymbol{X}) \frac{1}{e^{\mathcal{Q}}_a(\boldsymbol{X})} \mathbb{E}_{\mathcal{Q}}\left\{\left.\mathbb{I}(A = a) Y \right| \boldsymbol{X}\right\}\right]; \quad & \text{(pulling out }e^{\mathcal{Q}}_a\text{)} \\
    &= \mathbb{E}_{\mathcal{Q}}\left[w(\boldsymbol{X}) \frac{1}{e^{\mathcal{Q}}_a(\boldsymbol{X})} \mathbb{E}_{\mathcal{Q}}\left\{\left.\mathbb{I}(A = a) \mathbb{E}_{\mathcal{Q}}[Y\mid A, \boldsymbol{X}] \right| \boldsymbol{X}\right\}\right]; \quad & \text{(tower property)} \\
    &= \mathbb{E}_{\mathcal{Q}}\left[w(\boldsymbol{X}) \frac{1}{e^{\mathcal{Q}}_a(\boldsymbol{X})} e^{\mathcal{Q}}_a(\boldsymbol{X}) g_a(\boldsymbol{X})\right]; \quad & \text{(since } \mathbb{E}_{\mathcal{Q}}[\mathbb{I}(A = a) \mid \boldsymbol{X}] = e^{\mathcal{Q}}_a(\boldsymbol{X})) \\
    &= \mathbb{E}_{\mathcal{Q}}\left[w(\boldsymbol{X}) g_a(\boldsymbol{X})\right]; \quad & \text{(=} \mu(a) \text{ from above).}
\end{aligned}
\]
This shows the equivalence between Main Eq.~(4) and $\mathbb{E}_{\mathcal{P}}\left[\mathbb{E}(Y \mid A = a, \boldsymbol{X})\right]$.

\subsubsection{Inverse Probability Weighting 1}

We now come back to the Inverse Probability Weighting 1 estimator (IPW1). We define $e^{\mathcal{P}}_a(\boldsymbol{X}) = \mathcal{P}(A = a \mid \boldsymbol{X})$, where $\mathcal{P}$ denotes the population law. Further, $\pi_a(\boldsymbol{X}) = \Pr(S = 1 \mid A = a, \boldsymbol{X})$, so that $\pi(\boldsymbol{X}) = \Pr(S = 1 \mid \boldsymbol{X}) = \sum_{a} e^{\mathcal{P}}_{a}(\boldsymbol{X}) \pi_{a}(\boldsymbol{X})$. As before, $w(\boldsymbol{X}) = \tfrac{\Pr(S = 1)}{\pi(\boldsymbol{X})}$, and we now also define $w_a(\boldsymbol{X}) = \tfrac{\Pr(S = 1)}{\pi_a(\boldsymbol{X})}$. By Bayes' rule,
\[
    e^{\mathcal{Q}}_a(\boldsymbol{X}) = \frac{\Pr(S = 1 \mid A = a, \boldsymbol{X}) \mathcal{P}(A = a \mid \boldsymbol{X})}{\Pr(S = 1 \mid \boldsymbol{X})} = \frac{\pi_a(\boldsymbol{X}) e^{\mathcal{P}}_a(\boldsymbol{X})}{\pi(\boldsymbol{X})}.
\]
Consequently,
\[
    \frac{w(\boldsymbol{X})}{e^{\mathcal{Q}}_a(\boldsymbol{X})} = \frac{\Pr(S = 1) / \pi(\boldsymbol{X})}{\pi_a(\boldsymbol{X}) e^{\mathcal{P}}_a(\boldsymbol{X}) / \pi(\boldsymbol{X})} = \frac{\Pr(S = 1)}{\pi_a(\boldsymbol{X}) e^{\mathcal{P}}_a(\boldsymbol{X})} = \frac{w_a(\boldsymbol{X})}{e^{\mathcal{P}}_a(\boldsymbol{X})}.
\]
Again, assumptions (i) and (ii) ensure that $\mu(a)$ is well-defined. Consider $\mathbb{E}_{\mathcal{Q}}\left[w_a(\boldsymbol{X}) \frac{\mathbb{I}(A = a)}{e^{\mathcal{P}}_a(\boldsymbol{X})} Y\right]$. By the identity above, 
\[
\begin{aligned}
    \mathbb{E}_{\mathcal{Q}}\left[w_a(\boldsymbol{X}) \frac{\mathbb{I}(A = a)}{e^{\mathcal{P}}_a(\boldsymbol{X})} Y\right] &= \mathbb{E}_{\mathcal{Q}}\left[w(\boldsymbol{X}) \frac{\mathbb{I}(A = a)}{e^{\mathcal{Q}}_a(\boldsymbol{X})} Y \right]; \quad & \text{(algebraically equivalent)} \\
    &= \mathbb{E}_{\mathcal{Q}}\left[w(\boldsymbol{X}) g_a(\boldsymbol{X})\right]; \quad & \text{(=} \mu(a) \text{ from above).}
\end{aligned}
\]
This shows the equivalence between Main Eq.~(3) and $\mathbb{E}_{\mathcal{P}}\left[\mathbb{E}(Y \mid A = a, \boldsymbol{X})\right]$.

\subsection{Identification of the Potential Outcome Means}
\label{sec:a:ate}

Though not applicable to our motivating research question, in many scenarios the group membership variable of interest is a `treatment' or `exposure' in the more traditional sense, and we seek to target a causal quantity. To that end, we can adopt a counterfactual framework and, under stronger assumptions, show that the proposed estimators equivalently identify such a parameter. We will introduce some additional notation. Take $Y^{a}$ to be the random variable representing the {\it potential counterfactual outcome} within each group~\citep{rubin2005causal}. That is, despite only observing one outcome per individual, we know that each study participant had two potential outcomes -- the outcome that would be observed if the individual was in comparison group $a = 0$, denoted by $Y^{0}$, and the potential outcome that would be observed if the participant instead was in comparison group $a = 1$, denoted by $Y^{1}$. By considering an individual's {\it potential outcome values} based on their observed and counterfactual group membership, an inherent missing data problem arises. Importantly, there are two layers to the missingness mechanism in our setup. Firstly, we observe no data when $S = 0$. Secondly, due to the missingness of counterfactual values, we cannot fully observe all potential outcomes, $Y^{a}$. Further, as before, $\mathbb{E}_{\mathcal{Q}}\left[Y^{a}\right] \neq \mathbb{E}_{\mathcal{P}}\left[Y^{a}\right]$ in general, as the sampled study participants may not be representative of the larger target population. We can, however, make assumptions to estimate the unobserved, counterfactual outcome in the population. With the assumptions of the previous section, the following additional conditions are sufficient to identify the population potential outcome means, $\mathbb{E}_{\mathcal{P}}[Y^{a}]\ \forall a\in\mathcal{A}$:

\begin{enumerate}[label={(\roman*)}]
    \setcounter{enumi}{3}
    \item[(iii')] {\it Weak Selection Exchangeability for Potential Outcomes:} $\mathbb{E}_{\mathcal{P}}[Y^a \mid A = a, \boldsymbol{X}] = \mathbb{E}_{\mathcal{Q}}[Y^a \mid A = a, \boldsymbol{X}]$, that is, selection bias comes from $A, \boldsymbol{X}$. This is implied by, but does not imply, strong generalizability (in distribution), $Y^{a} \perp S \mid A, \boldsymbol{X}$.
    \item {\it Stable Unit Treatment Value (SUTVA):} $Y_i = Y_i^{a}$ for all observations $i = 1, \ldots, n$ and for all $A_i = a \in \mathcal{A}.$ 
    \item {\it Mean (Weak) Exchangeability:} $\mathbb{E}_{\mathcal{P}}[Y^a \mid \boldsymbol{X}] = \mathbb{E}_{\mathcal{P}}[Y^a \mid A = a, \boldsymbol{X}]$, that is, confounding bias comes from $\boldsymbol{X}$. This is implied by, but does not imply, strong exchangeability, i.e., $Y^{a} \perp A \mid \boldsymbol{X}$.
\end{enumerate}

\noindent This is thematically similar to \citet{dahabreh2020extending}, who study the generalization of causal inferences from individuals in randomized trials. The authors also differentiate between conditional exchangeability in the mean over $A$ in a randomized population and conditional exchangeability in the mean over $S$ in the context of randomized trials.

\subsubsection{Outcome Modeling and Direct Standardization} 

As before, $\mu(a)$ is well-defined under (i) and (ii). SUTVA  underpins the existence of the counterfactual outcomes, $Y^a$, and (iv) captures two aspects of SUTVA: `no interference,' which states that the $i$th observation's outcome is unaffected by the treatment of other observations and `no hidden variation,' which states that the potential outcome is well-defined for each treatment level. By (iv), $\mathbb{E}_{\mathcal{Q}}[Y^a \mid A = a, \boldsymbol{X}] = g_a(\boldsymbol{X})$. Further, by (iii'), $ \mathbb{E}_{\mathcal{P}}[Y^a \mid A = a,\boldsymbol{X}] = g_a(\boldsymbol{X})$. To identify $\mathbb{E}_{\mathcal{P}}\left[Y^a\right]$,
\[
\begin{aligned}
    \mathbb{E}_{\mathcal{P}}[Y^{a}] &= \mathbb{E}_{\mathcal{P}}\left[\mathbb{E}_{\mathcal{P}}(Y^{a} \mid \boldsymbol{X})\right]; \quad & \text{(iterated expectation)} \\
    &= \mathbb{E}_{\mathcal{P}}\left[\mathbb{E}_{\mathcal{P}}(Y^{a} \mid A = a, \boldsymbol{X})\right]; \quad & \text{(v)} \\
    &= \mathbb{E}_{\mathcal{P}}\left[g_a(\boldsymbol{X})\right]; \quad & \text{(iii')} \\
    &= \mathbb{E}_{\mathcal{Q}}\left[w(\boldsymbol{X}) g_a(\boldsymbol{X})\right]; \quad & \text{(change of measure)}
\end{aligned}
\]
This shows the equivalence between Main Eq.~(1) and $\mathbb{E}_{\mathcal{P}}[Y^{a}]$.

\subsubsection{Inverse Probability Weighting 2}

Under assumptions (i) - (v), we further show that the observed data functional for our inverse probability weighted 2 estimator also identifies the potential outcome mean in the target population, $\mathbb{E}_{\mathcal{P}}[Y^a]$. As before, $\mu(a)$ is well-defined under (i) and (ii). Using the tower property under $\mathcal{Q}$,
\[
\begin{aligned}
    \mathbb{E}_{\mathcal{Q}}\left[w(\boldsymbol{X}) \frac{\mathbb{I}(A = a)}{e^{\mathcal{Q}}_a(\boldsymbol{X})} Y\right] &= \mathbb{E}_{\mathcal{Q}}\left[w(\boldsymbol{X}) \mathbb{E}_{\mathcal{Q}}\left\{\left.\frac{\mathbb{I}(A = a)}{e^{\mathcal{Q}}_a(\boldsymbol{X})} Y \right| \boldsymbol{X}\right\}\right]; \quad & \text{(law of iterated expectations under }\mathcal{Q}\text{)} \\
    &=\mathbb{E}_{\mathcal{Q}}\left[w(\boldsymbol{X}) \frac{1}{e^{\mathcal{Q}}_a(\boldsymbol{X})} \mathbb{E}_{\mathcal{Q}}\left\{\left.\mathbb{I}(A = a) Y \right| \boldsymbol{X}\right\}\right]; \quad & \text{(pulling out }e^{\mathcal{Q}}_a \text{)} \\
    &= \mathbb{E}_{\mathcal{Q}}\left[w(\boldsymbol{X}) \frac{1}{e^{\mathcal{Q}}_a(\boldsymbol{X})} \mathbb{E}_{\mathcal{Q}}\left\{\left.\mathbb{I}(A = a) \mathbb{E}_{\mathcal{Q}}[Y\mid A, \boldsymbol{X}] \right| \boldsymbol{X}\right\}\right]; \quad & \text{(tower property)} \\    
    &= \mathbb{E}_{\mathcal{Q}}\left[w(\boldsymbol{X}) \frac{1}{e^{\mathcal{Q}}_a(\boldsymbol{X})} \underbrace{\mathbb{E}_{\mathcal{Q}}\left\{\left.\mathbb{I}(A = a)\right|\boldsymbol{X}\right\}}_{= e^{\mathcal{Q}}_a(\boldsymbol{X})} \underbrace{\mathbb{E}_{\mathcal{Q}}[Y\mid A = a, \boldsymbol{X}]}_{= g_a(\boldsymbol{X})}\right]; \quad & \text{(definitions of }e^{\mathcal{Q}}_a\text{ and }g_a\text{)} \\
    &= \mathbb{E}_{\mathcal{Q}}\left[w(\boldsymbol{X}) g_a(\boldsymbol{X})\right]; \quad & \text{(algebra; }e^{\mathcal{Q}}_a>0\text{ by positivity (i))} \\ 
    &= \mathbb{E}_{\mathcal{P}}\left[g_a(\boldsymbol{X})\right]; \quad & \text{(change of measure)} \\
    &= \mathbb{E}_{\mathcal{P}}\left[\mathbb{E}_{\mathcal{Q}}[Y\mid A = a,\boldsymbol{X}]\right]; \quad & \text{(definition of } g_a\text{)} \\
    &= \mathbb{E}_{\mathcal{P}}\left[\mathbb{E}_{\mathcal{Q}}[Y^{a}\mid A = a,\boldsymbol{X}]\right]; \quad & \text{(iv)} \\
    &= \mathbb{E}_{\mathcal{P}}\left[\mathbb{E}_{\mathcal{P}}[Y^{a}\mid A = a,\boldsymbol{X}]\right]; \quad & \text{(iii')} \\
    &= \mathbb{E}_{\mathcal{P}}\left[\mathbb{E}_{\mathcal{P}}(Y^{a}\mid \boldsymbol{X})\right]; \quad & \text{(v)} \\
    &= \mathbb{E}_{\mathcal{P}}[Y^{a}]; \quad & \text{(law of iterated expectations under }\mathcal{P}\text{)}.
\end{aligned}
\]
This shows the equivalence between Main Eq.~(4) and $\mathbb{E}_{\mathcal{P}}[Y^{a}]$.

\subsubsection{Inverse Probability Weighting 1}

Lastly, we show that under assumptions (i) - (v), the observed data functional for our inverse probability weighted 1 estimator also identifies the potential outcome mean in the target population, $\mathbb{E}_{\mathcal{P}}[Y^a]$. As before, $\mu(a)$ is well-defined under (i) and (ii). Using the tower property under $\mathcal{Q}$,
\[
\begin{aligned}
    \mathbb{E}_{\mathcal{Q}}\left[w_a(\boldsymbol{X}) \frac{\mathbb{I}(A = a)}{e_a^{\mathcal{P}}(\boldsymbol{X})} Y\right] &= \mathbb{E}_{\mathcal{Q}}\left[w_a(\boldsymbol{X}) \mathbb{E}_{\mathcal{Q}}\left\{\left.\frac{\mathbb{I}(A = a)}{e_a^{\mathcal{P}}(\boldsymbol{X})} Y \right| \boldsymbol{X}\right\}\right]; \quad & \text{(law of iterated expectations under } \mathcal{Q} \text{)} \\
    &= \mathbb{E}_{\mathcal{Q}}\left[w_a(\boldsymbol{X}) \frac{1}{e_a^{\mathcal{P}}(\boldsymbol{X})} \mathbb{E}_{\mathcal{Q}}\left\{\mathbb{I}(A = a) Y \mid \boldsymbol{X}\right\}\right]; \quad & \text{(pulling out } e_a^{\mathcal{P}} \text{)} \\
    &= \mathbb{E}_{\mathcal{Q}}\left[w_a(\boldsymbol{X}) \frac{1}{e_a^{\mathcal{P}}(\boldsymbol{X})} \mathbb{E}_{\mathcal{Q}}\left\{\mathbb{I}(A = a) g_a(\boldsymbol{X})\mid \boldsymbol{X}\right\}\right]; \quad & \text{(tower property; definition of } g_a \text{)} \\
    &= \mathbb{E}_{\mathcal{Q}}\left[w_a(\boldsymbol{X}) \frac{\mathcal{Q}(A = a \mid \boldsymbol{X})}{e_a^{\mathcal{P}}(\boldsymbol{X})} g_a(\boldsymbol{X})\right]; \quad & \text{(definition of } \mathcal{Q}(A = a \mid \boldsymbol{X}) \text{)} \\
    &= \mathbb{E}_{\mathcal{Q}}\left[w_a(\boldsymbol{X}) \frac{\pi_a(\boldsymbol{X}) e_a^{\mathcal{P}}(\boldsymbol{X})}{\pi(\boldsymbol{X}) e_a^{\mathcal{P}}(\boldsymbol{X})}  g_a(\boldsymbol{X})\right]; \quad &\text{(Bayes under } \mathcal{P}: e_a^{Q} = \pi_a e_a^{\mathcal{P}}/\pi \text{)} \\
    &= \mathbb{E}_{\mathcal{Q}}\left[\frac{\Pr(S = 1)}{\pi_a(\boldsymbol{X})}\cdot \frac{\pi_a(\boldsymbol{X})}{\pi(\boldsymbol{X})} g_a(\boldsymbol{X})\right]; \quad & \text{(substituting } w_a = \Pr(S = 1) / \pi_a \text{)} \\
    &= \mathbb{E}_{\mathcal{Q}}\left[w(\boldsymbol{X}) g_a(\boldsymbol{X})\right]; \quad & \text{(since } w = \Pr(S = 1)/\pi \text{)} \\
    &= \mathbb{E}_{\mathcal{P}}\left[g_a(\boldsymbol{X})\right]; & \text{(change of measure)} \\
    &= \mathbb{E}_{\mathcal{P}}\left[\mathbb{E}_{\mathcal{Q}}(Y^a \mid A = a, \boldsymbol{X})\right]; \quad & \text{(iv)} \\
    &= \mathbb{E}_{\mathcal{P}}\left[\mathbb{E}_{\mathcal{P}}(Y^a \mid A = a, \boldsymbol{X})\right]; \quad & \text{(iii')} \\
    &= \mathbb{E}_{\mathcal{P}}\left[\mathbb{E}_{\mathcal{P}}(Y^a \mid \boldsymbol{X})\right]; & \text{(v)} \\
    &= \mathbb{E}_{\mathcal{P}}[Y^a]; \quad & \text{(law of iterated expectations under } \mathcal{P} \text{).}
\end{aligned}
\]
This shows the equivalence between Main Eq.~(3) and $\mathbb{E}_{\mathcal{P}}[Y^{a}]$.

\subsection{Augmented Inverse Probability Weighting}

\subsubsection{Efficient Influence Function}

{\bf Data Structure:} Recall that we observe i.i.d.~draws, $\mathcal{O} = (\boldsymbol{X}, A, Y)$, from a joint law, $\mathcal{Q}$, where 
\[
    \mathcal{Q}(\cdot) = \Pr(\ \cdot \mid S = 1).
\]
We will derive the efficient influence function (EIF) for our setting under this conditional (sample) law, $\mathcal{Q}$, which restricts to $S = 1$, where $S \in \{0, 1\}$ is a selection/observation indicator, as this is where $\boldsymbol{X}, A, Y$ are observable. Throughout, expectations, $\mathbb{E}(\cdot)$, are taken under $\mathcal{Q}$, unless stated otherwise. \\

\noindent {\bf Nuisance Functions:} We further define the following nuisances for the propensity model, $e^{\mathcal{Q}}_a(\cdot)$, and the outcome model, $g_a(\cdot)$ under $\mathcal{Q}$, as
\[
    e^{\mathcal{Q}}_a(\boldsymbol{x}) = \mathcal{Q}(A = a \mid \boldsymbol{X} = \boldsymbol{x}),\quad
    g_a(\boldsymbol{x}) = \mathbb{E}_\mathcal{Q}[Y \mid A = a, \boldsymbol{X} = \boldsymbol{x}], 
\]
and we denote the design (i.e., selection) weights by
\[
    w(\boldsymbol{x}) = \frac{\Pr(S = 1)}{\Pr(S = 1 \mid \boldsymbol{X} = \boldsymbol{x})} = \frac{\Pr(S = 1)}{\pi(\boldsymbol{x})}.
\]
In practice, $w(\cdot)$ is either known from the design (preferred, as $\boldsymbol{X}$ is not observed for $S = 0$) or estimated from external information. In our setting, and throughout, we treat $w(\cdot)$ as a known nuisance function. \\

\noindent {\bf Target Parameter:} Under the assumption of weak selection exchangeability in the main text,
\[
    \mu(a) = \mathbb{E}_\mathcal{P}\left[g_a(\boldsymbol{X})\right] = \mathbb{E}_\mathcal{Q}\left[w(\boldsymbol{X}) g_a(\boldsymbol{X})\right],
\]
where $\mathcal{P}(\cdot) = \Pr(\cdot)$ is the `full law,' and the above reflects the Bayes change of measure,
\[
    f_{\boldsymbol{X}}(\boldsymbol{x}) = f_{\boldsymbol{X} \mid S = 1}(\boldsymbol{x})\frac{\Pr(S = 1)}{\pi(\boldsymbol{x})}.
\]

\noindent {\bf Tangent Space Under $\mathcal{Q}$:} Any regular parametric submodel, $Q_\varepsilon$, of $\mathcal{Q}$ has score
\[
    s(\mathcal{O}) = s_{\boldsymbol{X}}(\boldsymbol{X}) + s_{A \mid \boldsymbol{X}}(A \mid \boldsymbol{X}) + s_{Y \mid A, \boldsymbol{X}}(Y \mid A, \boldsymbol{X}),
\]
with the usual conditional mean-zero properties:
\[
    \mathbb{E}_\mathcal{Q}[s_{\boldsymbol{X}}(\boldsymbol{X})] = 0,\quad 
    \mathbb{E}_\mathcal{Q}[s_{A \mid \boldsymbol{X}}(A \mid \boldsymbol{X})\mid \boldsymbol{X}] = 0,\quad
    \mathbb{E}_\mathcal{Q}[s_{Y \mid A, \boldsymbol{X}}(Y \mid A, \boldsymbol{X}) \mid A, \boldsymbol{X}] = 0.
\]
Lastly, let $\mathcal{T}_\mathcal{Q}$ denote the tangent space (closure in $L^2_0(Q)$ of finite linear combinations of such scores). \\

\noindent {\bf Pathwise Derivative of $\mu(a) = \mathbb{E}_\mathcal{Q}[w(\boldsymbol{X}) g_a(\boldsymbol{X})]$:} Let $Q_\varepsilon$ be a submodel with score $s$. We write
\[
    g_{a,\varepsilon}(\boldsymbol{x}) = \mathbb{E}_{\mathcal{Q}_\varepsilon}[Y \mid A = a, \boldsymbol{X} = \boldsymbol{x}].
\]
Since $w(\cdot)$ is treated as fixed (i.e., design known), the map $\Psi(\mathcal{Q}) = \mathbb{E}_\mathcal{Q}[w(\boldsymbol{X}) g_a(\boldsymbol{X})]$ is smooth and
\[
    \left.\frac{\partial}{\partial \varepsilon} \Psi(\mathcal{Q}_\varepsilon)\right|_{\varepsilon = 0} = \underbrace{\mathbb{E}_\mathcal{Q}\left[w(\boldsymbol{X}) \dot{g}_a(\boldsymbol{X})\right]}_{\text{perturb }p_{Y \mid A, \boldsymbol{X}}} + \underbrace{\mathbb{E}_\mathcal{Q}\left[w(\boldsymbol{X}) g_a(\boldsymbol{X}) s_{\boldsymbol{X}}(\boldsymbol{X})\right]}_{\text{perturb }p_{\boldsymbol{X}}},
\]
where $\dot{g}_a(\boldsymbol{x}) = \left. \tfrac{\partial}{\partial\varepsilon}g_{a,\varepsilon}(\boldsymbol{x})\right|_0$. \\

\noindent {\bf Contribution from $p_{Y \mid A, \boldsymbol{X}}$:} For fixed $\boldsymbol{x}$, the standard identity for differentiating a conditional mean gives
\[
    \dot{g}_a(\boldsymbol{x}) = \mathbb{E}_\mathcal{Q}\left\{[Y - g_a(\boldsymbol{x})] s_{Y \mid A, \boldsymbol{X}}(Y \mid A = a, \boldsymbol{X} = \boldsymbol{x}) \mid A = a, \boldsymbol{X} = \boldsymbol{x}\right].
\]
Multiplying by $w(\boldsymbol{x})$, integrating over $\boldsymbol{x}$, and using iterated expectations,
\[
    \mathbb{E}_\mathcal{Q}\left[w(\boldsymbol{X}) \dot{g}_a(\boldsymbol{X})\right] = \mathbb{E}_\mathcal{Q}\left\{w(\boldsymbol{X}) \frac{\mathbb{I}(A = a)}{e^{\mathcal{Q}}_a(\boldsymbol{x})} [Y - g_a(\boldsymbol{X})] s_{Y\mid A, A}(Y\mid A, \boldsymbol{X})\right\}.
\]
Hence the Riesz representer for this component is
\[
    \phi_Y(\mathcal{O}) = w(\boldsymbol{X}) \frac{\mathbb{I}(A = a)}{e^{\mathcal{Q}}_a(\boldsymbol{x})} [Y - g_a(\boldsymbol{X})].
\]

\noindent {\bf Contribution from $p_{\boldsymbol{X}}$:} The second term is already in the inner product form,
\[
    \mathbb{E}_\mathcal{Q}\left[w(\boldsymbol{X}) g_a(\boldsymbol{X}) s_{\boldsymbol{X}}(\boldsymbol{X})\right] = \mathbb{E}_\mathcal{Q}\left\{[w(\boldsymbol{X}) g_a(\boldsymbol{X}) - \mu(a)] s_{\boldsymbol{X}}(\boldsymbol{X})\right\},
\]
since $\mathbb{E}_\mathcal{Q}[s_{\boldsymbol{X}}(\boldsymbol{X})] = 0$. Thus the Riesz representer for perturbations of $p_{\boldsymbol{X}}$ is
\[
    \phi_{\boldsymbol{X}}(\mathcal{O}) = w(\boldsymbol{X}) g_a(\boldsymbol{X}) - \mu(a).
\]

\noindent {\bf Orthogonality to $s_{A\mid \boldsymbol{X}}$:} We also need $\mathbb{E}_\mathcal{Q}[\phi s_{A \mid \boldsymbol{X}}(A \mid \boldsymbol{X})] = 0$. Using $\mathbb{E}_\mathcal{Q}[Y - g_a(\boldsymbol{X}) \mid A, \boldsymbol{X}] = 0$,
\[
    \mathbb{E}_\mathcal{Q}\left\{w(\boldsymbol{X}) \frac{\mathbb{I}(A = a)}{e^{\mathcal{Q}}_a(\boldsymbol{x})}[Y - g_a(\boldsymbol{X})] s_{A\mid \boldsymbol{X}}\right\} = 0,
\]
and $\mathbb{E}_\mathcal{Q}\left\{[w(\boldsymbol{X}) g_a(\boldsymbol{X}) - \mu(a)] s_{A \mid \boldsymbol{X}}\right\} = 0$ by $\mathbb{E}_\mathcal{Q}[s_{A\mid \boldsymbol{X}}\mid \boldsymbol{X}] = 0$. Hence the sum is orthogonal to $s_{A \mid \boldsymbol{X}}$. \\

\noindent {\bf EIF under $\mathcal{Q}$:} Combining the two components, the efficient influence function under the sample law, $\mathcal{Q}$, is
\[
\begin{aligned}
    \phi_a^{\mathcal{Q}}(\mathcal{O}) &= \underbrace{w(\boldsymbol{X}) \frac{\mathbb{I}(A = a)}{e^{\mathcal{Q}}_a(\boldsymbol{x})} [Y - g_a(\boldsymbol{X})]}_{\text{conditional\ outcome perturbation}} + \underbrace{\left[w(\boldsymbol{X}) g_a(\boldsymbol{X}) - \mu(a)\right]}_{\text{marginal }\boldsymbol{X}\text{ perturbation}} \\[2ex]
    &= w(\boldsymbol{X}) \left\{\frac{\mathbb{I}(A = a)}{e^{\mathcal{Q}}_a(\boldsymbol{x})} [Y - g_a(\boldsymbol{X})] + g_a(\boldsymbol{X})\right\} - \mu(a).
\end{aligned}
\]
This $\phi_a^{\mathcal{Q}} \in \mathcal{T}_\mathcal{Q}$ satisfies $\left. \frac{\partial}{\partial \varepsilon}\Psi(\mathcal{Q}_\varepsilon)\right|_0 = \mathbb{E}_\mathcal{Q}[\phi_a^{\mathcal{Q}}(\mathcal{O}) s(\mathcal{O})]$ for every regular submodel $\mathcal{Q}_\varepsilon$, and it is the unique element of $\mathcal{T}_\mathcal{Q}$ with minimal variance (canonical gradient). \\

\noindent {\bf Relation to a $\mathcal{P}$-law EIF:} As $A, \boldsymbol{X}, Y$ are only observed when $S = 1$, an EIF under the `full law,' $\mathcal{P}$, necessarily lives on the larger observed-data space $\mathcal{D} = (S, S\boldsymbol{X}, SA, SY)$. Deriving that EIF requires specifying whether $\pi(\boldsymbol{x}) = \Pr(S = 1 \mid \boldsymbol{X} = \boldsymbol{x})$, or $\pi_a(\boldsymbol{x}) = \Pr(S = 1 \mid A = a, \boldsymbol{X} = \boldsymbol{x})$, is:
\begin{itemize}
    \item {\bf Design-Known (e.g., known survey weights):} Here, we can embed the $\mathcal{Q}$-law result in $\mathcal{P}$ via the identity $\mu(a) = \mathbb{E}_\mathcal{P}\left[S w(\boldsymbol{X}) g_a(\boldsymbol{X})\right]$ and obtain an EIF in $\mathcal{P}$ that reduces, among $S = 1$ units, to $\phi_a^{\mathcal{Q}}$.
    \item {\bf Unknown and Unconstrained:} With $A,\boldsymbol{X}$ unobserved when $S = 0$, $\pi(\cdot)$ is not estimable from the internal data alone, so the canonical gradient under $\mathcal{P}$ includes the selection tangent component and is therefore model-dependent. In this setting, it is both standard and cleaner to work under $\mathcal{Q}$ (where $A, \boldsymbol{X}, Y$ are observed) and treat $w(\cdot)$ as external/known.
\end{itemize}
Given that we only observe $A, \boldsymbol{X}, Y$ when $S = 1$ in our design, we base estimation and inference on the $\mathcal{Q}$-law EIF above, which emits the usual doubly-robust estimator and influence function-based standard errors.

\subsubsection{Doubly-Robust Estimator}

Given this form of the EIF, an augmented inverse probability weighted (AIPW; i.e., doubly-robust) estimator is given by
\[
\begin{aligned}
    \hat{\mu}_{\rm DR}(a) &= \frac{1}{n}\sum_{i = 1}^n \left\{w(\boldsymbol{X}_i) \hat{g}_a(\boldsymbol{X}_i) + w(\boldsymbol{X}_i) \frac{\mathbb{I}(A_i = a)}{\hat{e}^\mathcal{Q}_a(\boldsymbol{X}_i)} [Y_i - \hat{g}_a(\boldsymbol{X}_i)]\right\} \\[2ex]
    &= \frac{1}{n}\sum_{i = 1}^n w(\boldsymbol{X}_i)\left\{\hat{g}_a(\boldsymbol{X}_i) + \frac{\mathbb{I}(A_i = a)}{\hat{e}^\mathcal{Q}_a(\boldsymbol{X}_i)} [Y_i - \hat{g}_a(\boldsymbol{X}_i)]\right\}.
\end{aligned}
\]
Under standard rate/positivity conditions (and with $w$ treated as fixed/known), we have that
\[
    \sqrt{n}[\hat{\mu}_{\rm DR}(a) - \mu(a)] \overset{d}{\to} \mathcal{N}(0, \mathbb{V}_\mathcal{Q}[\phi_a^{\mathcal{Q}}]),
\]
with the variance estimated by the sample variance of $\phi_a^{\mathcal{Q}}(O_i;\hat{\theta})$ with plug-in $\hat{g}_a$ and $\hat{e}_a$. \\

\subsubsection{Proof of Double Robustness}

From above, define the doubly-robust estimator under the sample law $\mathcal{Q} = \Pr(\ \cdot\mid S = 1)$ as
\[
    \hat{\mu}_{\mathrm{DR}}(a) = \frac{1}{n}\sum_{i = 1}^n w(\boldsymbol{X}_i)\left\{\hat{g}_a(\boldsymbol{X}_i) + \frac{\mathbb{I}(A_i = a)}{\hat{e}^{\mathcal{Q}}_a(\boldsymbol{X}_i)} \left(Y_i - \hat{g}_a(\boldsymbol{X}_i)\right)\right\}; \quad w(\boldsymbol{X})=\frac{\Pr(S = 1)}{\pi(\boldsymbol{X})},
\]
where $\hat{g}_a$ estimates $g_a(\boldsymbol{X}) = \mathbb{E}_{\mathcal{Q}}(Y \mid A = a, \boldsymbol{X})$, $\hat{e}^{\mathcal{Q}}_a$ estimates $e^{\mathcal{Q}}_a(\boldsymbol{X}) = \mathcal{Q}(A = a \mid \boldsymbol{X})$, and $\pi(\boldsymbol{X}) = \Pr(S = 1 \mid \boldsymbol{X})$ and $\Pr(S = 1)$ are known from the design. The estimator is {\it doubly robust} in the sense that it is consistent for $\mu(a)$ if either (1) the outcome regression, $g_a$, is correctly specified, regardless of $e^{\mathcal{Q}}_a$, or (2) the sample propensity, $e^{\mathcal{Q}}_a$, is correctly specified, regardless of $g_a$. We assume positivity, so all ratios are well-defined: $e^{\mathcal{Q}}_a(\boldsymbol{X}) > 0$ and $\pi(\boldsymbol{X}) > 0$ on the support of $(A, \boldsymbol{X})$.

\paragraph{Case 1. Correct Outcome Model:} 

Assume $\hat{g}_a(\boldsymbol{X}) \xrightarrow{p} g_a(\boldsymbol{X})$ and $w(\boldsymbol{X})$ known. Allow $\hat{e}^{\mathcal{Q}}_a$ to converge to an arbitrary (possibly misspecified) limit, $e^{\mathcal{Q},*}_a$. Consider the $\mathcal{Q}$-expectation of the summand:
\[
    \mathbb{E}_{\mathcal{Q}}\left[w(\boldsymbol{X}) \left\{g_a(\boldsymbol{X}) + \frac{\mathbb{I}(A = a)}{e^{\mathcal{Q},*}_a(\boldsymbol{X})} \left(Y - g_a(\boldsymbol{X})\right)\right\}\right].
\]
By the law of iterated expectations under $\mathcal{Q}$, and pulling out functions of $\boldsymbol{X}$,
\[
\begin{aligned}
    \mathbb{E}_{\mathcal{Q}}\left[w(\boldsymbol{X}) \left\{g_a(\boldsymbol{X}) + \frac{\mathbb{I}(A = a)}{e^{\mathcal{Q},*}_a(\boldsymbol{X})} \left(Y - g_a(\boldsymbol{X})\right)\right\}\right] &= \mathbb{E}_{\mathcal{Q}}\left[w(\boldsymbol{X}) g_a(\boldsymbol{X})\right] + \mathbb{E}_{\mathcal{Q}}\left[w(\boldsymbol{X})\frac{\mathbb{I}(A = a)}{e^{\mathcal{Q},*}_a(\boldsymbol{X})} \left(Y - g_a(\boldsymbol{X})\right)\right] \\
    &= \mathbb{E}_{\mathcal{Q}}\left[w(\boldsymbol{X}) g_a(\boldsymbol{X})\right] + \mathbb{E}_{\mathcal{Q}}\left[w(\boldsymbol{X}) \frac{1}{e^{\mathcal{Q},*}_a(\boldsymbol{X})} \underbrace{\mathbb{E}_{\mathcal{Q}}\left\{\left. \mathbb{I}(A = a)\left(Y - g_a(\boldsymbol{X})\right) \right| \boldsymbol{X}\right\}}_{e^{\mathcal{Q}}_a(\boldsymbol{X}) g_a(\boldsymbol{X}) - e^{\mathcal{Q}}_a(\boldsymbol{X}) g_a(\boldsymbol{X}) = 0}\right] \\
    &=\mathbb{E}_{\mathcal{Q}}\left[w(\boldsymbol{X}) g_a(\boldsymbol{X})\right].
\end{aligned}
\]
Using the change-of-measure identity $\mathbb{E}_{\mathcal{Q}}[w(\boldsymbol{X}) h(\boldsymbol{X})] = \mathbb{E}_{\mathcal{P}}[h(\boldsymbol{X})]$, we have that
\[
    \mathbb{E}_{\mathcal{Q}}\left[w(\boldsymbol{X}) g_a(\boldsymbol{X})\right]
=\mathbb{E}_{\mathcal{P}}\left[g_a(\boldsymbol{X})\right].
\]
For a causal target, invoke SUTVA (iv) and mean selection exchangeability for potential outcomes (iii') to equate $g_a(\boldsymbol{X})$ with $\mathbb{E}_{\mathcal{P}}[Y^a \mid A = a, \boldsymbol{X}]$, and then mean exchangeability (v), so that
\[
\begin{aligned}
    \mathbb{E}_{\mathcal{P}}[g_a(\boldsymbol{X})] &= \mathbb{E}_{\mathcal{P}}\left[\mathbb{E}_{\mathcal{P}}(Y^a \mid A = a, \boldsymbol{X}) \right] \\
    &= \mathbb{E}_{\mathcal{P}}\left[\mathbb{E}_{\mathcal{P}}(Y^a \mid \boldsymbol{X})\right] \\
    &= \mathbb{E}_{\mathcal{P}}[Y^a] \\
    &= \mu(a).  
\end{aligned}
\]
Thus, with $g_a$ correctly specified, $\hat{\mu}_{\mathrm{DR}}(a)$ is consistent, regardless of $e^{\mathcal{Q}}_a$.

\paragraph{Case 2. Correct Propensity (and Selection) Model:} 

Now assume $\hat{e}^{\mathcal{Q}}_a(\boldsymbol{X}) \xrightarrow{p} e^{\mathcal{Q}}_a(\boldsymbol{X})$ and $\pi(\boldsymbol{X})$ known. Let $g_a^*(\boldsymbol{X})$ denote an arbitrary (possibly misspecified) limit of $\hat{g}_a$. Then, conditioning on $\boldsymbol{X}$,
\[
    \mathbb{E}_{\mathcal{Q}}\left[\left.\frac{\mathbb{I}(A = a)}{e^{\mathcal{Q}}_a(\boldsymbol{X})}\left(Y - g_a^*(\boldsymbol{X})\right) \right| \boldsymbol{X}\right] = \underbrace{\mathbb{E}_{\mathcal{Q}}[Y \mid A = a, \boldsymbol{X}]}_{= g_a(\boldsymbol{X})} - g_a^*(\boldsymbol{X}).
\]
Therefore,
\[
\begin{aligned}
    \mathbb{E}_{\mathcal{Q}}\left[w(\boldsymbol{X})\left\{g_a^*(\boldsymbol{X}) + \frac{\mathbb{I}(A = a)}{e^{\mathcal{Q}}_a(\boldsymbol{X})}\left(Y - g_a^*(\boldsymbol{X})\right)\right\}\right] &= \mathbb{E}_{\mathcal{Q}}\left[w(\boldsymbol{X})g_a(\boldsymbol{X})\right] \\
    &= \mathbb{E}_{\mathcal P}\left[g_a(\boldsymbol{X})\right] \\
    &=\mu(a),
\end{aligned}
\]
where the last equalities use the same change-of-measure (and, for the causal target, assumptions (iii')–(v)) as above. Hence, with $e^{\mathcal{Q}}_a$ correctly specified, $\hat{\mu}_{\mathrm{DR}}(a)$ is consistent, regardless of $g_a$.

\newpage 


\section{Additional Simulation Results}
\label{sec:b}

\setcounter{table}{0}
\renewcommand{\thetable}{B\arabic{table}}

\setcounter{figure}{0}
\renewcommand{\thefigure}{B\arabic{figure}}

\subsection{Sensitivity Analysis: Single-Stage Unequal Probability Sampling}

{\small
\begin{longtable}{@{\extracolsep{\fill}}lrrrrrrr}
\caption{Sensitivity analysis comparing the proposed methods to the oracle estimator and standard analytic methods for calculating the population average controlled difference (ACD) in $Y$ across $A$, generating population-level data from a single stratum and single cluster. Results presented are the true ACD (ACD) estimated ACD (Est.~ACD), relative bias (Rel.~Bias), average of the estimated standard error (ASE), Monte Carlo standard error (MCSE), mean squared error (MSE) and coverage probability (Cov.) averaged over 500 samples from the population generated under each setting.} 
\label{tab:sensitivity_srs} \\
\toprule
Method & ACD & Est.~ACD & Rel.~Bias & ASE & MCSE & MSE & Cov. \\ 
\midrule
\multicolumn{8}{l}{\bf Setting 1: No Bias Expected ($\tau_{\boldsymbol{X}} = 0$, $\beta_A = 0$, $\beta_{\boldsymbol{X}} = 0$)} \\
\midrule
Oracle Estimator & 1.027 & 1.027 & 0.000 & 0.109 & 0.106 & 0.011 & 0.956 \\ 
Simple Reg. & 1.027 & 1.023 & -0.005 & 0.196 & 0.193 & 0.037 & 0.942 \\ 
Multiple Reg. & 1.027 & 1.028 & 0.000 & 0.109 & 0.107 & 0.011 & 0.952 \\ 
IPTW Estimator & 1.027 & 1.030 & 0.002 & 0.110 & 0.275 & 0.076 & 0.568 \\ 
Survey-Weighted Multiple Reg. & 1.027 & 1.028 & 0.000 & 0.109 & 0.107 & 0.011 & 0.962 \\
\end{longtable}
}

\newpage

\subsection{Sensitivity Analysis: Propensity/Outcome Model Misspecification}

{\small
\begin{longtable}{@{\extracolsep{\fill}}lrrrrrrr}
\caption{Sensitivity analysis comparing the proposed methods to the oracle estimator for calculating the population average controlled difference (ACD) in $Y$ across $A$ for varying degrees of propensity and/or outcome model misspecification. Results presented are the true ACD (ACD) estimated ACD (Est.~ACD), relative bias (Rel.~Bias), average of the estimated standard error (ASE), Monte Carlo standard error (MCSE), mean squared error (MSE) and coverage probability (Cov.) averaged over 500 samples from the population generated under each setting.} 
\label{tab:sensitivity_misspecification} \\
\toprule
Method & ACD & Est.~ACD & Rel.~Bias & ASE & MCSE & MSE & Coverage \\ 
\midrule
\multicolumn{8}{l}{Neither Misspecified} \\
\midrule
Oracle Estimator & 0.997 & 0.995 & -0.002 & 0.096 & 0.090 & 0.008 & 0.952 \\ 
Outcome Modeling and Direct Standardization & 0.997 & 0.991 & -0.006 & 0.100 & 0.059 & 0.004 & 0.990 \\ 
Inverse Probability Weighting 1 & 0.997 & 0.994 & -0.003 & 0.162 & 0.144 & 0.021 & 0.964 \\ 
Inverse Probability Weighting 2 & 0.997 & 1.007 & 0.010 & 0.219 & 0.218 & 0.048 & 0.968 \\ 
Augmented Inverse Probability Weighting & 0.997 & 0.996 & -0.001 & 0.158 & 0.135 & 0.018 & 0.962 \\
\bottomrule
\end{longtable}
}

\newpage

\subsection{Sensitivity Analysis: Limited Overlap}

{\small 
\begin{longtable}{@{\extracolsep{\fill}}lrrrrrrr}
\caption{Sensitivity analysis comparing the proposed methods to the oracle estimator for calculating the population average controlled difference (ACD) in $Y$ across $A$ for varying degrees of propensity score overlap ($\tau_A = 0.5$; good overlap versus 2.0; poor overlap). Results presented are the true ACD (ACD) estimated ACD (Est.~ACD), relative bias (Rel.~Bias), average of the estimated standard error (ASE), Monte Carlo standard error (MCSE), mean squared error (MSE) and coverage probability (Cov.) averaged over 500 samples from the population generated under each setting.} 
\label{tab:sensitivity_overlap} \\
\toprule
Method & ACD & Est.~ACD & Rel.~Bias & ASE & MCSE & MSE & Cov. \\ 
\midrule
\multicolumn{8}{l}{\bf Setting 1: Good Overlap $(\tau_A = 0.5)$} \\
\midrule
Oracle Estimator & 1.010 & 1.006 & -0.004 & 0.095 & 0.094 & 0.009 & 0.954 \\ 
Outcome Modeling and Direct Standardization & 1.010 & 1.008 & -0.002 & 0.113 & 0.095 & 0.009 & 0.982 \\ 
Inverse Probability Weighting 1 & 1.010 & 1.017 & 0.007 & 0.153 & 0.148 & 0.022 & 0.960 \\ 
Inverse Probability Weighting 2 & 1.010 & 1.038 & 0.028 & 0.213 & 0.223 & 0.050 & 0.986 \\ 
Augmented Inverse Probability Weighting & 1.010 & 1.018 & 0.008 & 0.150 & 0.144 & 0.021 & 0.968 \\ 
\midrule
\multicolumn{8}{l}{\bf Setting 2: Poor Overlap $(\tau_A = 2.0)$} \\
\midrule
Oracle Estimator & 0.999 & 1.013 & 0.013 & 0.108 & 0.106 & 0.011 & 0.952 \\ 
Outcome Modeling and Direct Standardization & 0.999 & 1.014 & 0.014 & 0.129 & 0.109 & 0.012 & 0.972 \\ 
Inverse Probability Weighting 1 & 0.999 & 0.904 & -0.095 & 0.542 & 0.966 & 0.940 & 0.918 \\ 
Inverse Probability Weighting 2 & 0.999 & 1.024 & 0.025 & 0.352 & 0.714 & 0.510 & 0.906 \\ 
Augmented Inverse Probability Weighting & 0.999 & 0.990 & -0.010 & 0.223 & 0.414 & 0.171 & 0.962 \\ 
\bottomrule
\end{longtable}
}

\newpage

\subsection{Sensitivity Analysis: Estimated Selection Mechanism}

{\small
\begin{longtable}{@{\extracolsep{\fill}}lrrrrrrr}
\caption{Sensitivity analysis comparing the proposed methods to the oracle estimator and standard analytic methods for calculating the population average controlled difference (ACD) in $Y$ across $A$, treating the selection mechanism as unknown for the proposed methods. Results presented are the true ACD (ACD) estimated ACD (Est.~ACD), relative bias (Rel.~Bias), average of the estimated standard error (ASE), Monte Carlo standard error (MCSE), mean squared error (MSE) and coverage probability (Cov.) averaged over 500 samples from the population generated under each setting.} 
\label{tab:sensitivity_estsel} \\
\toprule
Method & ACD & Est.~ACD & Rel.~Bias & ASE & MCSE & MSE & Cov. \\ 
\midrule
\multicolumn{8}{l}{\bf Setting 1: No Bias Expected ($\tau_{\boldsymbol{X}} = 0$, $\beta_A = 0$, $\beta_{\boldsymbol{X}} = 0$)} \\
\midrule
Oracle Estimator & 1.024 & 1.026 & 0.002 & 0.109 & 0.109 & 0.012 & 0.954 \\ 
Simple Reg. & 1.024 & 1.018 & -0.006 & 0.205 & 0.204 & 0.042 & 0.936 \\ 
Multiple Reg. & 1.024 & 1.026 & 0.001 & 0.109 & 0.109 & 0.012 & 0.954 \\ 
IPTW Estimator & 1.024 & 1.022 & -0.003 & 0.111 & 0.281 & 0.079 & 0.576 \\ 
Survey-Weighted Multiple Reg. & 1.024 & 1.025 & 0.001 & 0.110 & 0.109 & 0.012 & 0.956 \\
\end{longtable}
}

\newpage

\subsection{Sensitivity Analysis: Sampling Covariate Omission}

{\small
\begin{longtable}{@{\extracolsep{\fill}}lrrrrrrr}
\caption{Sensitivity analysis comparing the proposed methods to the oracle estimator and standard analytic methods for calculating the population average controlled difference (ACD) in $Y$ across $A$, with the addition of two unmeasured covariates that affect sample selection. Results presented are the true ACD (ACD) estimated ACD (Est.~ACD), relative bias (Rel.~Bias), average of the estimated standard error (ASE), Monte Carlo standard error (MCSE), mean squared error (MSE) and coverage probability (Cov.) averaged over 500 samples from the population generated under each setting.} 
\label{tab:sensitivity_omit} \\
\toprule
Method & ACD & Est.~ACD & Rel.~Bias & ASE & MCSE & MSE & Cov. \\ 
\midrule
\multicolumn{8}{l}{\bf Setting 1: No Bias Expected ($\tau_{\boldsymbol{X}} = 0$, $\beta_A = 0$, $\beta_{\boldsymbol{X}} = 0$)} \\
\midrule
Oracle Estimator & 0.958 & 0.954 & -0.005 & 0.060 & 0.060 & 0.004 & 0.966 \\ 
Simple Reg. & 0.958 & 0.956 & -0.003 & 0.109 & 0.107 & 0.011 & 0.952 \\ 
Multiple Reg. & 0.958 & 0.954 & -0.004 & 0.060 & 0.061 & 0.004 & 0.956 \\ 
IPTW Estimator & 0.958 & 0.957 & -0.002 & 0.060 & 0.121 & 0.015 & 0.692 \\ 
Survey-Weighted Multiple Reg. & 0.958 & 0.972 & 0.014 & 0.139 & 0.168 & 0.028 & 0.896 \\
\end{longtable}
}

\newpage


\section{Additional Data Example Results}
\label{sec:c}

\setcounter{table}{0}
\renewcommand{\thetable}{C\arabic{table}}

\setcounter{figure}{0}
\renewcommand{\thefigure}{C\arabic{figure}}

\vspace{2ex}

\subsection{Main Analysis of Race, SES, and Telomere Length}

\begin{longtable}{lp{1in}p{1.5in}p{1.5in}p{0.5in}}
\caption{Characteristics of the $n = 5,270$ non-Hispanic Black and non-Hispanic White National Health and Nutrition Examination Survey (NHANES) participants in our analytic sample; overall and stratified by self-reported race. Descriptive statistics are reported as median (IQR) for continuous measures or number (\%) for categorical measures. $P$-values denote the univariate tests of significance (Wilcoxon rank sum test, Pearson’s Chi-squared test, or Fisher’s exact test) comparing the distribution of each measure among non-Hispanic Black versus non-Hispanic White participants.} 
\label{tab:nhanesdesc} \\
\toprule
\textbf{Characteristic} & \textbf{Overall}, N = 5,298\textsuperscript{\textit{1}} & \textbf{Non-Hispanic White}, N = 3,965\textsuperscript{\textit{1}} & \textbf{Non-Hispanic Black}, N = 1,333\textsuperscript{\textit{1}} & \textbf{p-value}\textsuperscript{\textit{2}} \\
\midrule
Telomere Length, Mean T/S & 1.00 (0.84, 1.18) & 0.98 (0.83, 1.16) & 1.05 (0.88, 1.25) & $<$0.001 \\
Age, Years & 50 (35, 67) & 52 (35, 70) & 45 (34, 62) & $<$0.001 \\
Sex &  &  &  & 0.4 \\
\quad Male & 2,574 (49\%) & 1,939 (49\%) & 635 (48\%) &  \\
\quad Female & 2,724 (51\%) & 2,026 (51\%) & 698 (52\%) &  \\
Education &  &  &  &  \\
\bottomrule
\end{longtable}
\begin{minipage}{\linewidth}
\textsuperscript{\textit{1}}Median (IQR); n (\%)\\
\textsuperscript{\textit{2}}Wilcoxon rank sum test; Pearson's Chi-squared test; Fisher's exact test\\
\end{minipage}

\begin{longtable}{@{\extracolsep{\fill}}lcccccc}
\caption{Results from our (a) unweighted and (b) survey-weighted `propensity for race' models. Propensity scores were constructed by fitting a logistic regression of self-reported race (Black = 1, White = 0) on twelve socioeconomic covariates. Results presented are the odds ratios (OR), corresponding 95\% confidence intervals (95\% CI) and $p$-values. Em-dashes (—) denote the reference group.} 
\label{tab:nhanespropscore} \\
\toprule
& \multicolumn{3}{c}{\bf a. Unweighted} & \multicolumn{3}{c}{\bf b. Survey-Weighted} \\
\cmidrule(lr){2-4} \cmidrule(lr){5-7}
\textbf{Characteristic} & \textbf{OR}\textsuperscript{\textit{1}} & \textbf{95\% CI}\textsuperscript{\textit{1}} & \textbf{p-value} & \textbf{OR}\textsuperscript{\textit{1}} & \textbf{95\% CI}\textsuperscript{\textit{1}} & \textbf{p-value} \\ 
\midrule
Education &  &  &  &  &  &  \\
\quad High School or GED & — & — &  & — & — &  \\
\quad Some College & 0.75 & 0.64, 0.89 & $<$0.001 & 0.80 & 0.67, 0.96 & 0.016 \\
\quad College Graduate & 0.49 & 0.39, 0.61 & $<$0.001 & 0.49 & 0.38, 0.62 & $<$0.001 \\ 
\quad Refused/Unknown & 0.96 & 0.12, 5.42 & $>$0.9 & 0.29 & 0.02, 3.60 & 0.3 \\
\bottomrule
\end{longtable}
\begin{minipage}{\linewidth}
\textsuperscript{\textit{1}}OR = Odds Ratio, CI = Confidence Interval\\
\end{minipage}

\begin{figure}[!ht]
    \centerline{\includegraphics[width=\textwidth]{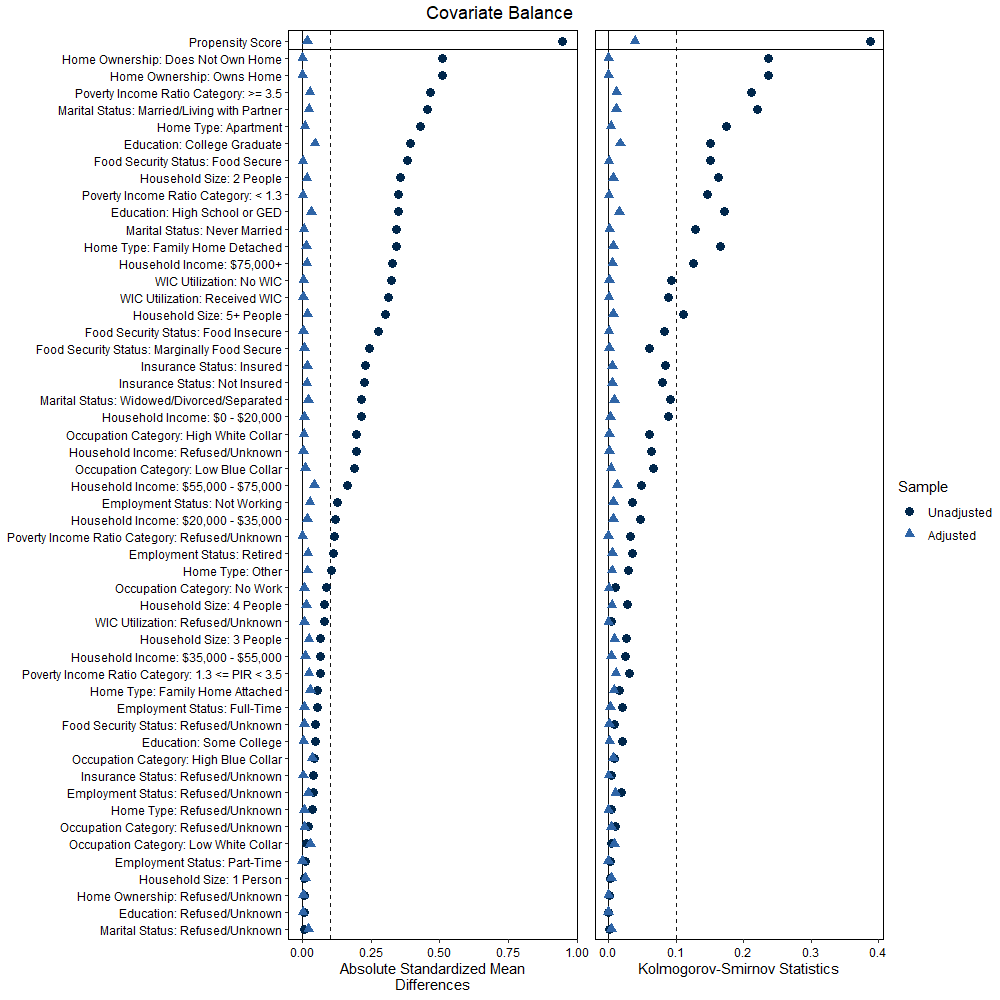}}
    \caption{Absolute standardized mean differences and Kolmogorov-Smirnov statistics for the distribution of socioeconomic status factors before and after adjustment by an unweighted propensity score constructed via a generalized linear model on self-reported race. Dark blue circles represent pre-adjustment statistics, while light blue triangles represent post-adjustment differences. The solid vertical line represents a reference value of 0.0, while the dotted vertical line represents a chosen threshold value of 0.1.}
    \label{fig:smd_unweighted}
\end{figure}

\begin{figure}[!ht]
    \centerline{\includegraphics[width=\textwidth]{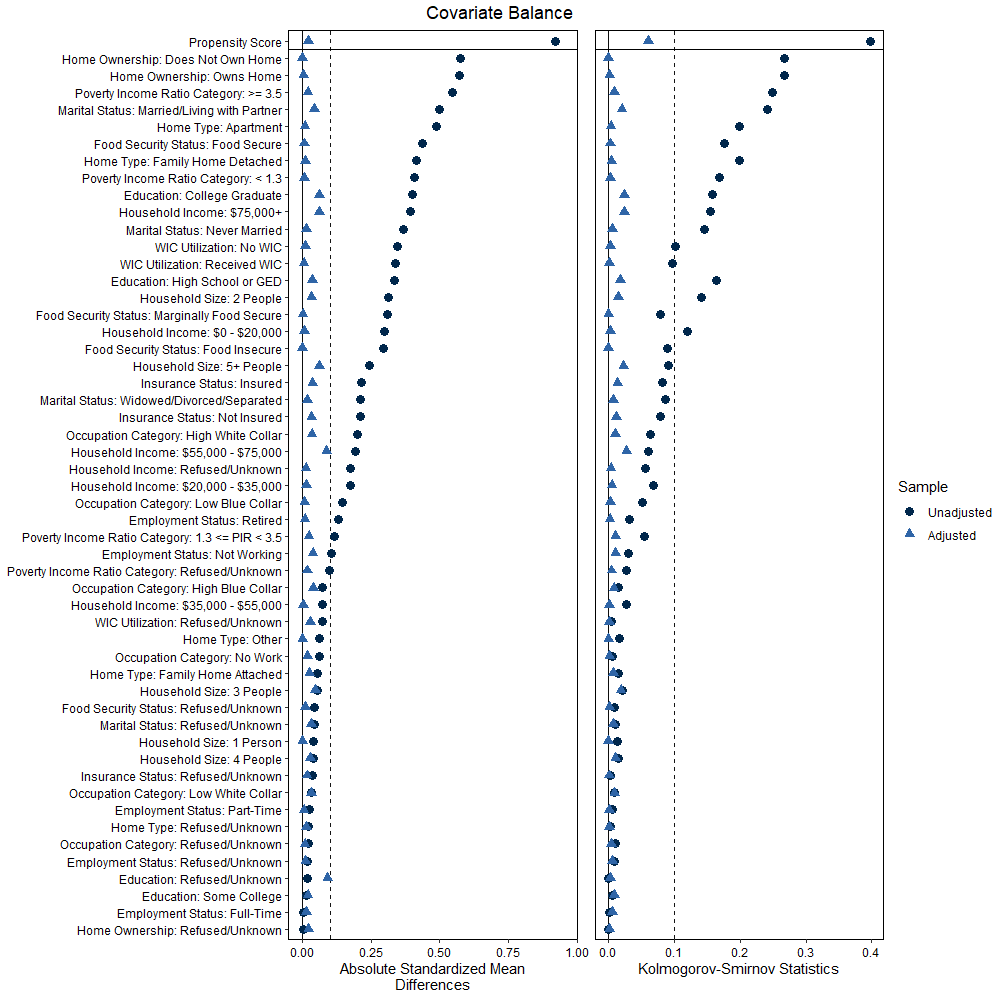}}
    \caption{Absolute standardized mean differences and Kolmogorov-Smirnov statistics for the distribution of socioeconomic status factors before and after adjustment by a survey-weighted propensity score constructed via a generalized linear model on self-reported race. Dark blue circles represent pre-adjustment statistics, while light blue triangles represent post-adjustment differences. The solid vertical line represents a reference value of 0.0, while the dotted vertical line represents a chosen threshold value of 0.1.}
    \label{fig:smd_weighted}
\end{figure}

\clearpage

\subsection{Supplemental Analysis of Lead Exposure, SES, and Telomere Length}

In a supplementary analysis, we present a different example from NHANES which may be amenable to a causal interpretation (in the sense that counterfactual/potential outcomes and plausibly changing exposures are more meaningful). To illustrate how the proposed methods can, under stronger assumptions, identify the average difference in potential outcome means ({\it population average treatment effect}; PATE), we study the effect of exposure to a particular environmental contaminant, lead, on telomere length~\citep{zota2015associations}. Lead exposure is associated with negative health outcomes, such as cardiovascular and kidney diseases~\citep{navas2007lead, navas2009blood}, and biological markers of oxidative stress and inflammation~\citep{lin2016systematic}. Recent works suggest that lead exposure may be associated with telomere shortening~\citep{lin2016systematic, wu2012high, zota2015associations}, however, evidence from these studies is mixed and depends on the analytic strategy.  Thus, further exploring the estimated effect of lead exposure on telomere length with our proposed approaches may be of interest.

Blood lead levels were measured at the CDC's National Center for Environmental Health (Atlanta, Georgia) using a simultaneous multi-element atomic absorption spectrometer (SIMAA 6000; PerkinElmer, Norwalk, Connecticut) with Zeeman background correction~\citep{cdc1999lab, cdc2001lab}. The limit of detection (LOD) was 0.3 $\mu$g/dL for blood lead, with < 1\% of study participants having blood lead concentrations below the LOD. Among these patients, we imputed their blood lead concentrations by LOD divided by the square root of 2~\citep{hornung1990estimation}. To define our exposure, we dichotomized LOD-imputed blood lead concentrations at the median $\mu$g/dL to study the difference in potential outcome means between `higher' and `lower' exposure groups. 

We repeated our the setup of our main analysis, treating the dichotomized lead variable as our exposure $(A)$. As before, we fit a propensity score for lead exposure conditional on the twelve socioeconomic indicators $(\boldsymbol{X})$ described in the main text. We further included self-reported race using two strategies -- first as a precision covariate along with age and blood composition (i.e., not in $\boldsymbol{X}$) and second, in a sensitivity analysis, as a part of our confounders, $\boldsymbol{X}$. The results of this analysis are presented in Supplemental Table C3. While the point estimates and findings differed across methods, we found a significant exposure effect for both the traditional and proposed IPW estimators, with an estimated 3.3-5.0\% reduction in telomere length. For the remaining methods, including the proposed outcome modeling and AIPW estimators, we failed to reject the null hypothesis. Further, there were no appreciable differences in these estimates or the corresponding conclusions regarding significance when comparing our results with versus without including self-reported race in the propensity model (where applicable).

\begin{longtable}{@{\extracolsep{\fill}}lccc}
\caption{Population Average Treatment Effect (PATE) estimates and corresponding 95\% confidence intervals (95\% CI) across various analytic approaches for the comparison of effect of lead exposure (high versus low exposure, dichotomized at median $\mu$g/dL) on log-transformed telomere length among $n = 5,267$ participants in the National Health and Nutrition Examination Survey (NHANES). IPTW: Inverse Probability of Treatment Weighting.}
\label{tab:supp_nhanes_results} \\
\toprule
{\bf Method} & {\bf PATE} & \multicolumn{2}{c}{\bf 95\% CI} \\ 
\midrule
Multiple Regression & 0.0002 & -0.0139 & 0.0142 \\ 
IPTW Estimator & -0.0478 & -0.0800 & -0.0156 \\ 
IPTW Estimator w/ Race in Propensity Model & -0.0504 & -0.0809 & -0.0199 \\ 
Survey-Weighted Multiple Regression & -0.0032 & -0.0229 & 0.0165 \\ 
IPTW Multiple Regression & -0.0024 & -0.0157 & 0.0109 \\ 
IPTW Multiple Regression w/ Race in Propensity Model & -0.0029 & -0.0161 & 0.0103 \\ 
IPTW + Survey-Weighted Multiple Regression & -0.0051 & -0.0250 & 0.0149 \\ 
IPTW + Survey-Weighted Multiple Regression w/ Race in Propensity Model & -0.0053 & -0.0255 & 0.0149 \\ 
Weighted IPTW + Survey-Weighted Multiple Regression & -0.0049 & -0.0253 & 0.0155 \\ 
Weighted IPTW + Survey-Weighted Multiple Regression w/ Race in Propensity Model & -0.0052 & -0.0257 & 0.0153 \\ 
Outcome Modeling and Direct Standardization & 0.0001 & -0.0121 & 0.0123 \\ 
Outcome Modeling and Direct Standardization w/ Race in Propensity Model & 0.0001 & -0.0121 & 0.0123 \\ 
Inverse Probability Weighting 1 & -0.0358 & -0.0469 & -0.0248 \\ 
Inverse Probability Weighting 1 w/ Race in Propensity Model & -0.0368 & -0.0479 & -0.0257 \\ 
Inverse Probability Weighting 2 & -0.0328 & -0.0433 & -0.0223 \\ 
Inverse Probability Weighting 2 w/ Race in Propensity Model & -0.0332 & -0.0440 & -0.0224 \\ 
Augmented Inverse Probability Weighting & -0.0007 & -0.0121 & 0.0106 \\ 
Augmented Inverse Probability Weighting w/ Race in Propensity Model & -0.0009 & -0.0124 & 0.0107 \\ 
\bottomrule
\end{longtable}


\end{document}